\documentclass[10pt,preprint]{aastex}
\shorttitle{H$_2$ excitation around AGN}
\shortauthors{Davies et al.}

\newcommand{\kms}{\,\hbox{\hbox{km}\,\hbox{s}$^{-1}$}}
\newcommand{\wm}{\,\hbox{\hbox{W}\,\hbox{m}$^{-2}$}}

\begin{document}

\title{Molecular Hydrogen Excitation around Active Galactic 
Nuclei\footnote{Based on observations at the European Southern
Observatory VLT (69.B-0075).}}

\author{R.I. Davies}
\affil{Max-Planck-Institut f\"ur extraterrestrische Physik, 
Postfach 1312, 85741, Garching, Germany}
%\email{davies@mpe.mpg.de}

\author{A. Sternberg}
\affil{School of Physics and Astronomy, Tel Aviv University, Tel Aviv
69978, Israel}

\author{M.D. Lehnert}
\affil{Max-Planck-Institut f\"ur extraterrestrische Physik, 
Postfach 1312, 85741, Garching, Germany}

\and

\author{L.E. Tacconi-Garman}
\affil{European Southern Observatory, Karl Schwarzschildstrasse 2,
85748 Garching, Germany}

\begin{abstract}

We report $R\sim3000$ Very Large Telescope ISAAC K-band spectroscopy
of the nuclei (i.e. central 100--300\,pc) of nine galaxies hosting an
active galactic nucleus.
For five of these we also present spectra of the circumnuclear region out
to 1\,kpc.
We have measured a number of molecular hydrogen lines in the
$\nu=1-0$, $2-1$, and $3-2$ vibrational transitions, as well as the
Br$\gamma$ and He{\sc\,i} recombination lines, and the Na{\sc\,i}
stellar absorption feature.
Although only three of the galaxies are classified as type~1 Seyferts in
the literature,
broad Br$\gamma$ (FWHM $\gtrsim$1000\,\kms) is seen in seven
of the objects.
The $\nu=1-0$ emission appears thermalised at temperatures
$T\sim1000$\,K.
However, the $\nu=2-1$ and $\nu=3-2$ emission show evidence of being
radiatively excited by far-ultraviolet photons.
The photo-dissociation region models that fit the data best are, as
for the ultraluminous infrared galaxies in \cite{dav03}, those for
which the H$_2$ emission arises 
 in dense clouds illuminated by intense FUV radiation.
The Na{\sc\,i} stellar absorption line is clearly seen in six of the
nuclear spectra of these AGN, indicating the presence of a significant
population of late type stars.
It is possible that these stars are a result of the same episode
of star formation that gave rise to the stars heating the PDRs.
It seems unlikely that the AGN is the dominant source of excitation for the
near infrared H$_2$ emission: in two of the nuclear spectra H$_2$ was
not detected at all, and in general we find no evidence of
suppression of the 2-1\,S(3) line, which may occur in X-ray irradiated
gas.
Our data do not reveal any significant difference between the nuclear and
circumnuclear line ratios, suggesting that the physical conditions of
the dominant excitation mechanism are similar both near the AGN and in the
larger scale environment around it, and that star formation is an
important process even in the central 100\,pc acround AGN.

\end{abstract}


\section{Introduction}
\label{sec:intro}

Molecular hydrogen (H$_2$) emission lines include some of
the brightest lines in the near infrared K-band and are therefore
commonly targeted as diagnostic lines.
Comparing the ratios in several or more rotational and
vibrational
transitions allows one to probe the physical conditions of molecular
clouds which are illuminated by incident far ultraviolet (FUV)
radiation or heated by shocks.
A number of authors have also suggested that X-ray irradiation from
active galactic nuclei (AGN) may
excite bright H$_2$ line emission, a conclusion based primarily on the
ratio of the H$_2$ 1-0\,S(1) line to other emission lines or
quantities \citep{kaw87,kaw90,mou89,mou90}.
\cite{vei97} thought that this could be important, but that
shocks excited by nuclear outflows is a likely mechanism for heating
H$_2$ in many Seyfert~2 galaxies.
\cite{qui99} rule out X-ray heating as the dominant source of H$_2$
emission due to the lack of correlation between 1-0\,S(1) and hard
X-ray flux.
Instead, because they spatially resolve the H$_2$ in a number of
Seyfert~2 galaxies and one Seyfert~1,
they suggest that a variety of different processes are
probably involved depending on the object.
The extended nature of the H$_2$ emission in this Seyfert~1 was
confirmed by \cite{dav05}, who also found extended H$_2$ emission in
other type~1 AGN \citep{dav04a,dav04b}.

In this paper, we address the issue of H$_2$ excitation in and
 around AGN from a different perspective: using the ratios of the
 H$_2$ lines themselves across several vibrational levels to constrain
 the dominant excitation mechanism.
This can be a very powerful tool as shown, for example, by our
 application of H$_2$ excitation models to ultraluminous infrared
 galaxies (ULIRGs) \citep{dav03}.
The evidence above that H$_2$ can be extended in both type~1 and
 type~2 AGN suggests that active star formation, which typically
 occurs over large (0.1-1\,kpc) scales, could be a good
 candidate for its origin.
This premise is strengthened by the increasing evidence for active
 star formation in the tens to hundreds of parsecs around both
 Seyfert~2s \citep{cid04,sto01,gon01} and 
 Seyfert~1s \citep{dav04a,dav04b,dav05}.
We therefore address whether:
(1) the dominant H$_2$ excitation mechanism either close around an AGN or in
 its wider environment is consistent with excitation in
 photo-dissocation regions, as one might expect if it originates from 
star formation; and 
(2) whether suppression of the 2-1\,S(3) can be taken as an indicator
of X-ray irradiated gas in AGN, as has been proposed 
theoretically and observationally \citep{bla87,dra90,kra00}.
To do so, we have undertaken longslit observations of the 9
Seyfert galaxies listed in Table~\ref{tab:basdat}, including both
 type~1 and type~2 Seyferts, although the 
classification is often an indistinct issue.
In order to analyse these, we have divided them into two groups: 
five which are near enough ($\langle D \rangle = 25$\,Mpc) to study
both the nuclear and circumnuclear H$_2$ line ratios separately;
four which are further away ($\langle D \rangle = 75$\,Mpc),
and for which only a nuclear spectrum was extracted.

After describing the observations and data reduction in
Section~\ref{sec:obs}, we discuss how the stellar continuum was
removed and the line fluxes measured in Section~\ref{sec:contfit}.
Notes on the individual objects in our sample are given in
Section~\ref{sec:notes} before addressing the issues of star formation around
the AGN and H$_2$ excitation
in Sections~\ref{sec:starform} and~\ref{sec:poplev} respectively.
Finally, our conclusions are presented in Section~\ref{sec:conc}.

\section{Observations and Data Reduction}
\label{sec:obs}

Our data were obtained as part of the service observing programme at
the {\em Very Large Telescope} between May and August 2002 in seeing
of typically 0.8\arcsec\ %($\pm0.25$\arcsec) 
in the K-band.

\begin{figure}
\epsscale{.18}
\plotone{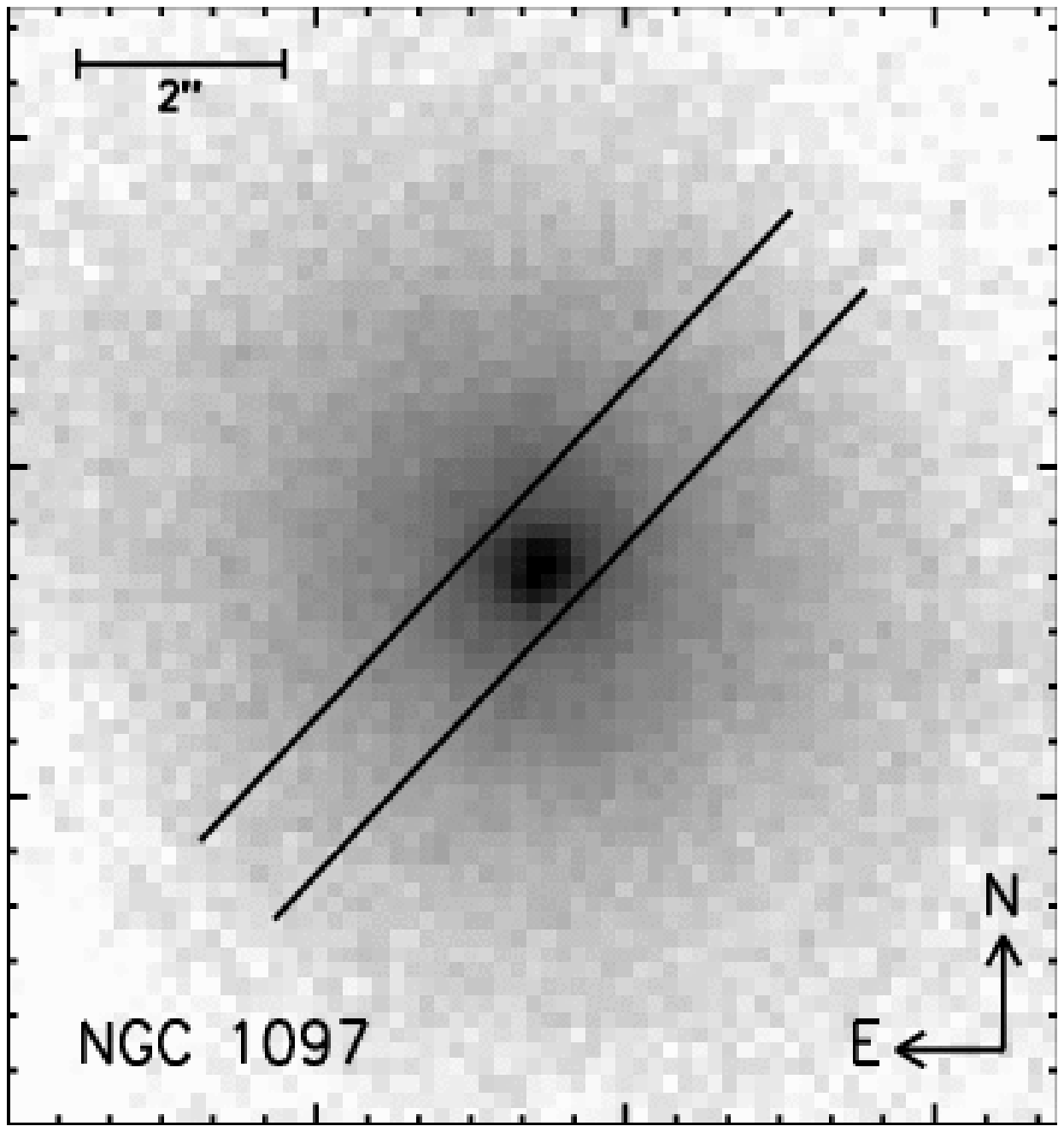}\plotone{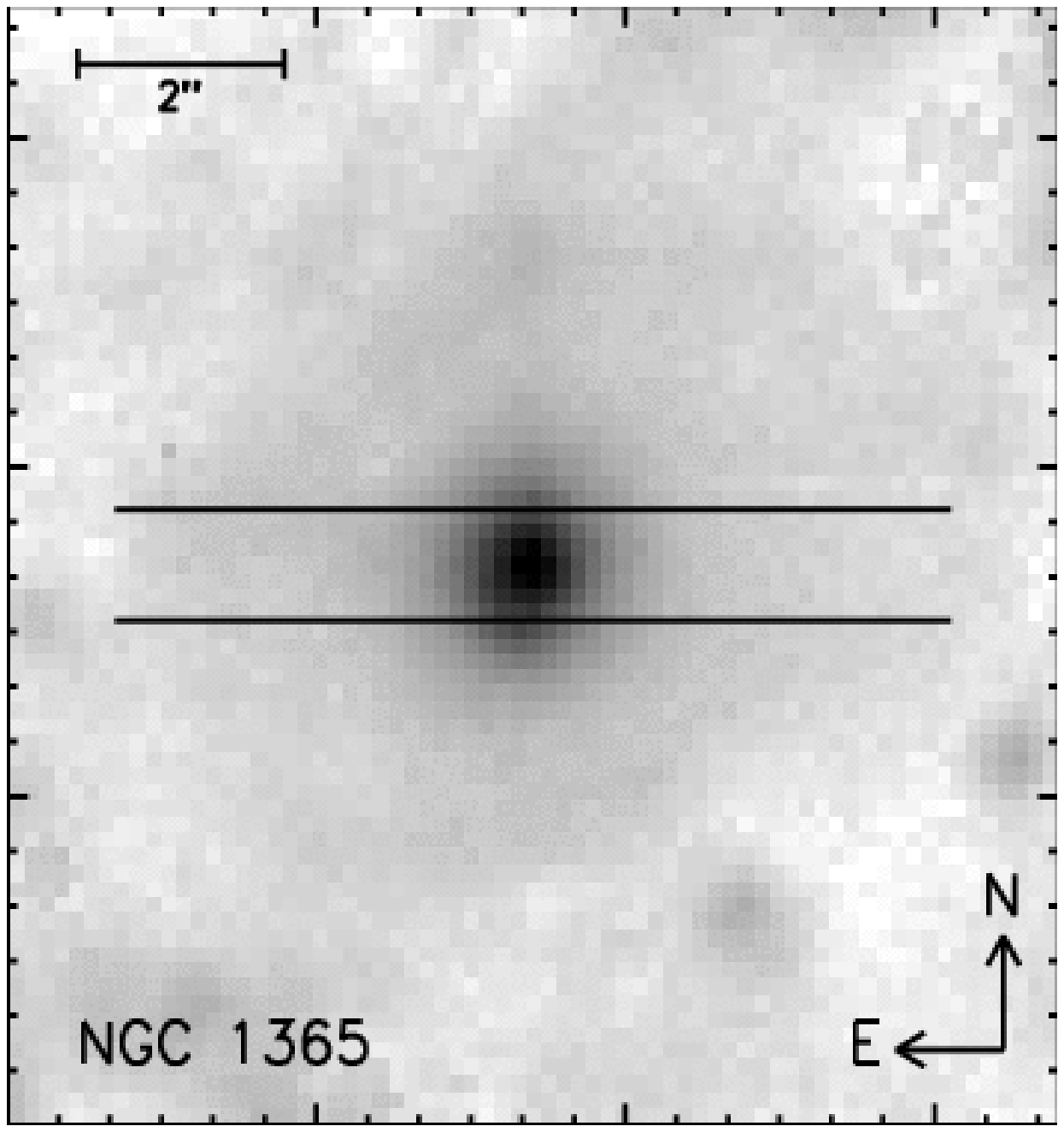}\plotone{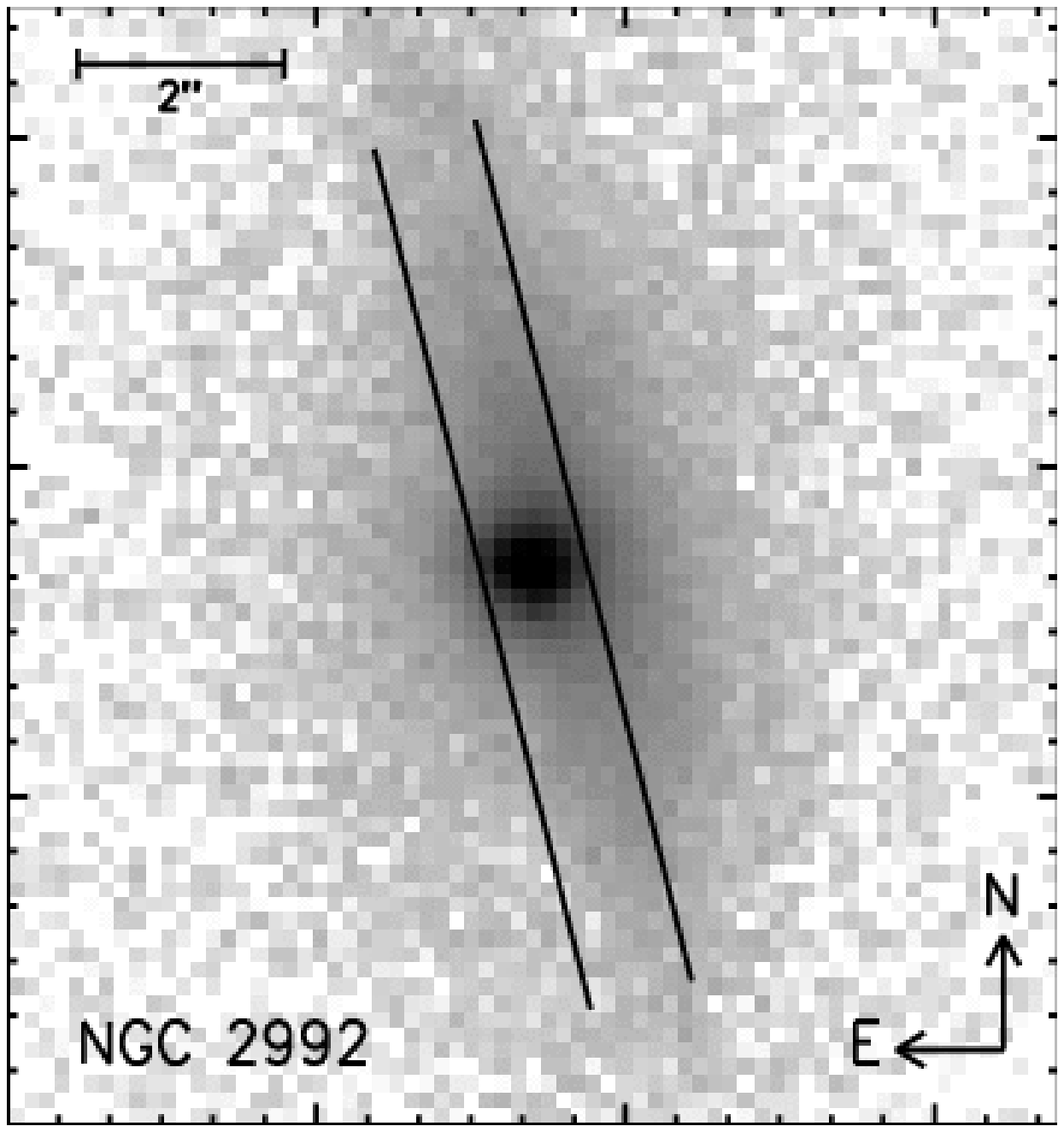}\plotone{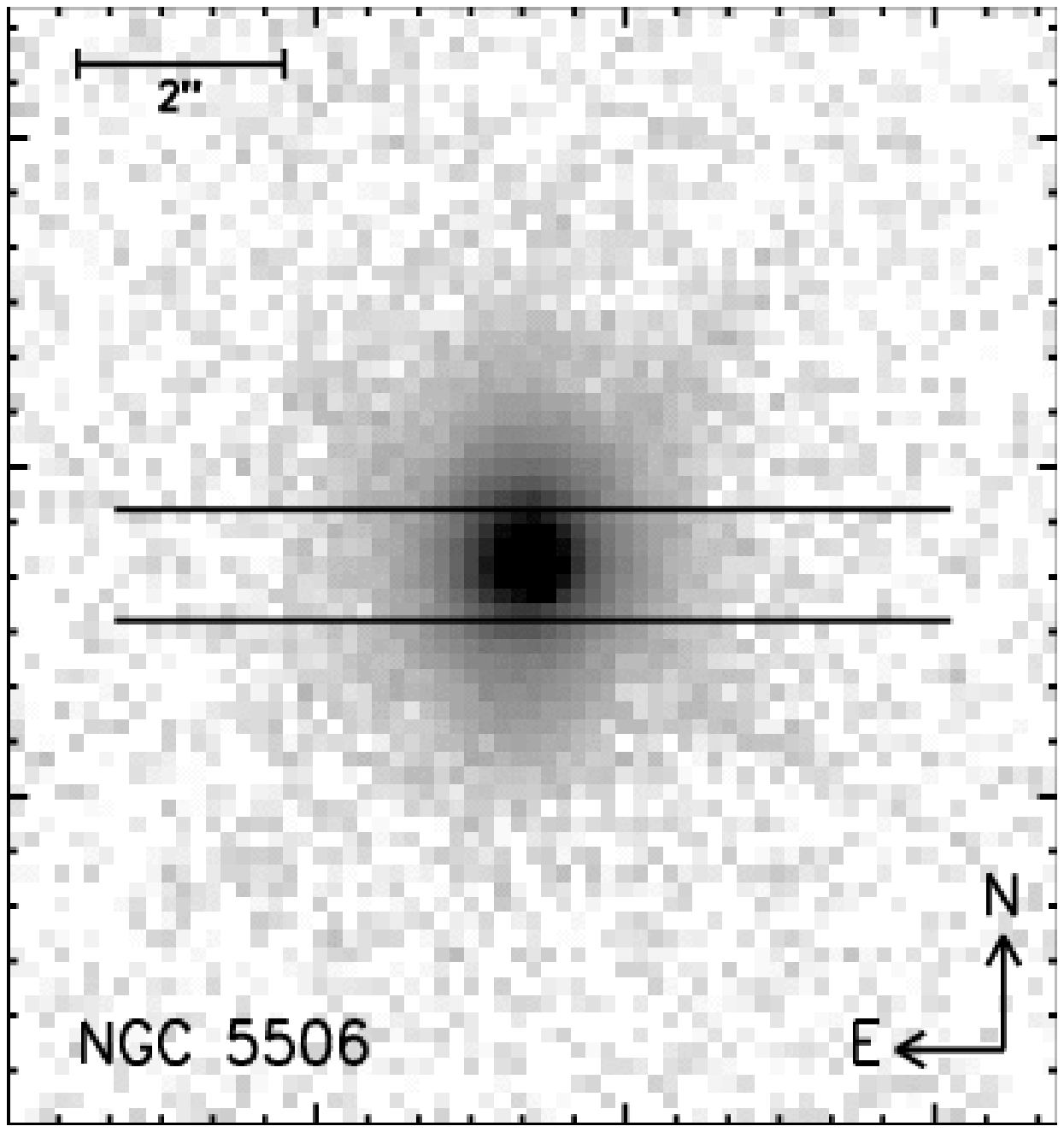}\plotone{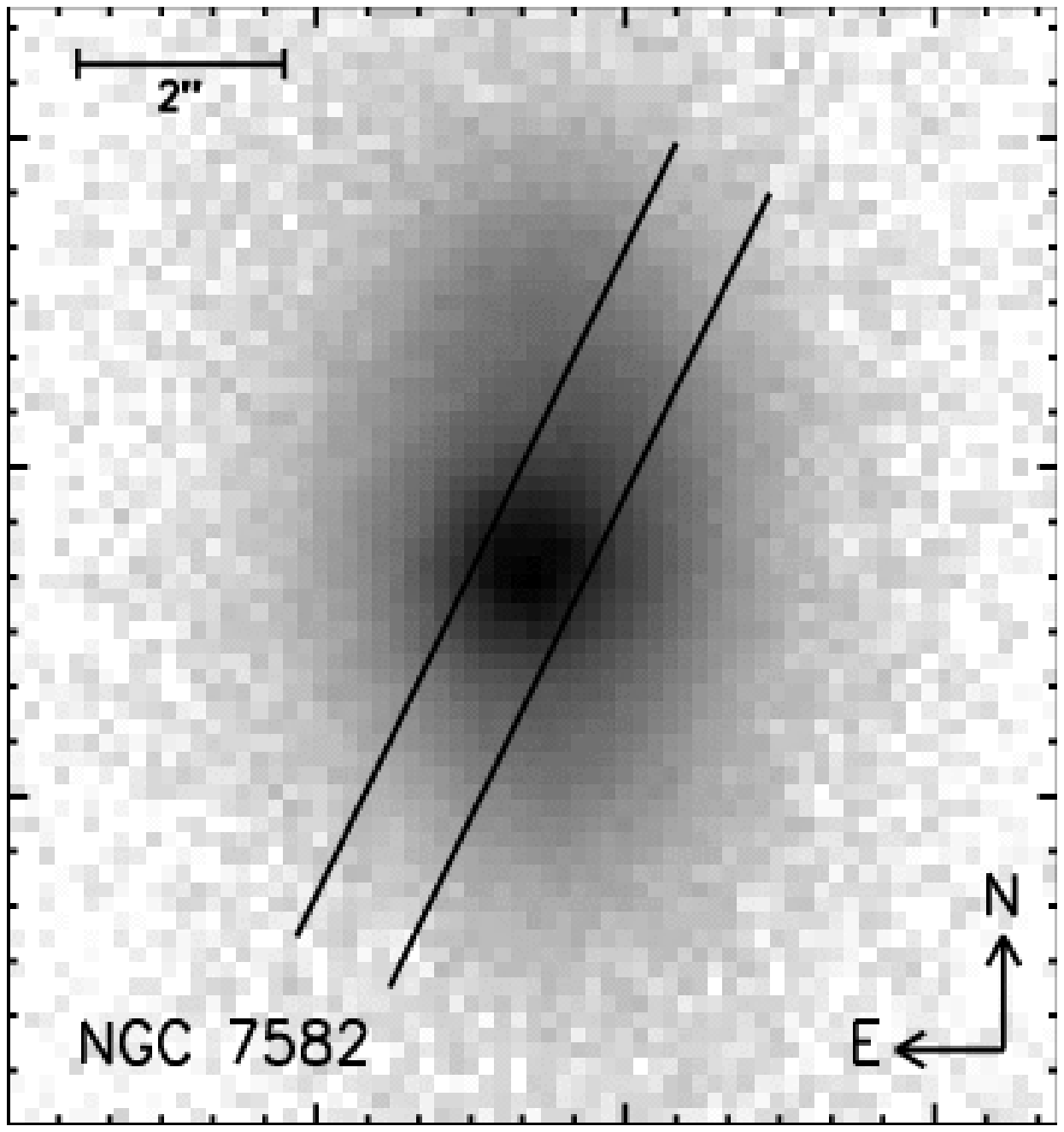}\plotone{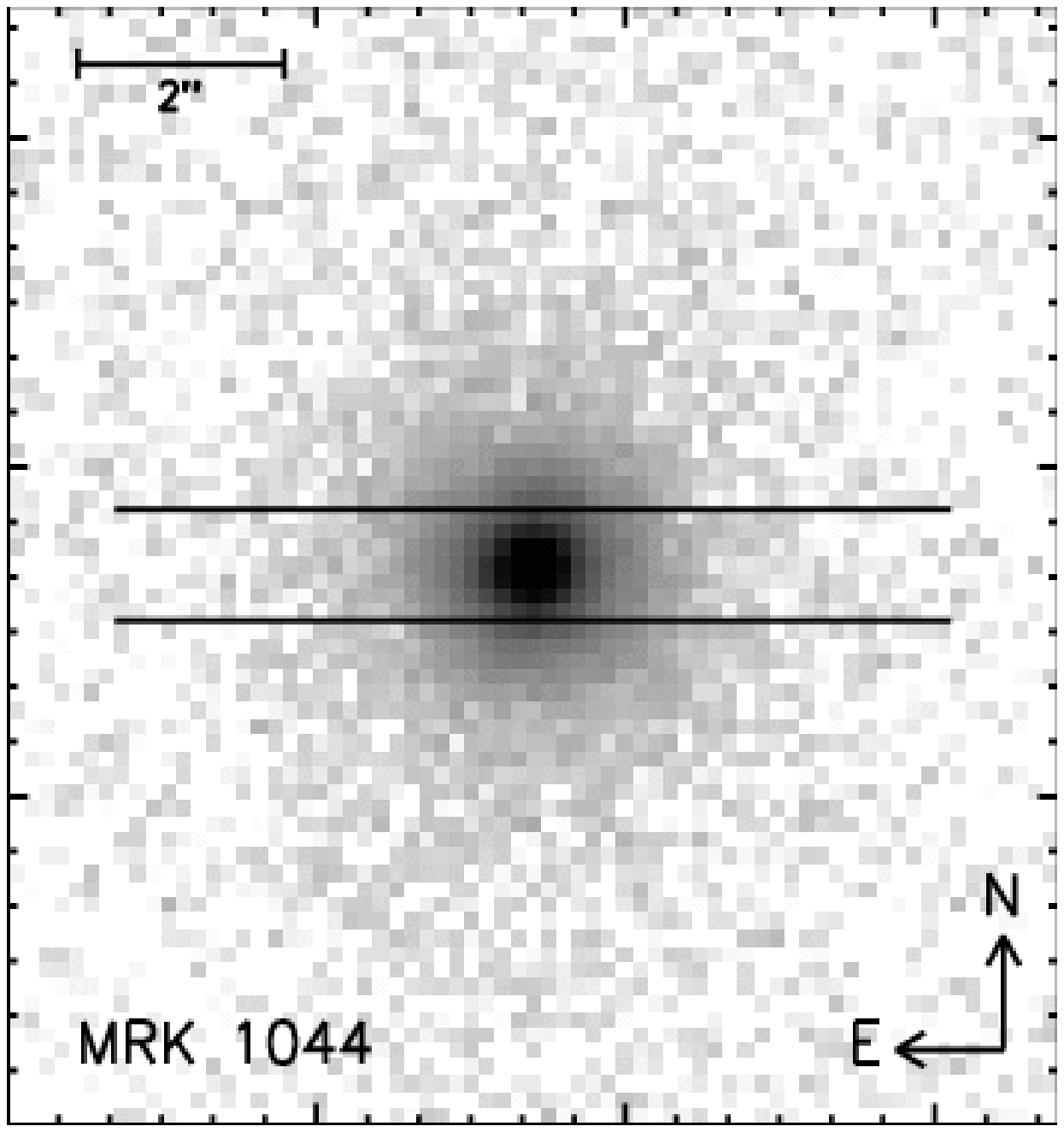}\plotone{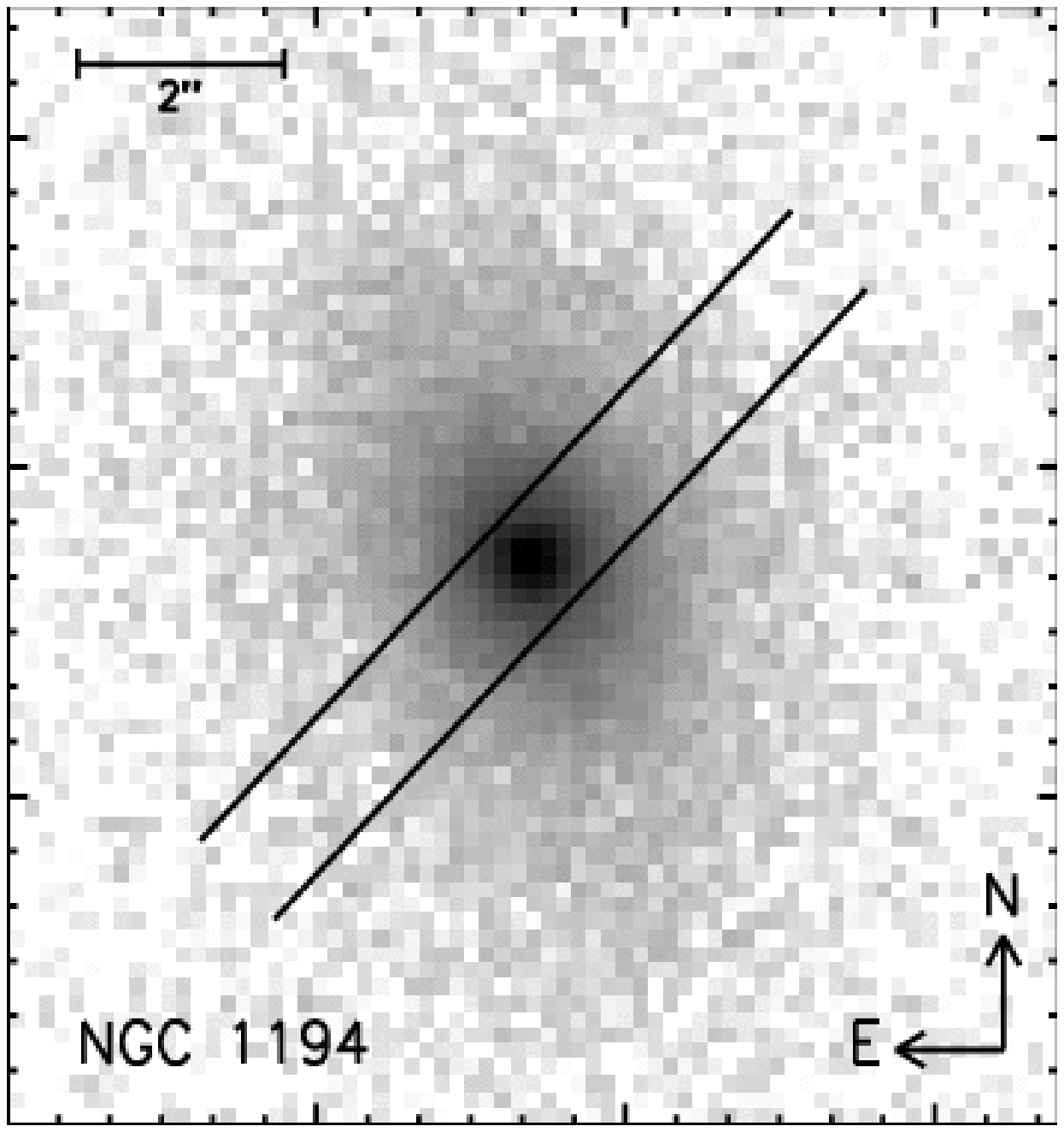}\plotone{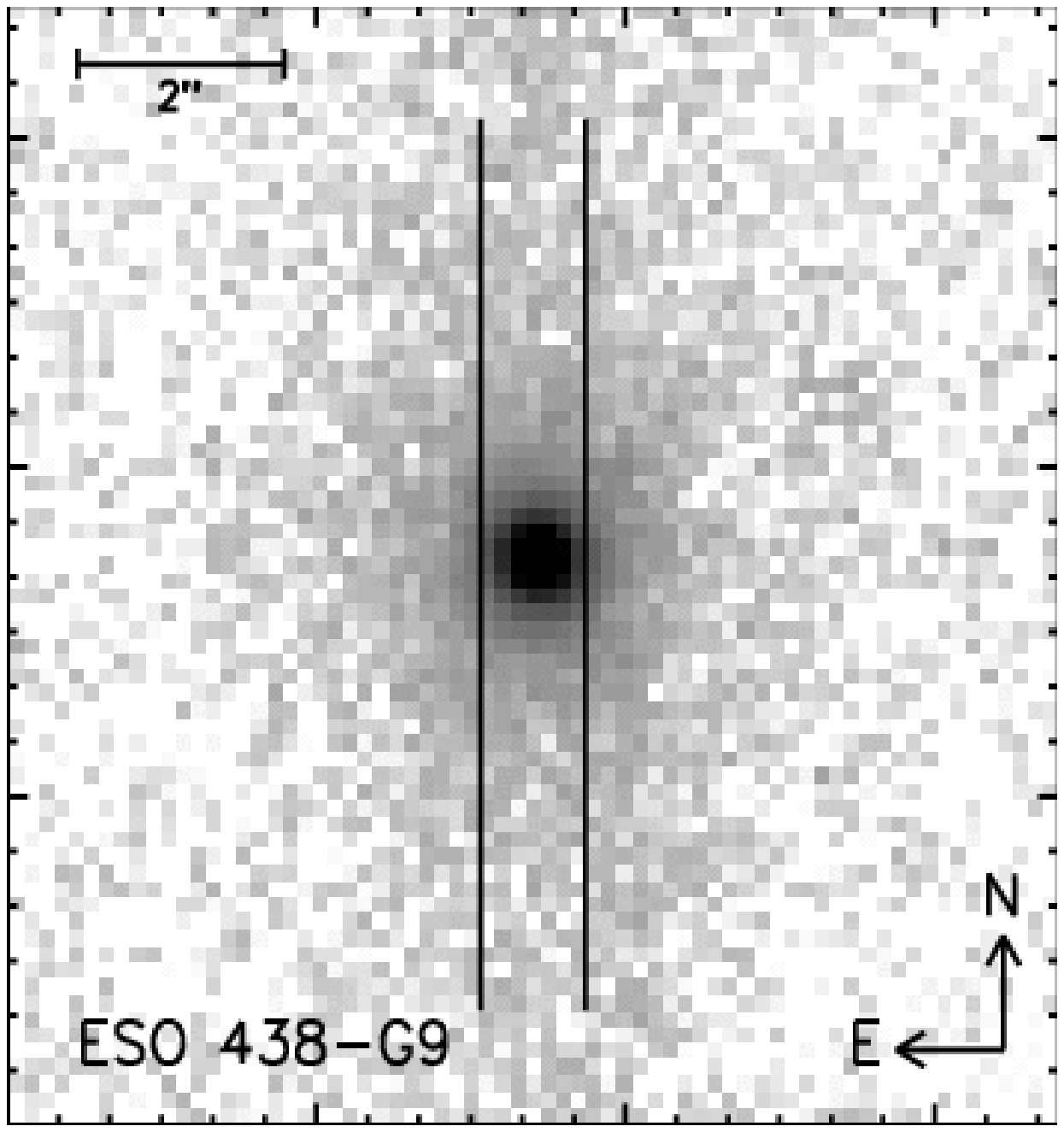}\plotone{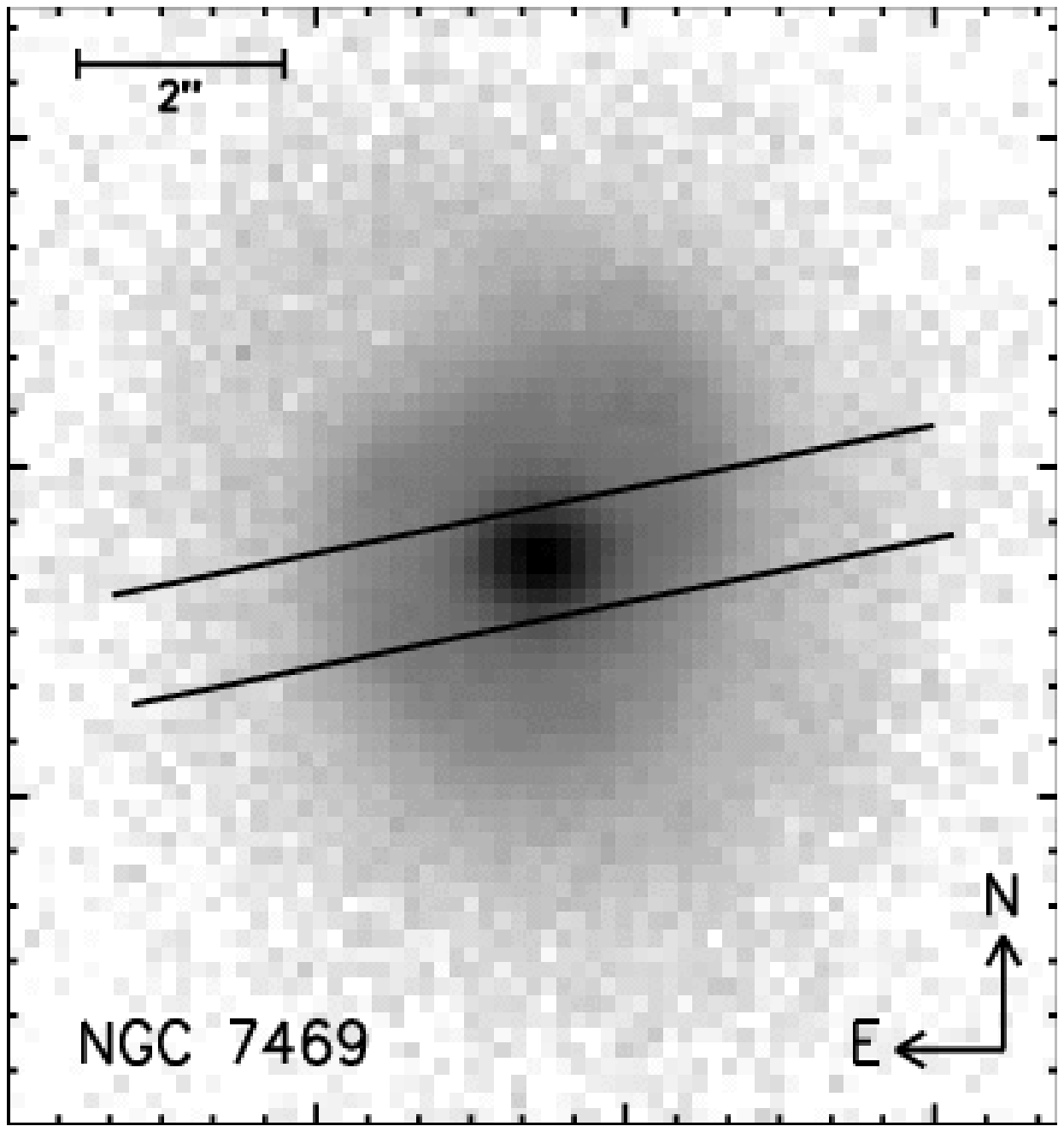}
%\centerline{\psfig{file=f01a.eps,width=3.2cm}\hspace{1mm}\psfig{file=f01b.eps,width=3.2cm}\hspace{1mm}\psfig{file=f01c.eps,width=3.2cm}\hspace{1mm}\psfig{file=f01d.eps,width=3.2cm}\hspace{1mm}\psfig{file=f01e.eps,width=3.2cm}}
%\vspace{2mm}
%\centerline{\psfig{file=f01f.eps,width=3.2cm}\hspace{1mm}\psfig{file=f01g.eps,width=3.2cm}\hspace{1mm}\psfig{file=f01h.eps,width=3.2cm}\hspace{1mm}\psfig{file=f01i.eps,width=3.2cm}}
\caption{Acquisition images of the targets taken through (crossed) narrow band
  filters close to 2.1\,\micron\ with integration times of a few seconds.
Each field is 10\arcsec$\times$10\arcsec, and is displayed with
logarithmic scaling.
The parallel lines indicate the position angle and width of the slit.
Top row shows the 5 nearby objects ($\langle D \rangle = 25$\,Mpc; mostly
  classified as Seyfert~2);
Bottom row shows the 4 more distant objects 
($\langle D \rangle = 75$\,Mpc; mostly classified as Seyfert~1).
}
\label{fig:acq}
\epsscale{1.0}
\end{figure}

Acquisition images with integration times of a few seconds were taken
before each spectrum through a narrow band filter near to 2.1\,\micron. 
These are shown in Fig.~\ref{fig:acq} together with the orientation of
the slit across the nucleus of each object.
The position angle of the slit was set to lie along the major axis of
the object as determined from 2MASS K-band or red
Digitised Sky Survey images. 
Where there was no clear axis, the slit was set east-west.

ISAAC was used in Medium Resolution mode in the K-band, providing a
nominal resolution of $R\sim3000$ with a 1\arcsec\ slit.
Since the wavelength coverage in this mode is only 0.122\,\micron, two
settings were used with an overlap of 5--10\%.
The total integration time was 2000\,seconds per setting.
Standard calibrations were performed, including atmospheric standard
stars (type B or G2V), arcs, flatfields, and dark frames.
The data were reduced using PC-IRAF~2.11.3 using standard techniques.
Wavelength calibration was achieved by tracing the sky lines to create
a transformation which removed curvature, at the same time as applying
the final wavelength scale.
It was found that the spectral slope in each row of the object spectra
had a small residual curvature which depended on the position of the row.
This curvature was removed without affecting the overall slope of each
spectrum, allowing a much better merging of the two wavelength
segments.
However, this does mean there could be some uncertainty in the slope
of the resulting spectra.
The merging was performed on the 2-dimensional spectra, convolving
the observation which had better seeing with a Moffat function to match
the one which had poorer seeing. 
At the same time the fluxes of the two spectra were scaled according
to their overlapping regions.

The flux calibration was derived from the standard stars,
the K-band magnitudes of which were extracted from the 2MASS Point
Source Catalogue.
There was a very good correspondance between these magnitudes and
those estimated from the colour of the spectral type and V-band
magnitude, the difference being on average 0.15\,mag.
Slit losses are compensated to zero order by the standard star for
each observation.
No further correction was applied since the main theme of the
analysis involves line ratios rather than absolute fluxes.
Since also the change in atmospheric dispersion in the K-band is small,
the data are unaffected by differential slit losses.

\begin{figure}
\epsscale{.3}
\plotone{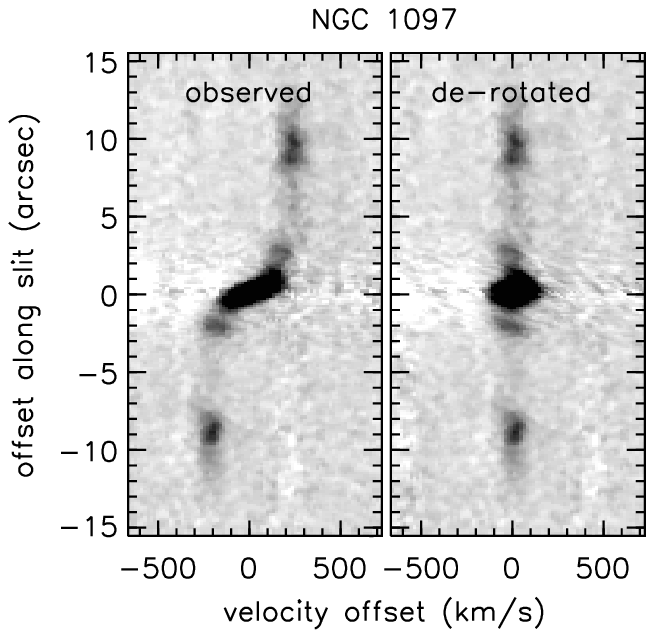}\plotone{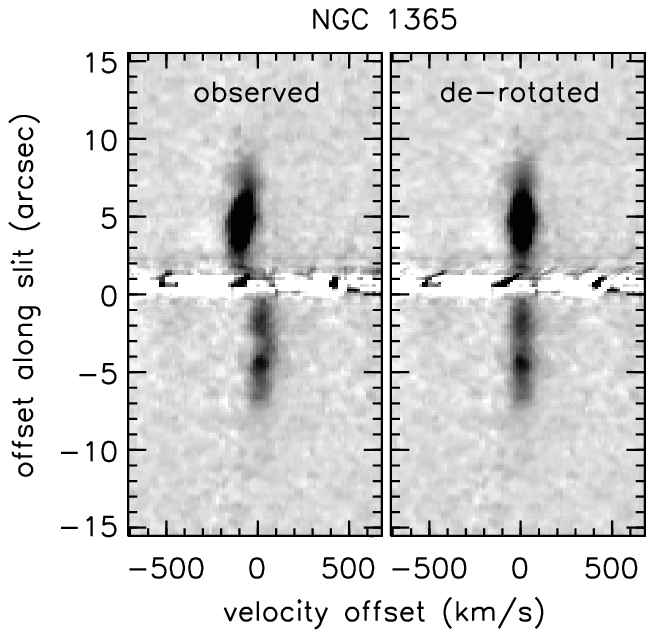}\plotone{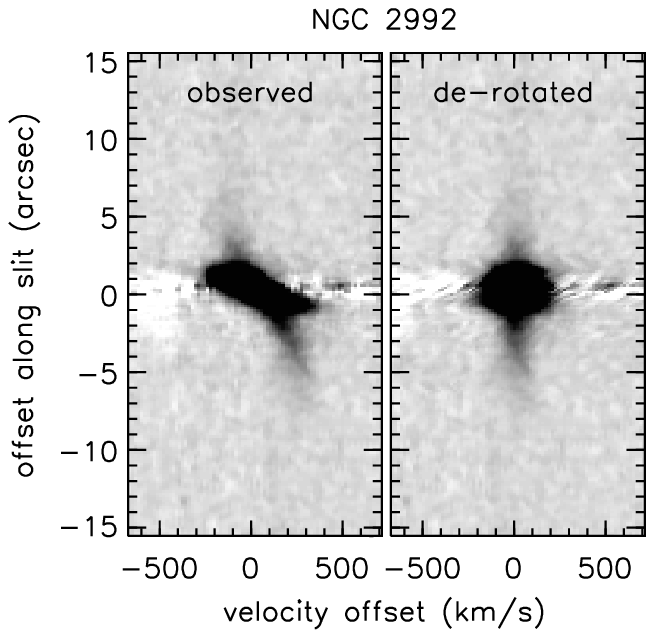}\plotone{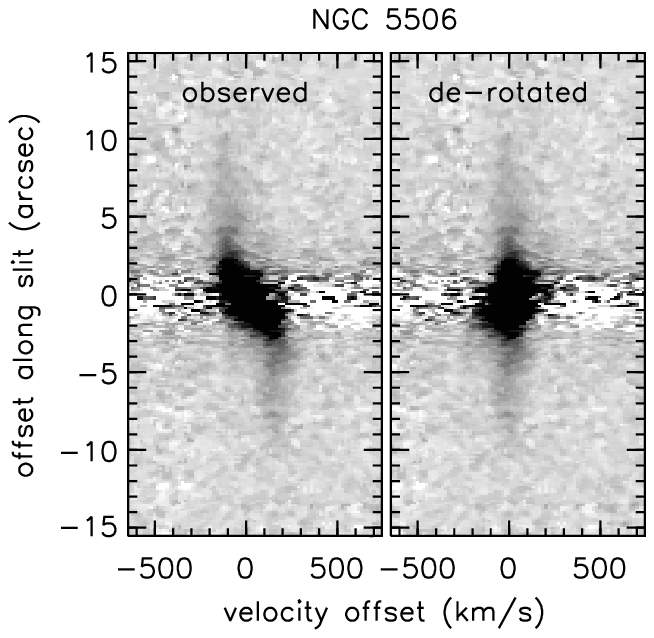}\plotone{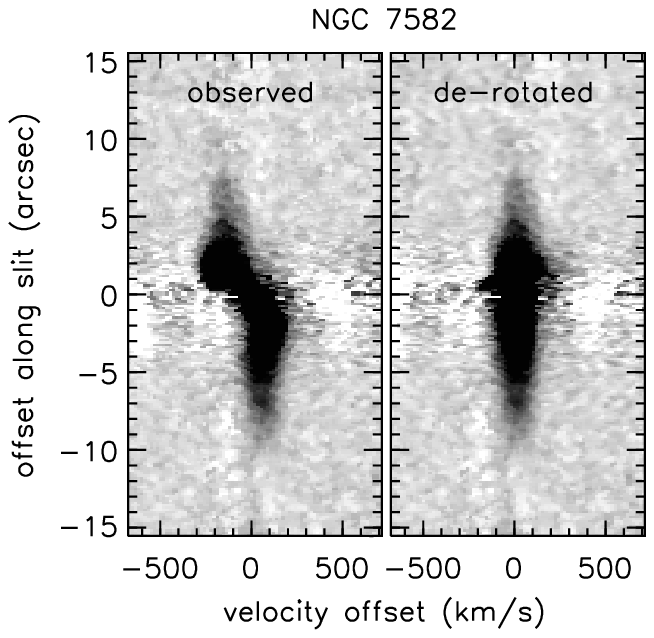}
%\centerline{\psfig{file=f02a.eps,width=50mm}\psfig{file=f02b.eps,width=50mm}\psfig{file=f02c.eps,width=50mm}}
%\centerline{\psfig{file=f02d.eps,width=50mm}\psfig{file=f02e.eps,width=50mm}}
\caption{Position velocity diagrams of the H$_2$ 1-0\,S(1) line in the
  nearby ($\langle D \rangle = 25$\,Mpc) objects.
Left: the observed spatial line profile, which traces the rotation curve.
Right: after applying a de-rotation to straighten the line profile.
The de-rotation has two effects which act to increase the
  signal-to-noise of a spectrum extracted across a number of spatial
  rows:
it reduces the width of the spectral 
  features and at the same time decorrelates the systematic noise of
  imperfect telluric correction.
}
\label{fig:derotate}
\epsscale{1.0}
\end{figure}

The final step in the data reduction was to de-rotate the spectra, as
shown in Fig.~\ref{fig:derotate}, to straighten the rotation curves.
The advantage this brings is twofold.
It restricts the spectral width of all the lines (emission and
absorption) in a spectrum extracted over a long aperture;
and it decorrelates the noise introduced by systematic residuals from
imperfect telluric correction, since these features are no longer
aligned with the spatial axis.
Thus the signal to noise in the spectral features that we wish to
measure is significantly increased.
The procedure was performed on the 5 nearby objects,
since their rotation curves could be traced out beyond the nucleus.
For each of these, the velocity shift of the 1-0\,S(1) line was measured in
each spatial
row where the line was sufficiently well detected.
These rows were shifted to a zero offset position.
All other spatial rows were shifted according to a function 
(the error function, since it provides a
simple smooth approximation to a typical rotation curve)
which was fit to the rows where the line offset could be measured.

\section{Spectral Decomposition: Continuum Fitting and Line Fluxes}
\label{sec:contfit}

For all the galaxies listed in Table~\ref{tab:basdat}, we have
extracted spectra from the nuclear
region, which we define as the central 1\arcsec\ (to match the typical
seeing).
This includes everything within 60\,pc (for the nearby galaxies) or
190\,pc (for the more distant galaxies) of the AGN.
Additionally, for the nearby galaxies we have extracted a
circumnuclear spectrum, which we define as everything outside the
central 2\arcsec\ (to reject any nuclear emission) and up to 1\,kpc
from the AGN.
Estimations of the contamination of the circumnuclear flux by the
nuclear flux in each case and for the ambient seeing indicate that
this is less than 10\% in the worst case and typically 1\% or less.

As for our previous work on ultraluminous galaxies \citep{dav03}, we
have fit stellar templates (F--M
type supergiants from the library of \citealt{wal97})
to the continuum in order to be able to measure the line emission
accurately.
However, because of the broad Br$\gamma$ in most of the nuclear
spectra, it was not practical to mask out regions with line emission.
Instead, spectra were fit simultaneously with stellar templates and
both broad and narrow emission lines.
The wavelengths of the lines were fixed, and they were assumed to have
the same redshift as the stellar continuum.
However, the minimisation was allowed to adjust the velocity width of
the lines (one width for all the narrow lines, and one for all the
broad lines).
While this procedure can match the continuum well, it
provides poor estimates of (particularly the weak) line fluxes --
since these are affected by local over-/underestimates in the relative
level of the continuum.
The line fluxes were therefore remeasured by fitting Gaussian
functions of fixed width to the spectrum after the fitted continuum
had been subtracted.
The uncertainties were estimated from the standard deviation of the
local residual (covering a spectral range 10 times the FWHM of the line)
and the line widths using Monte Carlo techniques.
Specifically, a spectral segment was constructed to have the same
statistical noise properties as the actual residual spectrum, to which
a Gaussian profile of the measured flux and FWHM was added.
A new Gaussian profile was then fitted to the segment, producing
slightly different parameters.
This was repeated 10000 times, yielding the uncertainties.
Limits were derived in a similar way.
We caution that while this treatment accurately estimates uncertainty
from random noise, it cannot account for systematic error such as

The absolute flux scaling is given in Table~\ref{tab:measdat}, and
the line ratios and uncertainties are tabulated in Table~\ref{tab:ratios}.

\begin{figure}
\epsscale{.4}
\plotone{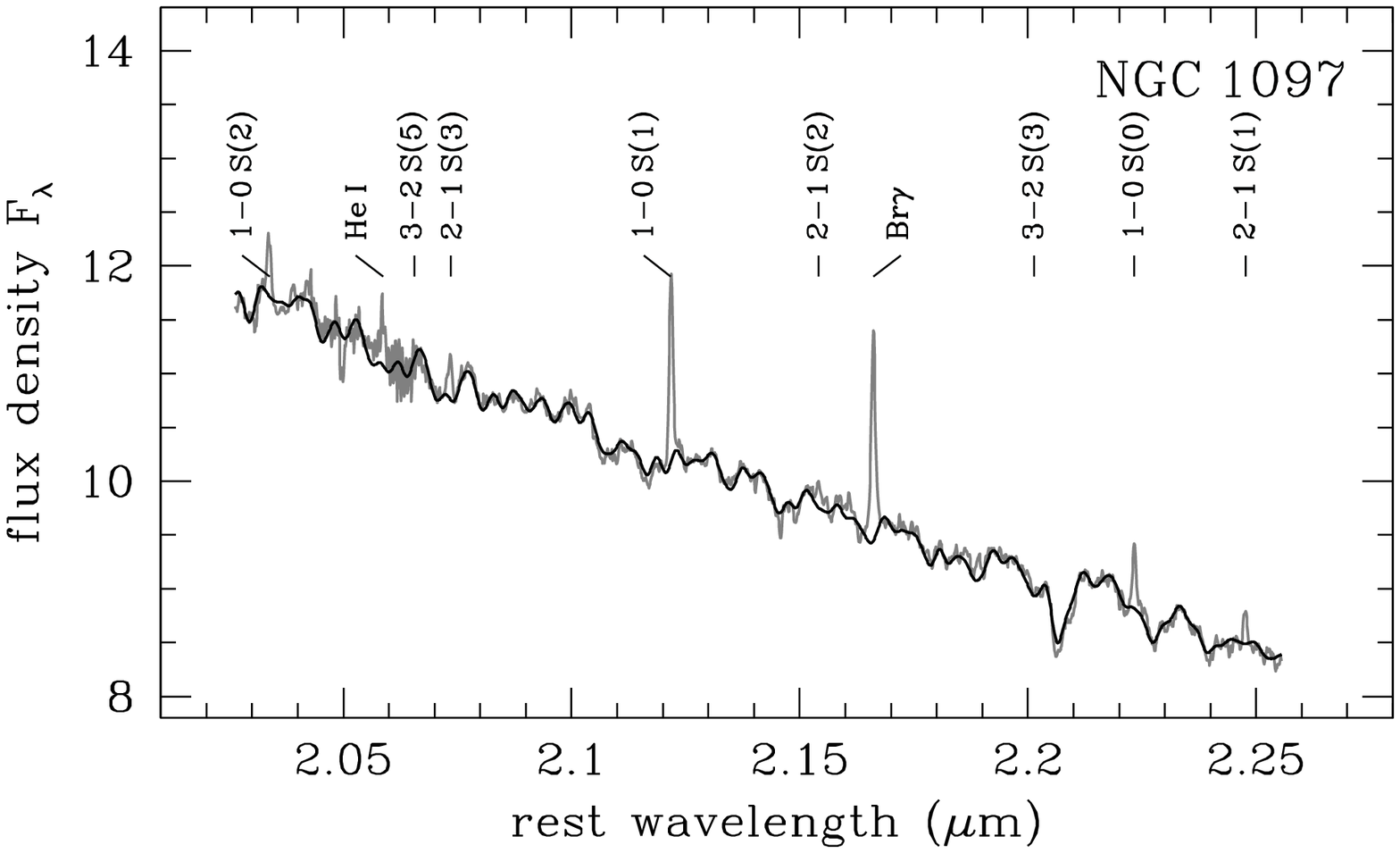}\plotone{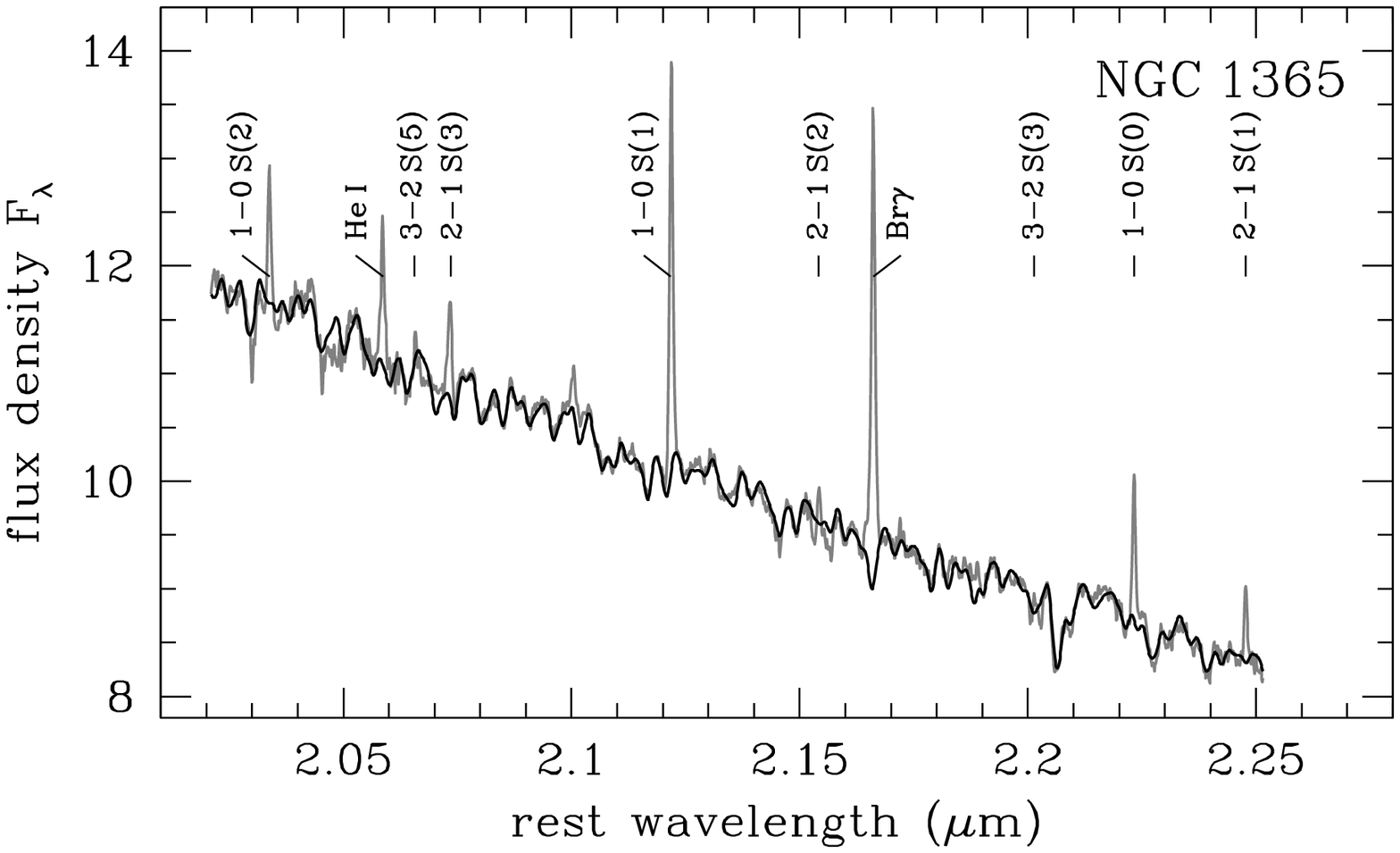}\plotone{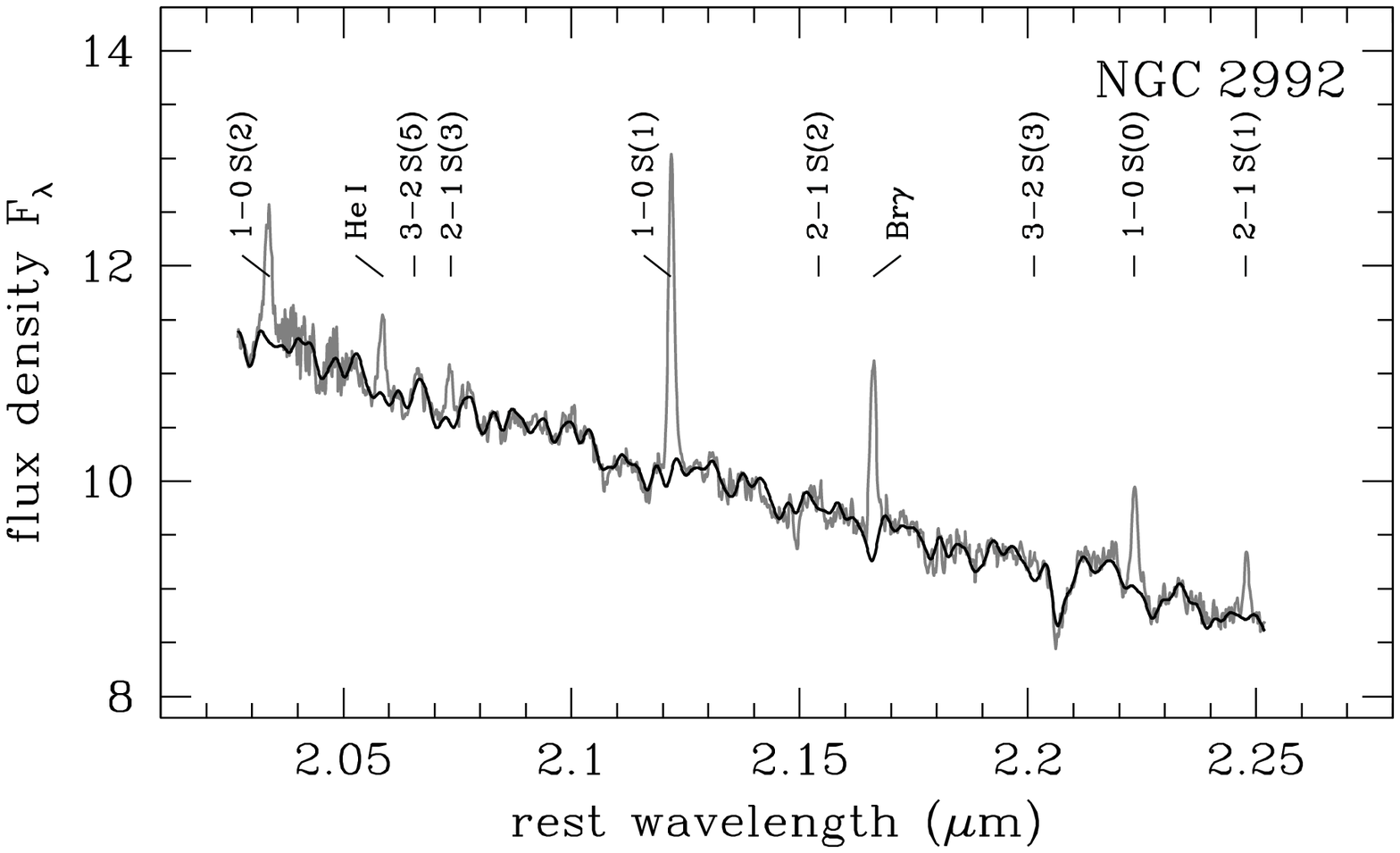}\plotone{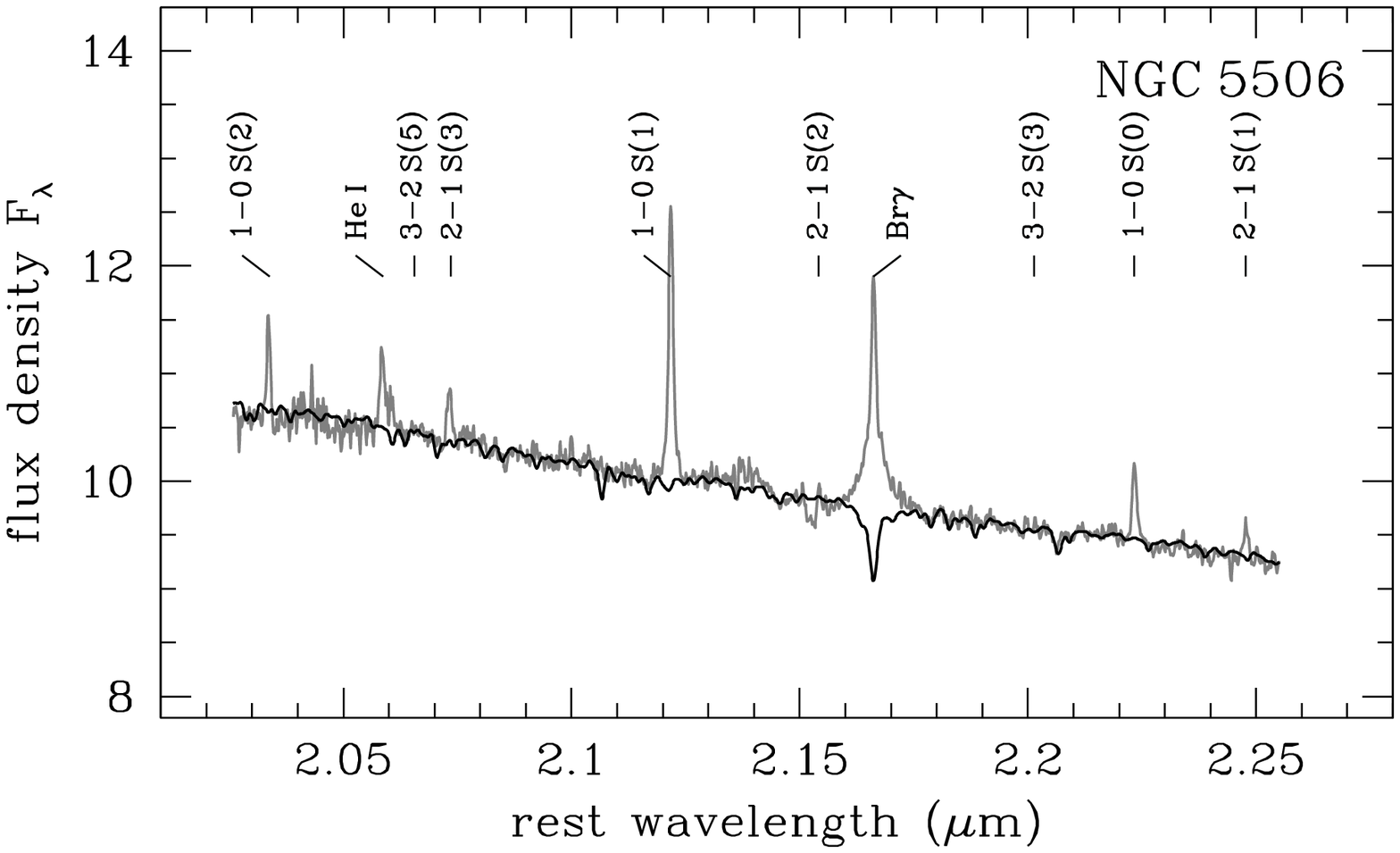}\plotone{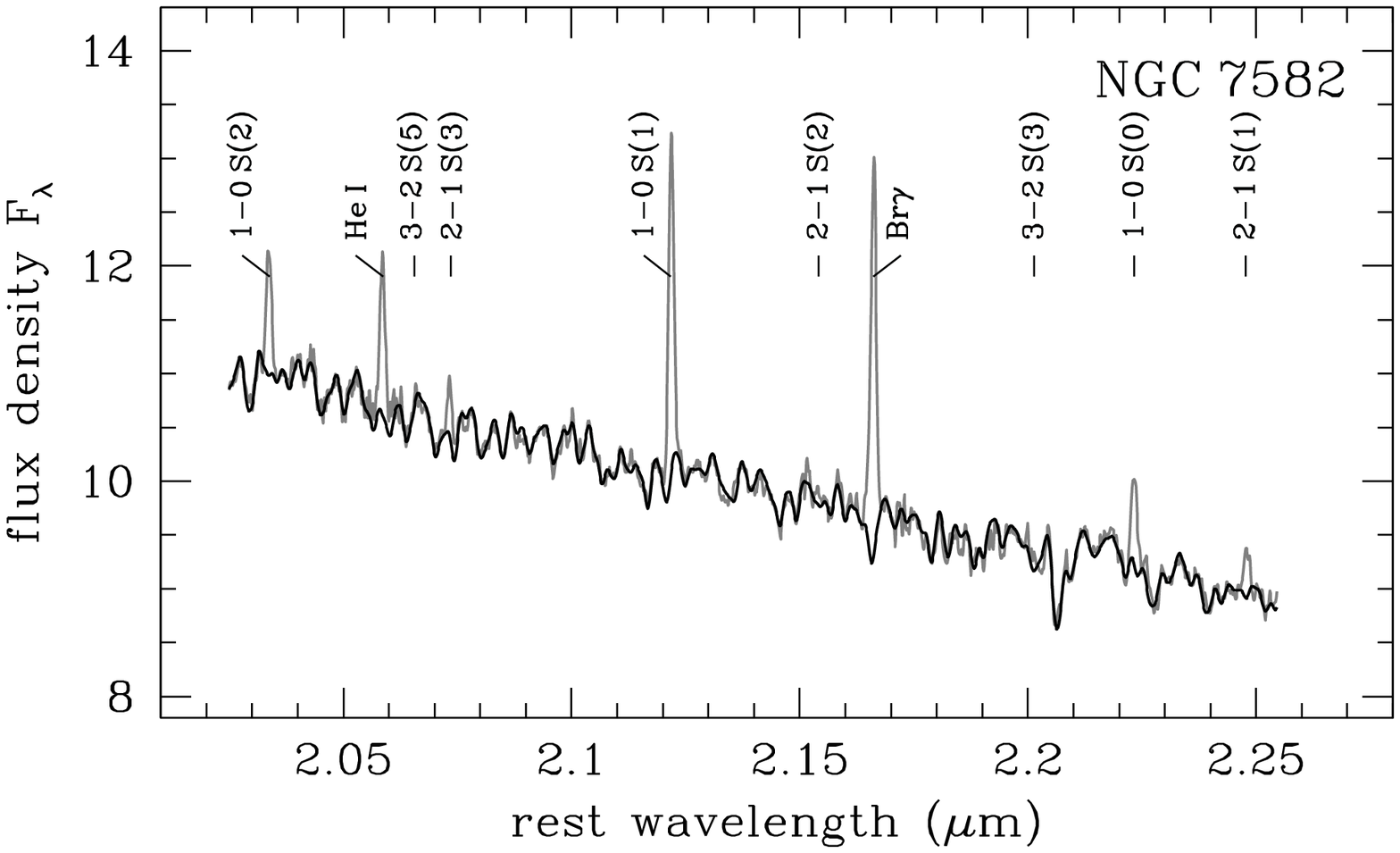}
%\centerline{\psfig{file=f03a.eps,width=90mm}\psfig{file=f03b.eps,width=90mm}}
%\vspace{1mm}
%\centerline{\psfig{file=f03c.eps,width=90mm}\psfig{file=f03d.eps,width=90mm}}
%\vspace{1mm}
%\centerline{\psfig{file=f03e.eps,width=90mm}}
\caption{Normalised spectra of the circumnuclear (i.e. to a
  radius of 1\,kpc, excluding everything within the central 1\,arcsec)
  region of each of the nearby ($\langle D \rangle = 25$\,Mpc)
  objects, plotted at rest wavelength and 
  normalised to their mean values.
The dark overlaid line is the best fitting stellar continuum
  constructed from spectra of template stars (see text for details).
The positions (but not necessarily detections) of important emission
  lines are marked.
}
\label{fig:circum-spec-near}
\epsscale{1.0}
\end{figure}

The circumnuclear spectra of all the targets except NGC\,5506 had deep
absorption features.
For these we assumed there was no contribution from hot dust, but
allowed the extinction to vary in order to match the spectral slope.
For NGC\,5506, the very weak Na{\sc\,i} absorption at 2.206\,\micron\ suggests
that the near infrared is not dominated by late-type stars, perhaps because
there is a significant hot dust component -- even away from the nucleus.
The residual broad Br$\gamma$ also suggests that there may be some
contamination from the nucleus.
For this spectrum we not only allowed the extinction to vary, but
added a thermal component at $\sim1000$\,K representing hot
dust.
Although there is some degeneracy between these two effects since they
both change the spectral slope, the primary aim is to remove the
effect of the absorption features rather than uniquely identify the
components comprising the continuum; 
and in this respect the process is successful.
The results of this continuum fitting procedure for the circumnuclear
regions are shown in Fig.~\ref{fig:circum-spec-near}.

\begin{figure}
\epsscale{.4}
\plotone{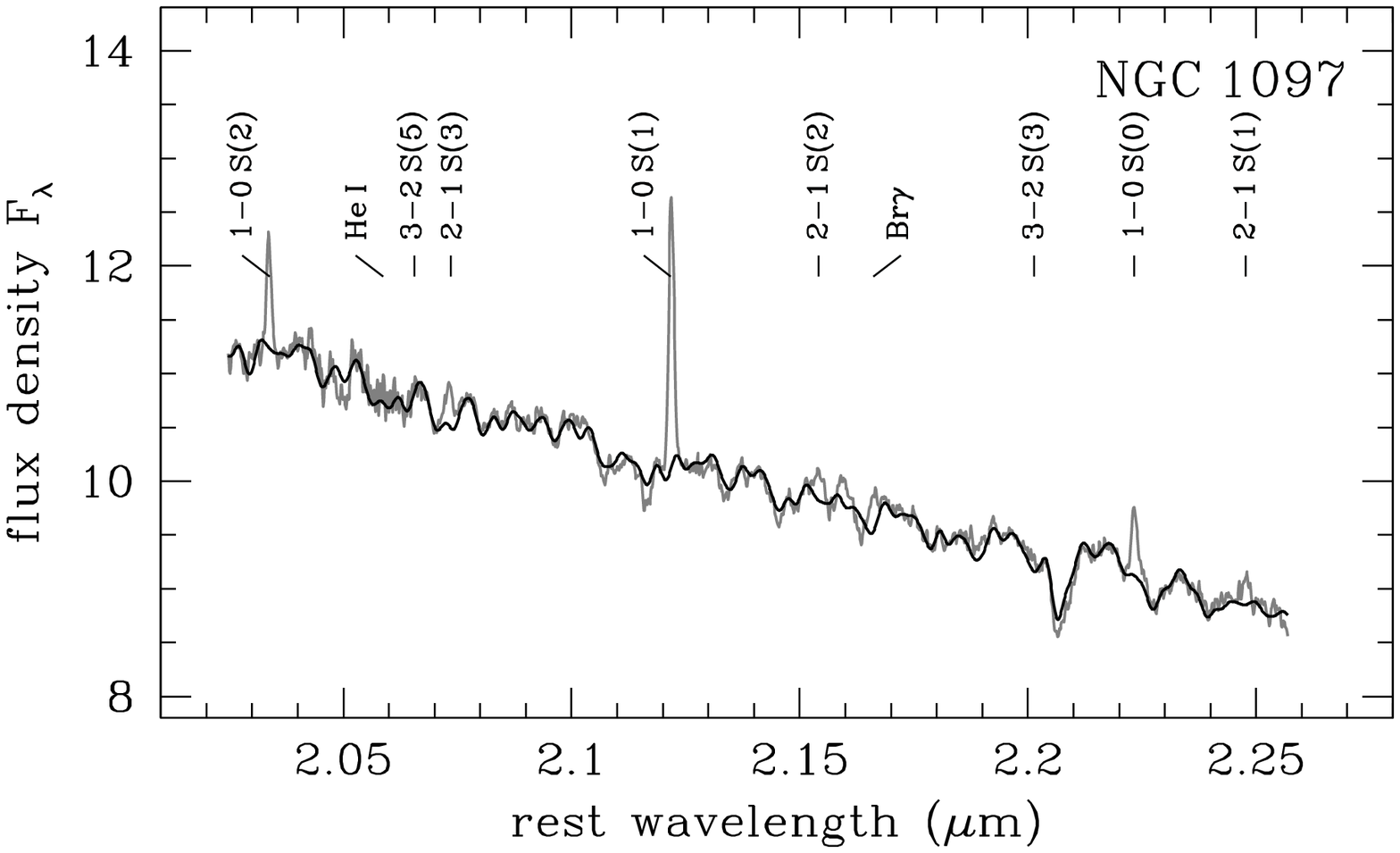}\plotone{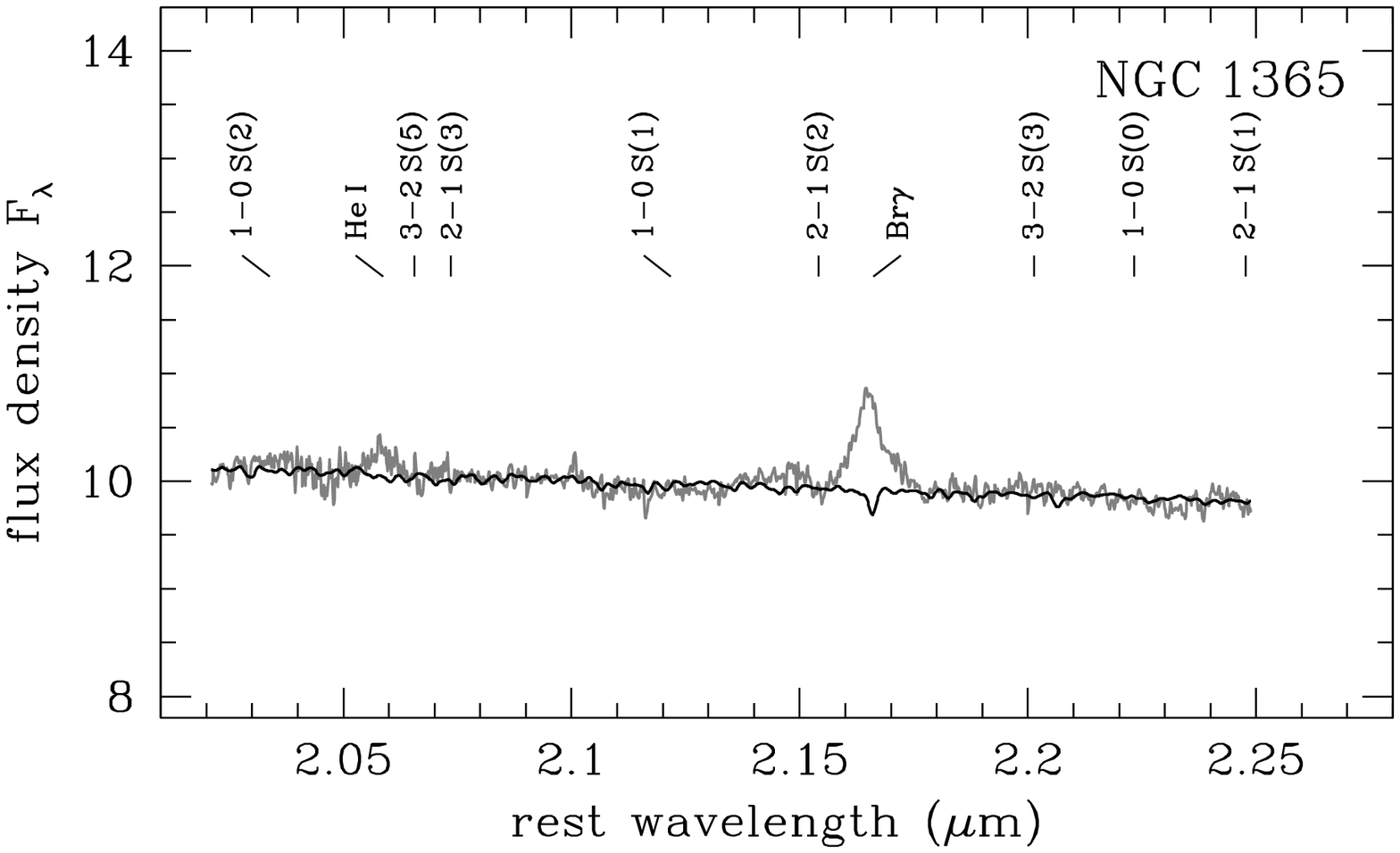}\plotone{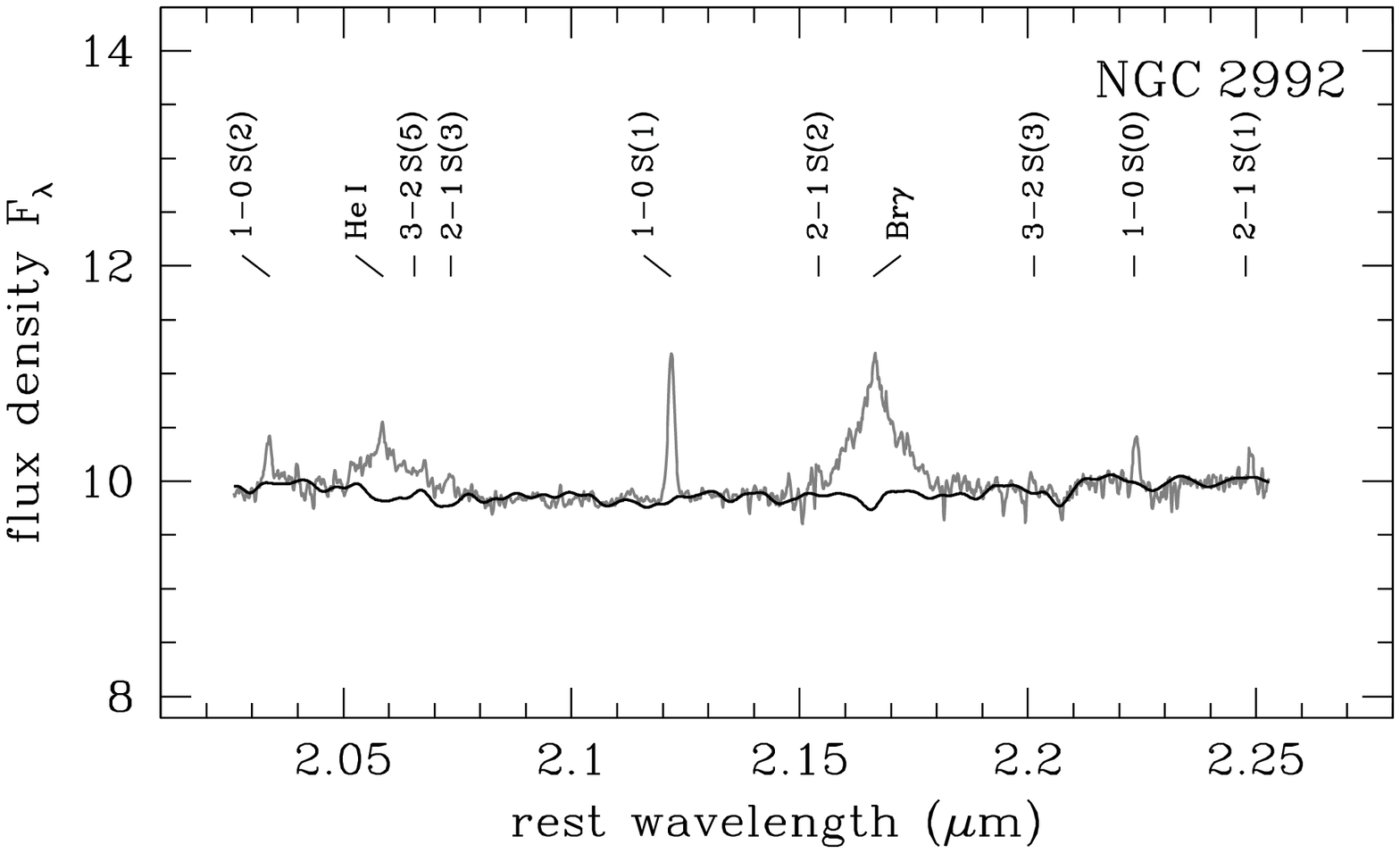}\plotone{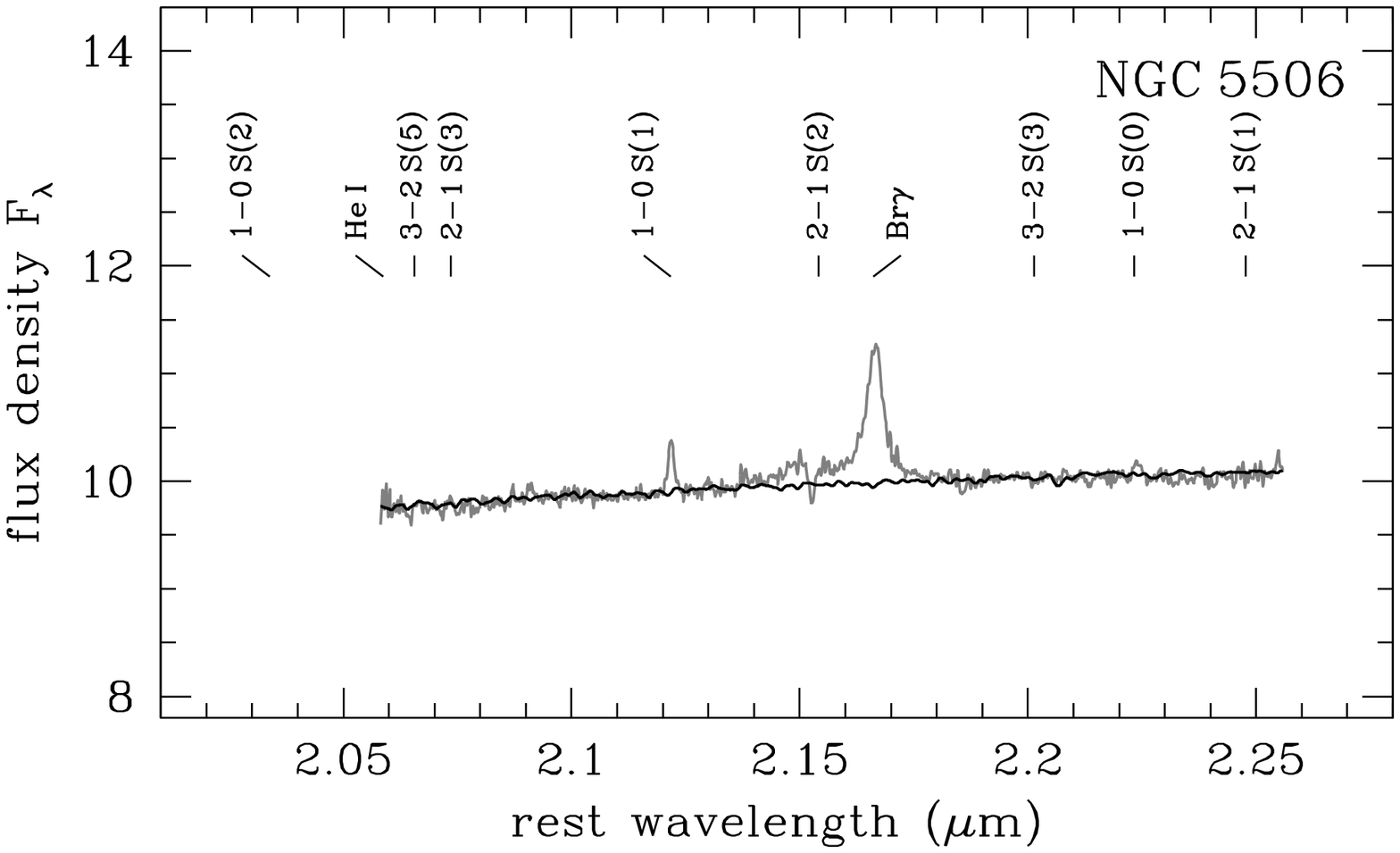}\plotone{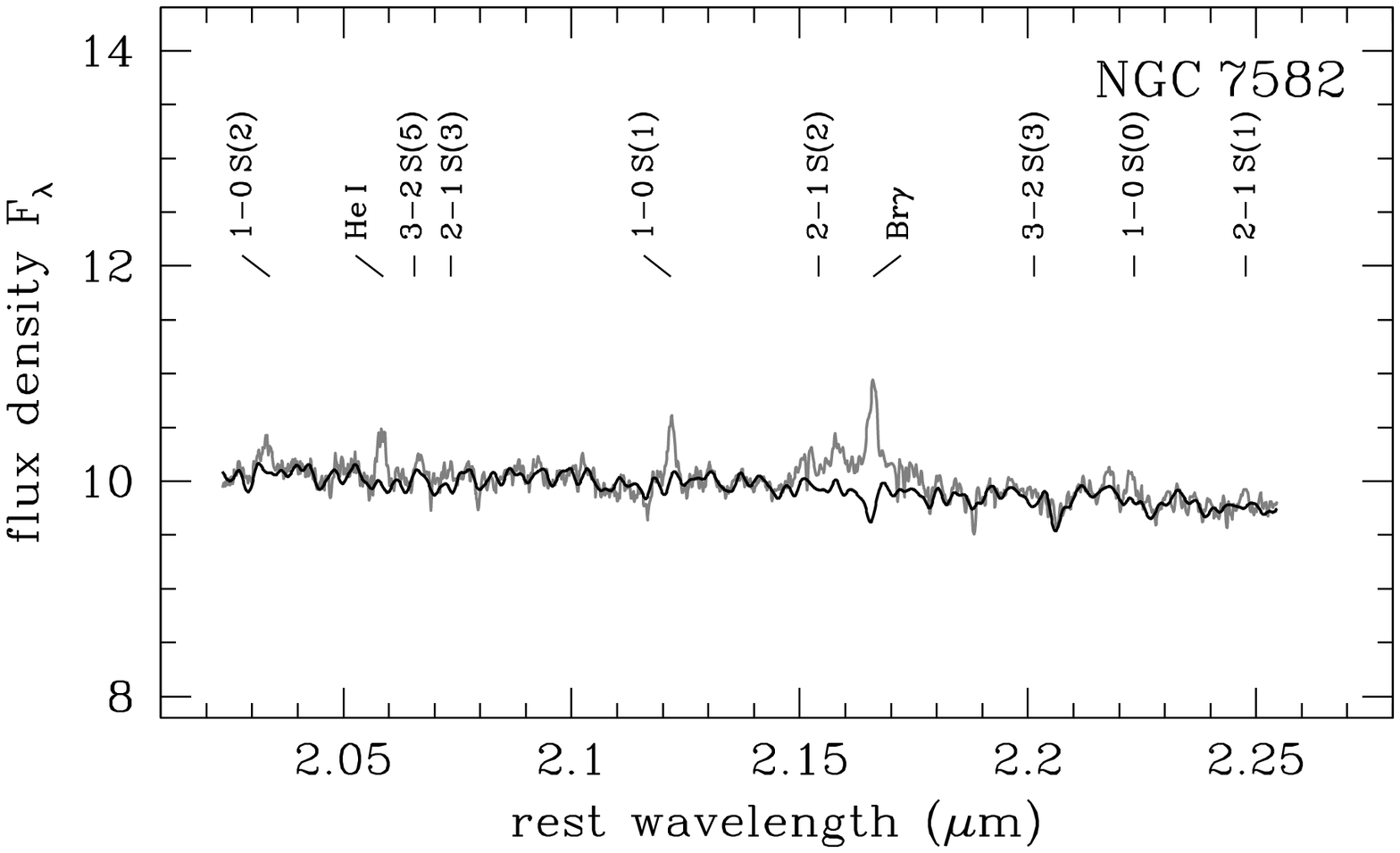}
%\centerline{\psfig{file=f04a.eps,width=90mm}\psfig{file=f04b.eps,width=90mm}}
%\vspace{1mm}
%\centerline{\psfig{file=f04c.eps,width=90mm}\psfig{file=f04d.eps,width=90mm}}
%\vspace{1mm}
%\centerline{\psfig{file=f04e.eps,width=90mm}}
\caption{Normalised spectra of the nuclear region (i.e. central
  1\arcsec) of each of the nearby ($\langle D \rangle = 25$\,Mpc)
  objects, plotted at rest wavelength and 
  normalised to their mean values.
The dark overlaid line is the best fitting stellar continuum
  constructed from spectra of template stars (see text for details).
The positions (but not necessarily detections) of important emission
  lines are marked.
}
\label{fig:nuclear-spec-near}
\epsscale{1.0}
\end{figure}

In all cases except NGC\,1097, the stellar features in the nuclear
spectra are much less clearly defined. 
There are two important reasons for this:
1) the redder spectral slope indicates that there is a significant
contribution from hot dust, which dilutes the stellar continuum and
adds to the noise;
2) one would expect a higher velocity dispersion in the nucleus
(particularly if the stellar population is dynamically hot) which,
combined with the steep rotation curve (albeit corrected) and seeing
lead to a large broadening of the features.
In fitting the continuum we have, in each case except NGC\,1097, left
both the $\sim1000$\,K hot dust contribution and the extinction as
free parameters.
The results of the continuum fitting procedure for the nuclear regions are
shown in Fig.~\ref{fig:nuclear-spec-near} for the nearby objects and
Fig.~\ref{fig:nuclear-spec-far} for the more distant objects.

\section{Object Notes}
\label{sec:notes}

In this section we summarise some notes about the objects in our
sample, in order to put into context our data and the results we
discuss later.

\subsubsection*{NGC\,1097}

The classification of NGC\,1097 as a Seyfert~1 is based on its broad
H$\alpha$ line \citep{sto93} and hard X-ray excess \citep{iyo96},
although the AGN is low luminosity.
The black hole mass is nevertheless expected to be
$10^7$--$10^8$\,M$_\odot$, based on the stellar velocity dispersion and
M$_{\rm BH}-\sigma$ relation \citep{sto93} -- an order of magnitude
more massive than that in the very active NGC\,7469 \citep{col98}.
Monitoring the broad H$\alpha$ over a timespan of 11\,yrs,
\cite{sto03} indicated that the luminosity decreased to 1/3 of
its initial value over this period.
During our observations there was no evidence of broad Br$\gamma$ and,
assuming a velocity width of 7500\,\kms\ FWHM (similar to that for the
total broad H$\alpha$ line), we can impose a
3$\sigma$ limit of $8.5\times10^{-19}$\wm\ (1\arcsec\ aperture) just
consistent with the broad H$\alpha$ flux measured towards the end of
2001 if extinction is negligible.
However, it is plausible and more likely that the broad component
continued dimming towards the middle of 2002 when our data were taken.
The narrow Br$\gamma$ is also weak in the nucleus, with a flux of
$7.6\times10^{-19}$\wm\ in a 1\arcsec\ aperture, indicating that there
is little current star formation.
The last episode may have ceased relatively recently, as suggested by
the very deep absorption features which even with the use of
supergiant templates we have difficulty matching.
NGC\,1097 therefore offers an excellent opportunity to study the
distribution and dynamics of the stars in the region close around a
(quiescent) AGN.
Our detection of strong H$_2$ 1-0\,S(1) in the nuclear region is in
stark contrast to the lack of such emission in the spectrum of
\cite{sos01}, and we can offer no explanation for this difference.

NGC\,1097 exhibits a very clear and well studied circumnuclear ring
\citep{ger88,sto96,kot00}, which is also apparent in our data from the
increase in Br$\gamma$ flux at radii of 8--9\arcsec.

\subsubsection*{NGC\,1365}

NGC\,1365, a strongly barred galaxy in the Fornax cluster, has been
reviewed extensively by \cite{lin99}.
Its AGN exhibits both broad and narrow components to the H$\alpha$
line, the FWHM of the former being 1800\,\kms\ \citep{all81}, somewhat
wider than the $\sim1000$\,\kms\ we find for the Br$\gamma$ line.
Assigning 1800\,\kms\ to the broad component in our data would only be
possible if the narrow component had FWHM 500\,\kms, in
conflict with \cite{all81} reporting it as unresolved.
The soft X-ray (0.1--2.5\,keV) luminosity, which can in principle be
explained entirely in terms of star formation, suggests the AGN is of
rather low luminosity \citep{ste99}.
On the other hand, the presence of a relatively unabsorbed power law
component at higher (up to 10\,keV) energies and a 6.6\,keV Fe~K emission
line indicate that there must be an AGN \citep{iyo97}.
The ionisation cone, seen in [O{\sc\,iii}] and with kinematics
indicative of outflowing gas, was modelled by \cite{hje96}, who found
that the cone axis lay close to the galaxy's rotation axis and the
opening angle was sufficient to include the line of sight to the
partially obscured Seyfert nucleus.
The galaxy has a prominent circumnuclear ring at a distance of 5--10\arcsec\
from the nucleus, which is actively forming stars.
It has been 
observed in both recombination line and optical/infrared/radio
continuum emission and is also clearly seen in our data.

\subsubsection*{NGC\,2992}

NGC\,2992 exhibits an extended biconical narrow line region on kpc
scales \citep{all99}, which traces outflowing gas.
\cite{vei01} were able to model the kinematics in terms of gas
rotating in the disk plane and a biconical outflow, and argue that the
latter is a thermal wind driven by the
AGN rather than vigorous star formation.
On smaller scales, there is a figure-8 loop of radio
continuum and H$\alpha$ emission, centered on the nucleus and
extending to about 500\,pc \citep{weh88}.
Using adaptive optics to reach a resolution $<0.2$\arcsec, \cite{cha00}
proposed that the loop structure is in fact a superposition of
emission from star formation along a spiral arm in the disk and
outflow bubbles coincident with the larger scale structures.
In contrast to \cite{cha00}, our data show a very significant
reduction in the equivalent width of the stellar absorption features
(whether 2.293\,\micron\ CO\,2-0 or 2.206\,\micron\ Na{\sc\,i}) at the
nucleus, which is most easily explained in terms of dilution due to
hot dust.
The generally accepted classification of NGC\,2992 as a Seyfert~2 (or
1.9) is perhaps surprising given that it exhibits broad emission lines
in the optical and near infrared.
However, given that in the model of \cite{vei01} the line of sight is
very close to the edge of (but not within) the opening angle of the
outflow from the AGN, a classification close to type~2 makes sense
with respect to the standard orientation dependent model.

\subsubsection*{NGC\,5506}

In hard X-rays (2--10\,keV), NGC\,5506 is one of the most luminous
local Seyferts \citep{mat01}, making it rather unusual amongst type~2s.
However, its X-ray properties are more consistent with those of narrow
line Seyfert~1s which, based on its J-band spectrum, NGC\,5506 appears
to be \citep{nag02}.
Our data tend to support this classification.
A detailed study of the kinematics of the double peaked lines by
\cite{mai94} suggested that the object is a nearly edge-on Seyfert.
Their model, a refined version of that proposed by \cite{wil85}, has
the disk plane inclined by 75$^\circ$, above and below which ionised
gas is outflowing in cones extend with opening angle $\sim80^\circ$.
A highly inclined geometry is consistent with the hard X-ray spectrum,
for which the lack of evidence for an accretion disk from the ionised
component of the 6.4\,keV iron line suggests that it must be nearly
edge-on \citep{mat01}. 
\cite{mai94} also argue that, based on the observed trends in
emission line ratios, there is significant star formation occuring at
distances of 300--400\,pc from the nucleus that can account for about
half of the H$\alpha$ emission.
Our data confirm the evidence for active star formation close to the
AGN, showing symmetrical dips in the 
1-0\,S(1)/Br$\gamma$ ratio at offsets of $\pm$2--2.5\arcsec\
(250--300\,pc) from the nucleus.

\subsubsection*{NGC\,7582}

NGC\,7582 is the brightest narrow line X-ray galaxy, and perhaps
prototypical of this class \citep{sch98}.
That it has an AGN cannot be doubted, due to the rapid and large
variations in hard (2--10\,keV) X-ray flux occuring while the soft
(0.5--2\,keV) X-ray flux remianed constant \citep{sch98,xue98}, and
the presence of an [O{\sc\,iii}] ionisation cone \citep{sto91}.
\cite{are99} noted the appearance and subsequent variations in a broad
component to several optical emission lines.
We also have detected a broad bump in the K-band spectrum, which most
likely corresponds to broad ($\sim4000$\,\kms) Br$\gamma$.
This is also apparent in the nuclear spectrum of \cite{sos01}, who
noted the complex morphologies of the emission lines.
\cite{are99} reached no
consensus on whether this type~2 to type~1 transition was due to the
appearance of type IIn supernovae or to patchy obscuration in the
torus aorund the AGN.
By examining the FUV flux, \cite{sto00} found evidence for young stars
in the nucleus (central 2\arcsec) of NGC\,7582, superimposed on the
older bulge population.
We also see relatively strong Na{\sc\,i} absorption in the spectrum of
the central 100\,pc. 
On the other hand, the equivalent width of this
absorption feature decreases steadily from a radius
of $\pm$3\arcsec\ (300\,pc) to a minimum at the nucleus of $\sim1$\,\AA.
This feature could imply a transition to a rather young
stellar population, or more likely the presence of hot dust grains --
although whether the heating on such extended scales can be attributed
to the AGN is unclear.

\subsubsection*{MRK\,1044}

An {\em HST} image of Mrk\,1044 shows it to be a spiral galaxy
inclined by about $40^\circ$ from face-on \citep{cre03}, and
it is included in the list of 59 narrow line Seyfert~1 galaxies
complied by \cite{ver01}.
As is typical of such objects, it has a steep soft X-ray photon index
and shows some variability \citep{bol96}.
A spectrum covering 0.8--2.4\,micron\ presented by \cite{rod02} shows
no emission lines in the H-band and, as also found by \cite{sos01} and
ourselves, only Br$\gamma$ in the K-band.
Although their spectrum becomes somewhat less sparsely populated
shortward of the H-band,
presumably because the dilution of features by hot dust continuum is
much less, they still do not detect any coronal lines.
In this galaxy we detect neither stellar absorption features, nor
H$_2$ line emission (compact or extended).
This paucity of emission or absorption features makes it
extremely difficult to draw any conclusions about excitation
mechanisms or dynamics.

\subsubsection*{NGC\,1194}

NGC\,1194 is a highly inclined spiral galaxy.
It was classified as a Seyfert~1 by \cite{gri92}, who were
surveying warm IRAS sources, based on optical line widths and ratios.
In a survey of 12\,\micron\ galaxies, \cite{ima03} (who
classed it as a type~2, as given in the original sample
definition) obtained a 2.8--4.1\,\micron\ spectrum of this object,
finding no 3.3\,\micron\ PAH emission, but evidence for coronal
[Si{\sc\,ix}] at 3.9\,\micron. 
Our detection of only narrow Br$\gamma$ would tend to support the
Seyfert~2 classification.
\cite{sos01} observed strong CO bandhead absorption longward of
2.3\,\micron\ and we too, as for more than half of the nuclear spectra
in our sample, find relatively strong stellar absorption 
features present in the nuclear spectrum suggesting that recent or
currently active star formation is an important component of its
nuclear energy budget.

\subsubsection*{ESO\,438-G9}

ESO\,438-G9 was discovered to be a Seyfert galaxy by \cite{kol83}, who
also noted that it has strong optical Fe{\sc\,ii} lines, weak or
absent higher ionisation lines (probably due to the steep FUV
spectrum), and double absorption features blueward of H$\alpha$ and
H$\beta$.
In their survey of warm infrared galaxies, \cite{kew01} noted that this
object had strong broad Balmer lines, and that particularly the
[O{\sc\,iii]} line had broad wings.
The R-band image of \cite{mal98} shows it to be a spiral galaxy with a
very bright unresolved nucleus.
We observe a broad pedestal to the Br$\gamma$ line with a FWHM of
2000\,\kms. 
There are also clear stellar absorption features in the nuclear
spectrum, indicating a significant contribution to the K-band
continuum from late type stars within a few hundred parsecs of the AGN.

\subsubsection*{NGC\,7469}

Our spectrum of NGC\,7469 is similar to that of \cite{sos01}, showing
fairly deep absorption features as well as both narrow and broad
Br$\gamma$ emission.
Although this galaxy is considered to be a proto-typical Seyfert~1, much
of the attention it has received is focussed on the circumnuclear
ring on scales of $\sim1.5$--2.5\arcsec\ which is responsible for
about half of the galaxy's bolometric luminosity.
Additionally, about one third of the K-band continuum in the
nucleus itself originates in star formation \citep{maz94,gen95,dav04a}.
Comparing the luminosity profile of the 1-0\,S(1) line with its
dynamics at a resolution of $<$0.1\arcsec\ (and also the dynamics of
the cold molecular gas at 0.7\arcsec\ resolution), \cite{dav04a}
concluded that the very central peak in the 1-0\,S(1) emission was not
associated with the gas or star forming distributions, and was
most likely to originate in gas irradiated by X-rays from the AGN.
Unfortunately, in a 1\arcsec\ aperture, the 1-0\,S(1) flux from this
component is expected to contribute only about 20--25\% of the total
1-0\,S(1) flux.
It may not be possible -- on seeing limited scales -- to
resolve the more extended dominant component, that is presumably
excited by star formation, from the nuclear point source.

\section{Star Formation around the AGN}
\label{sec:starform}

In the previous section we have noted several times that deep stellar
absorption features are often present even in the central 100\,pc around the
AGN.
In this section we assess the evidence for active star formation in these
regions, particularly since this will enable us to place the following
discussion of the H$_2$ excitation mechanisms in context.
High spatial resolution spectroscopy has already provided direct evidence for
intense star formation on scales as small as a few tens of parsecs, and
even a few parsecs, around several type~1 AGN,
namely NGC\,7469 \citep{dav04a}, Mkn\,231 \citep{dav04b}, 
NGC\,3227 \citep{dav05}.
Other authors \cite[e.g.][]{cid04,sto01,gon01} have demonstrated through
empirical population synthesis that in about 40\% of type~2 Seyferts there is
recent (i.e. $<$1\,Gyr) star formation within a few hundred parsecs of the
nucleus.
It is therefore expedient that we consider whether there could be active and
intense star formation in the central apertures of the AGN presented here.

\begin{figure}
\epsscale{.3}
\plotone{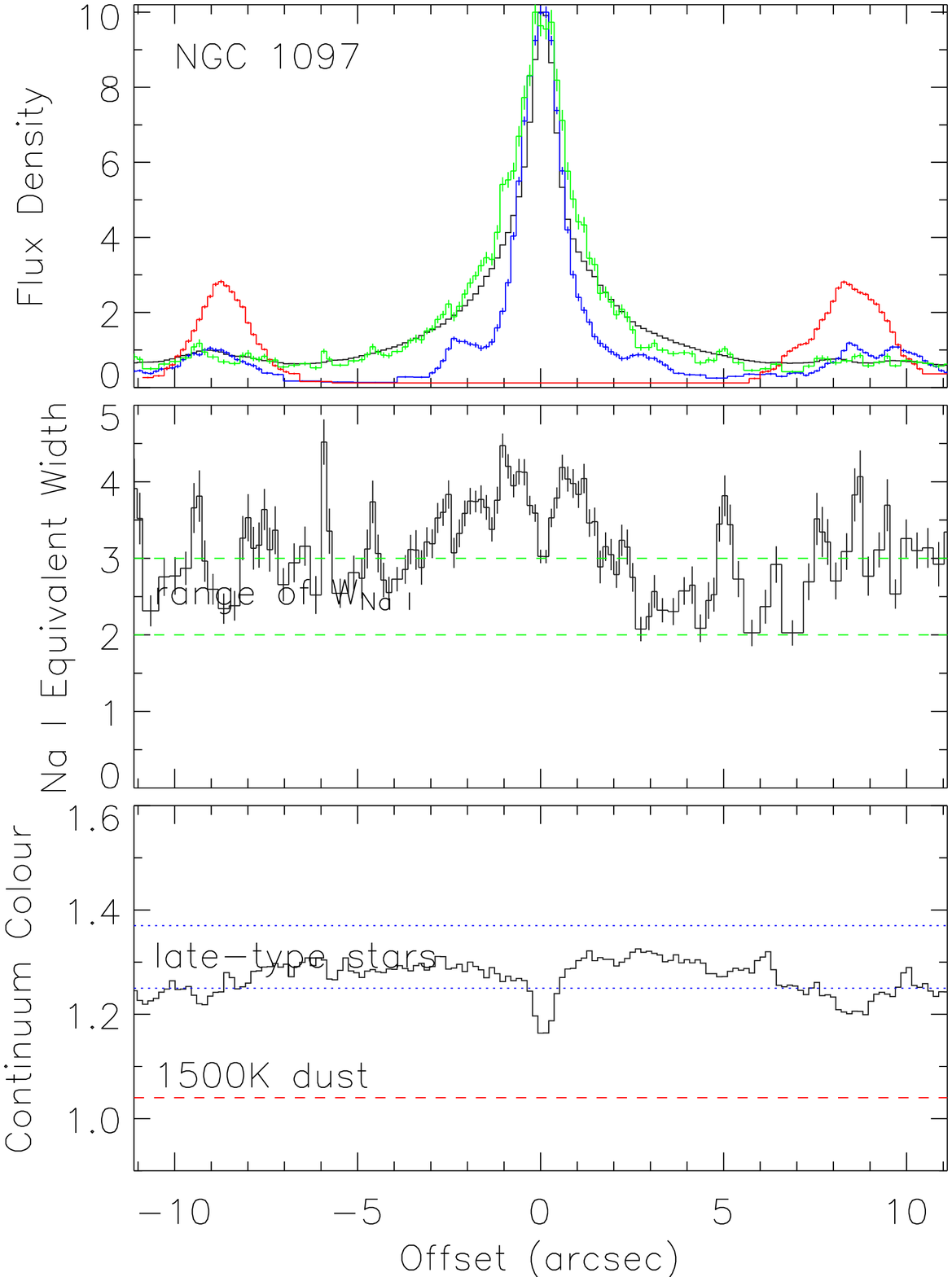}\plotone{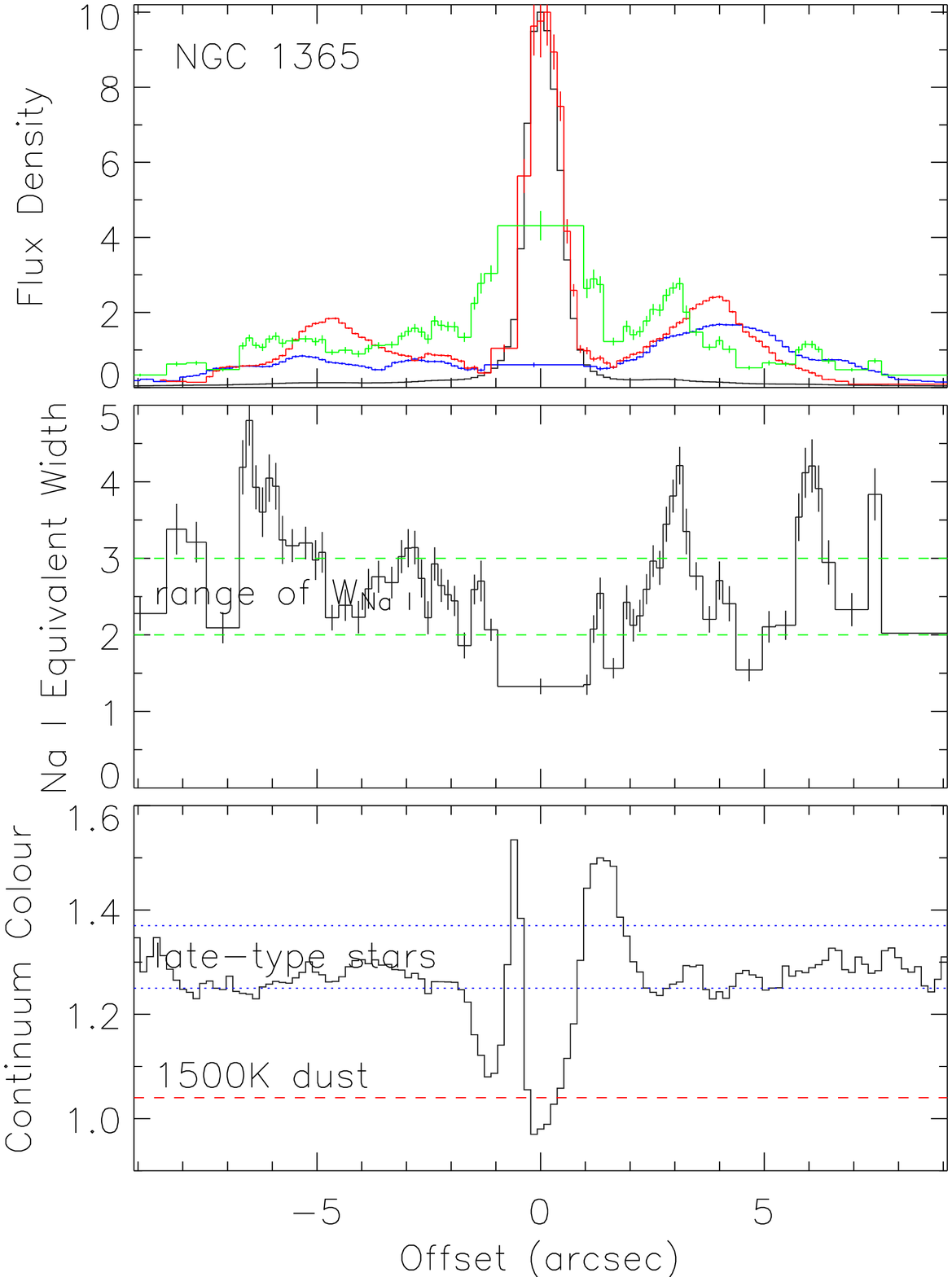}\plotone{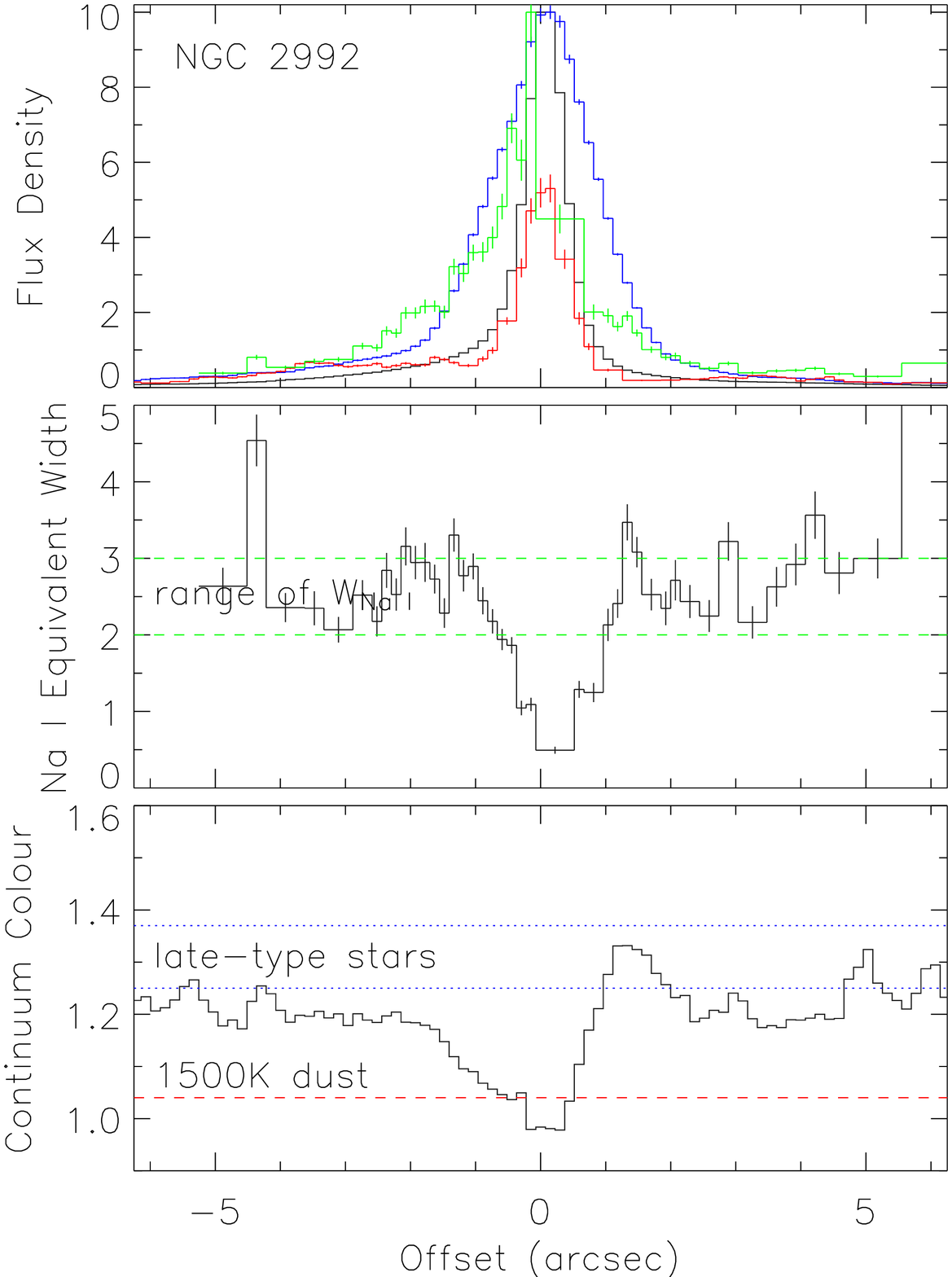}\plotone{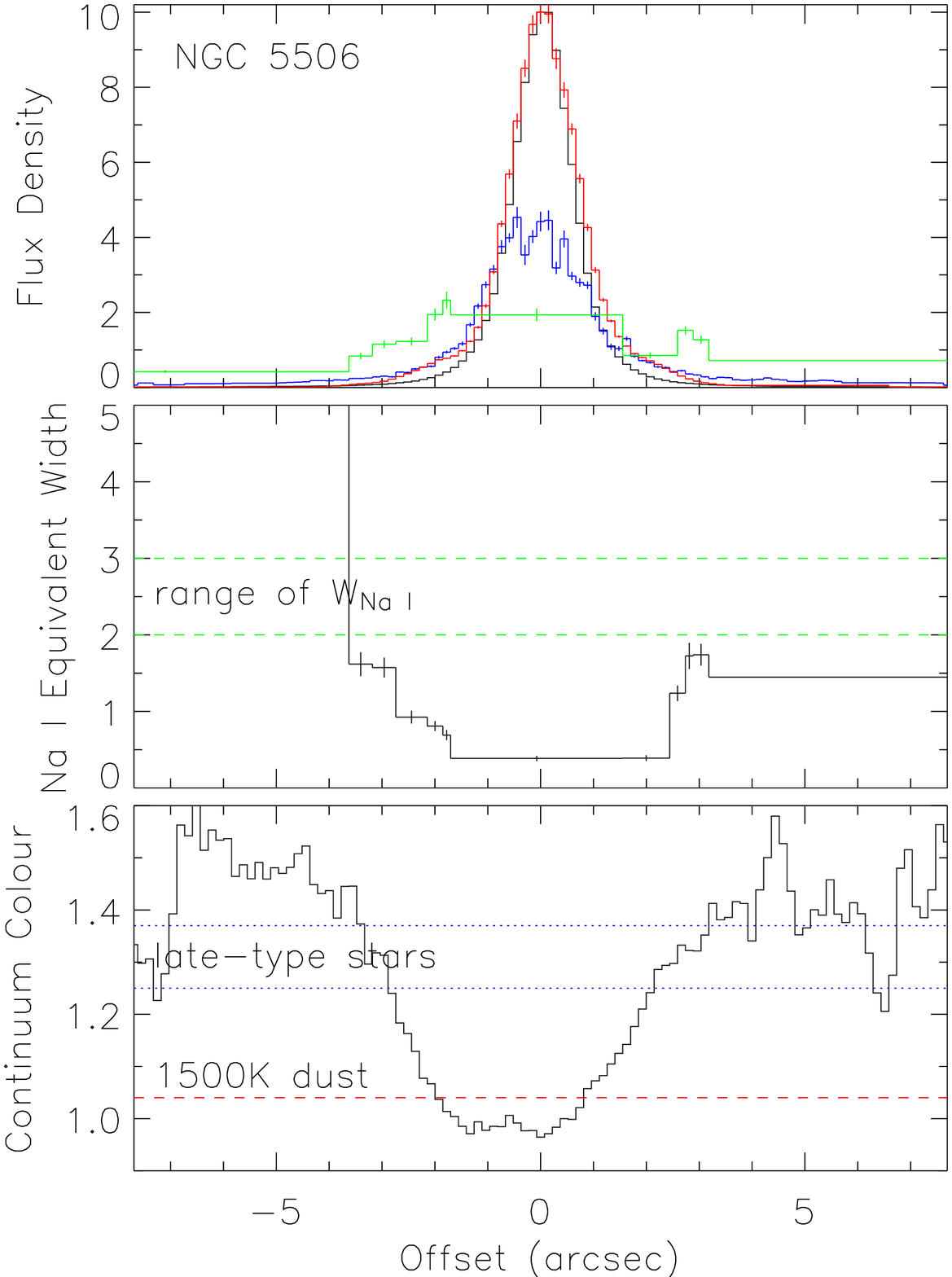}\plotone{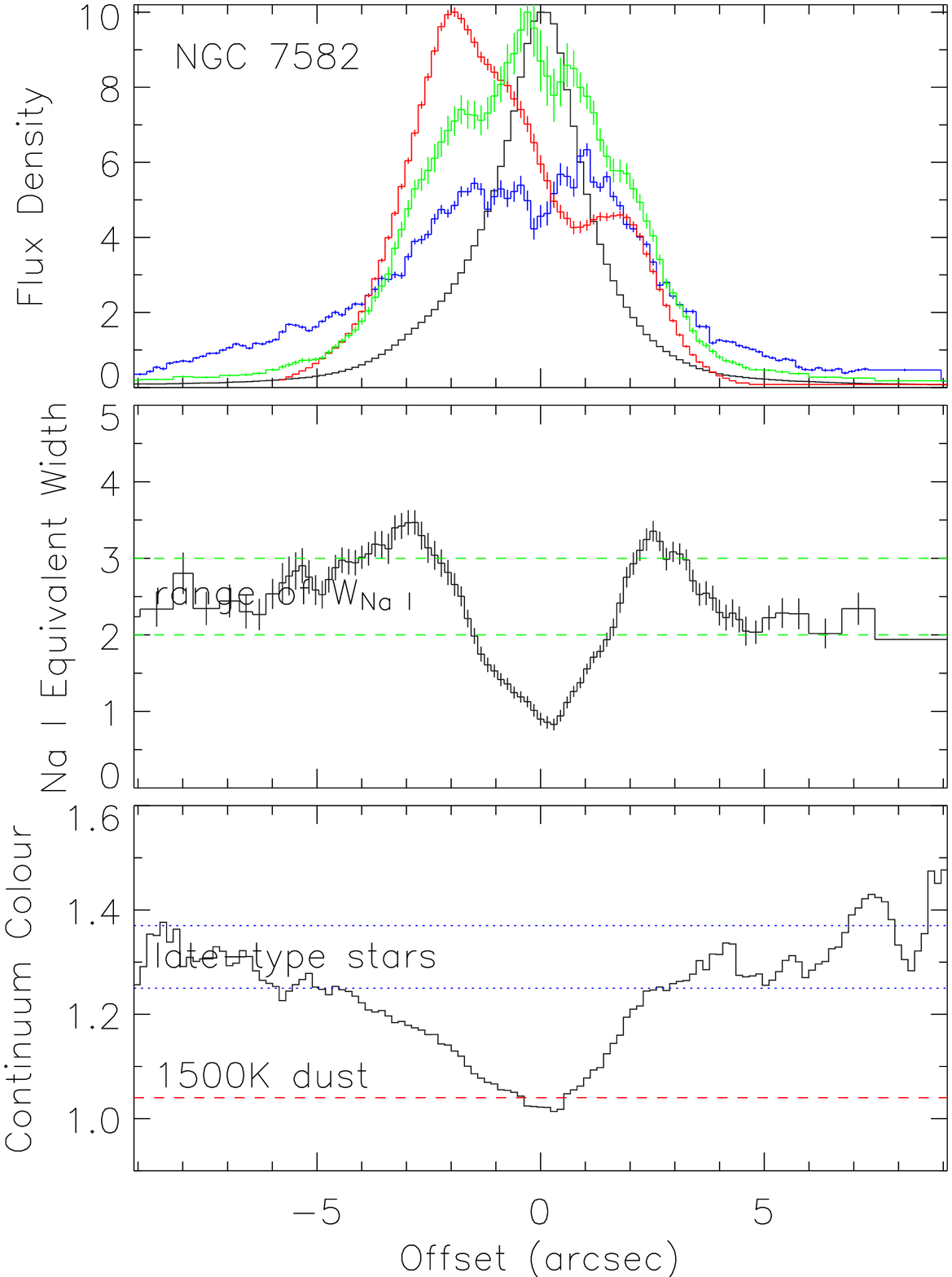}
%\centerline{\psfig{file=f06a.eps,width=6cm}\psfig{file=f06b.eps,width=6cm}\psfig{file=f06c.eps,width=6cm}}
%\vspace{5mm}
%\centerline{\psfig{file=f06d.eps,width=6cm}\psfig{file=f06e.eps,width=6cm}}
\caption{Spatial distributions out to a radius of 1\,kpc for each of of 
  the nearby ($\langle D \rangle = 25$\,Mpc) objects.
Top: total continuum (grey), H$_2$ 1-0\,S(1) (blue), Br$\gamma$ (red), and
  stellar continuum (green).
The latter three are binned so as to exceed a signal-to-noise
  of 10, but with a limit on the maximum number of pixels included in
  any one bin.
Centre: 2.206\,\micron\ Na{\sc\,i}
  equivalent width, binned so as to exceed a signal-to-noise
  of 10 as above.
Overlaid in the centre panel is the range of equivalent widths (2--3\,\AA) one
would expect for most star formation histories.
Bottom: continuum colour, defined here as the ratio between the flux densities
at 2.19\,\micron\ and 2.25\,\micron.
Overlaid in the bottom panel are the colours one would expect for late
  type stars and for pure hot dust emission at 1500\,K. If the late
  type stars were reddened by $A{_V}=10$, the colour would
  decrease from around 1.3 to 1.15.
}
\label{fig:spat-near}
\epsscale{1.0}
\end{figure}

\begin{figure}
\plotone{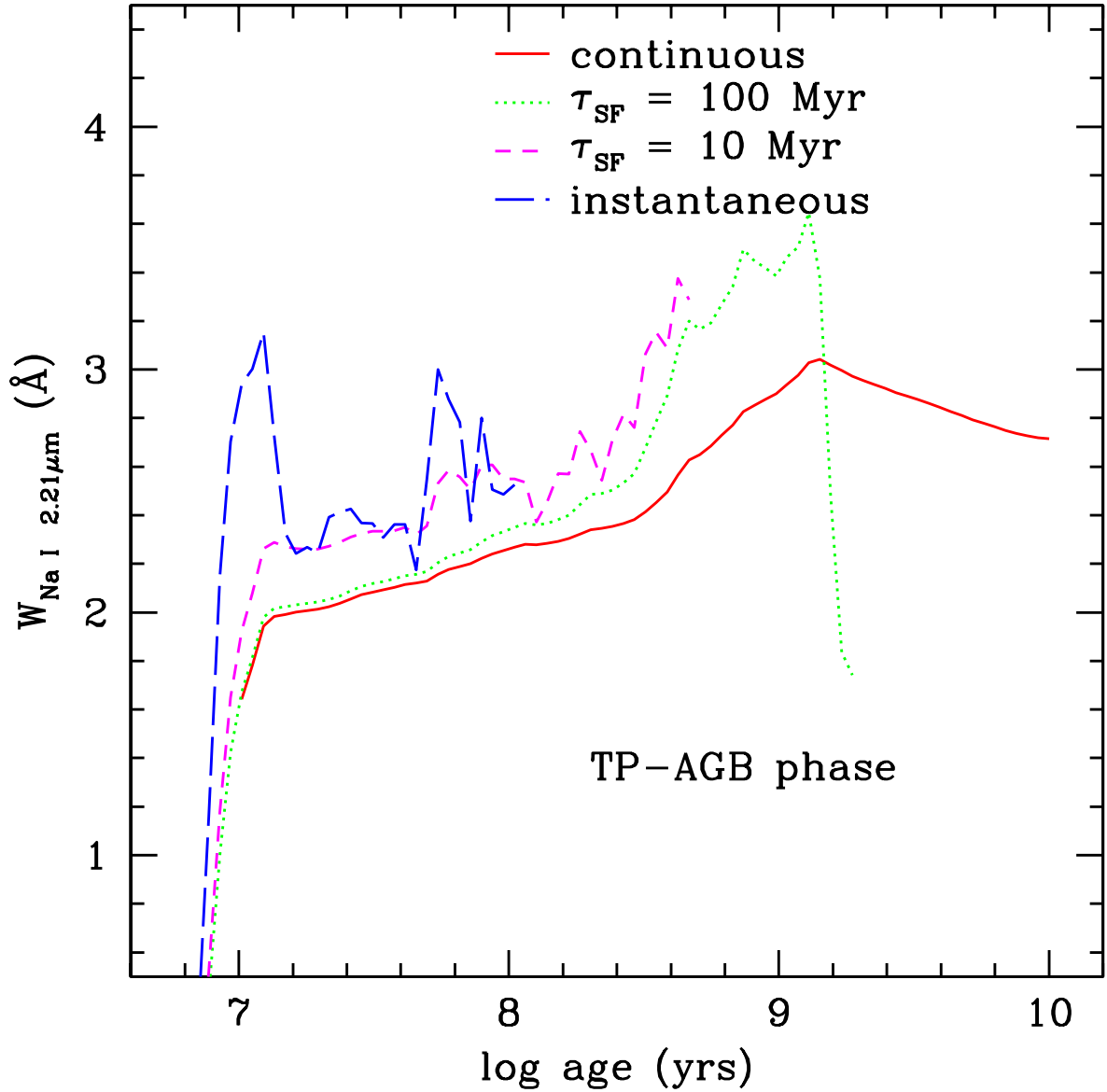}
%\centerline{\psfig{file=f07.eps,width=10cm}}
\caption{Equivalent width of the 2.206\,\micron\ Na{\sc\,i} line as a function
  of age, calculated using STARS (which includes the thermally pulsing
  asymptotic giant branch stars). 
  Several different star formation histories
  are shown: instantaneous, a decay time
  of 10\,Myr, a decay time of 100\,Myr, and continuous. For each history, data
  are plotted where the K-band luminosity is at least 1/15 of its
  maximum. This indicates that one expects $W_{\rm Na\,I}$ in the range
  2--3\,\AA\ for nearly all star forming scenarios.}
\label{fig:EWna}
\end{figure}

Fig.~\ref{fig:spat-near} shows, for each of the five nearer galaxies, the
spatial distributions of various quantities along the slit to a projected
distance of 1\,kpc from the centre.
The top panels trace the relative surface brightnesses of the total
K-band continuum together with the stellar continuum, 1-0\,S(1) and
Br$\gamma$ line emission, the latter two with the same normalisation.
We have attempted to measure only the narrow Br$\gamma$.
However, in several cases this is difficult to separate from the broad
emission (see Fig.~\ref{fig:nuclear-spec-near}) and hence over the central
arcsec should be taken as a guide only.
The centre panels show the equivalent width of the 2.206\,\micron\ Na{\sc\,i}
stellar absorption line, $W_{\rm Na\,I}$.
To assist in interpreting this, we show also the typical range of values
(2--3\,\AA) that one would expect it to take.
This is based on results of the STARS star formation models
\citep{ste98,tho00,dav03} shown in Fig.~\ref{fig:EWna}.
This code calculates the distribution of stars in the Hertzsprung-Russell
diagram as a function of age for an exponentially decaying star formation
rate.
Using empirically determined $W_{\rm Na\,I}$ from library spectra
\citep{for00}, the code then computes the 
time-dependent $W_{\rm Na\,I}$ for the entire cluster of stars.
The code includes the thermally pulsing asymptotic giant branch
(TP-AGB) stars which have a very significant impact on the 
depth of the absorption features at ages of 0.4-2\,Gyr \citep{for03,mar04}.
The stellar continuum in the top panel is derived assuming it exhibits
a constant equivalent width, which this figure shows to be a
reasonable zero-order approximation.
Finally, the bottom panels in Fig.~\ref{fig:spat-near} show how the colour of
the continuum, defined here as the ratio of the flux densities at
2.19\,\micron\ and 2.25\,\micron, varies along the slit.
Again, several reference values are superimposed: the range of colours typical
of late type stars, and the colour of a blackbody continuum at 1500\,K
representative of hot dust associated with an AGN.
Extinction is unlikely to change these colours significantly since the relevant
quantity is the differential extinction between 2.19 and 2.25\,\micron. This
only starts to become important for screen extinction models with
$A_V\gtrsim10$.

In general, away from the nucleus itself, $W_{\rm Na\,I}$ is consistent with
what one expects from the star formation models, indicating that there
is little or no dilution.
The nuclear Br$\gamma$ emission at these radii (e.g. NGC\,1097,
NGC\,1365) is indicative of young stars, and the spatial coincidence
of 1-0\,S(1) emission would suggest that it is a related 
phenomenon: either from photo-dissociation regions or associated with
supernovae. 
In this sense, the occasional peaks with $W_{\rm Na\,I} > 3$\,\AA\
could be due to the stars in individual clusters passing through the
late-type supergiant phase at $\sim$10\,Myr.

However, the regions close around the nucleus tell a rather different story.
The stellar continuum appears to increase steeply;
but in every case $W_{\rm Na\,I}$ exhibits a dramatic decrease,
which is also associated with a reddening of the continuum colour.
The simplest explanation is that the stellar continuum is strongly diluted by
continuum emission from hot dust.
If these features were unresolved, as in NGC\,1097, one would immediately
assign them to dust heated to 500--1500\,K by the AGN.
But more typically, they occur over an extent of 200\,pc or more and are
therefore unlikely to be directly associated with an AGN which is expected to
heat grains to such high temperatures over rather limited distances -- for
example the inner edge of a canonical torus on parsec scales or very small
grains transiently heated by a jet, as may be occuring in the central
few tens of parsecs in NGC\,1068 \citep{rou04}.
AGN often exhibit ionisation cones which can extend over distances
of hundreds of parsecs, and both NGC\,1365 and NGC\,7582 possess one
\citep{sto91}.
Our slits for these two objects are oriented through the ionisation
cones. 
But over the 5\arcsec\ regions corresponding to the extent of
the cones, the respective data sets show totally different properties
from each other -- suggesting that the observations are not tracing
processes associated with the ionisation cones themselves.
In addition, the ionisation cone in NGC\,4945 \citep{moo96}, shows that one
would not expect them to produce significant K-band continuum:
these authors only see the cone in continuum (J-band rather than
K-band) because it reduces the extinction to the starlight behind.
Based on this, one would expect the cone neither to dilute the stellar
features nor to redden the continuum.
We conclude that the AGN and its associated phenomena such as ionisation cones
are unlikely to produce the combination of properties we observe here.

On the other hand, the reduction in $W_{\rm Na\,I}$ and reddening of the
continuum colour are spatially coincident with the brighter
stellar continuum, and so -- as for the circumnuclear regions further
out -- could be associated with active star formation.
The approximate coincidence of Br$\gamma$ 
emission provides additional circumstantial evidence for a population of
very young stars, or at least on-going star formation.
The dilution of the stellar features and reddening of the continuum
colour are still most likely attributable to hot dust, although at
present one can only make conjectures about the heating source.
It has been suggested that a population of very small grains could be
heated to temperatures of 500--1000\,K by young stars
\cite[e.g.][]{sel84,dav02}.
However, there is little evidence for such
a process seen in galactic H{\sc ii} regions.
On the other hand, the environment of the region around the AGN -- in
terms of density, pressure,
turbulence, energetics, etc. -- may be sufficiently different from
that of galactic H{\sc ii} regions that it is able to produce a population
of sufficiently small grains, as well as drive them near enough to the
young stellar clusters so that they can be heated.
One could also envisage other processes which may be at work, such as
Compton heating by energetic photons from the AGN.
However, addressing the issue of the dust heating mechanism is beyond
both the data presented here and the scope of this work.

The pertinent issue is that for these extended ($\sim$200\,pc) nuclear
regions, one must consider all the observed characteristics together: 
the line emission, reduction in $W_{\rm Na\,I}$, reddening of the
continuum, and also the brightness of the stellar continuum.
A natural, and arguably the simplest, explanation to account for all
these simultaneously is the presence of active star formation, from
which the combination of direct emission (e.g. bright stellar continuum) and
indirect emission (e.g. recombination lines and plausibly hot dust
re-radiation) can account for what has been observed.

\section{H$_2$ Excitation Diagrams and PDR Models}
\label{sec:poplev}

For the H$_2$ line
fluxes listed in Table~\ref{tab:measdat} and~\ref{tab:ratios}, 
we have calculated the implied molecular
column densities, $N_{vj}\equiv 4\pi f/A\Omega$, in 
the upper rotational-vibrational ($vj$) levels of
the observed transitions, where $f$ is the observed flux, $A$ is the
radiative rate, and $\Omega=1.0^\Box\arcsec$ is our aperture size. 
We assume that the quadrupole transitions are optically thin
and we use the radiative
$A$ values as given by \citet{wol98}.
After normalising the population distributions to the
$N_{1,3}/g_{1,3}$ level (i.e. the 1-0\,S(1) line), we have plotted
them as 
$\log{N_{vj}/g_{vj}}$ (where $g_{vj}$ is the statistical weight)
 against energy (in K) of each observed $vj$ level.
In such excitation diagrams, the log$N/g$ points for isothermal
populations should lie on a straight line.

In order to interpret these diagrams, we consider purely thermal
emission as well as a representative set of
5 PDR models which are described by \cite{dav03}.
These models consist of
static, plane-parallel, semi-infinite clouds that are
exposed to isotropic FUV radiation fields \citep{ste89,stn99}. 
At each cloud depth we compute the equilibrium
atomic to molecular hydrogen density ratio,
$n({\rm H})/n({\rm H_2})$, and we solve for the 
steady-state population densities in the rotational
and vibrational H$_2$ levels in the ground electronic
state. In solving for the $vj$ populations
we include the effects of FUV-pumping via the Lyman
and Werner bands, collisional
processes with H$^+$, H, and H$_2$, 
and quadrupole radiative decays. 

\begin{figure}
\plotone{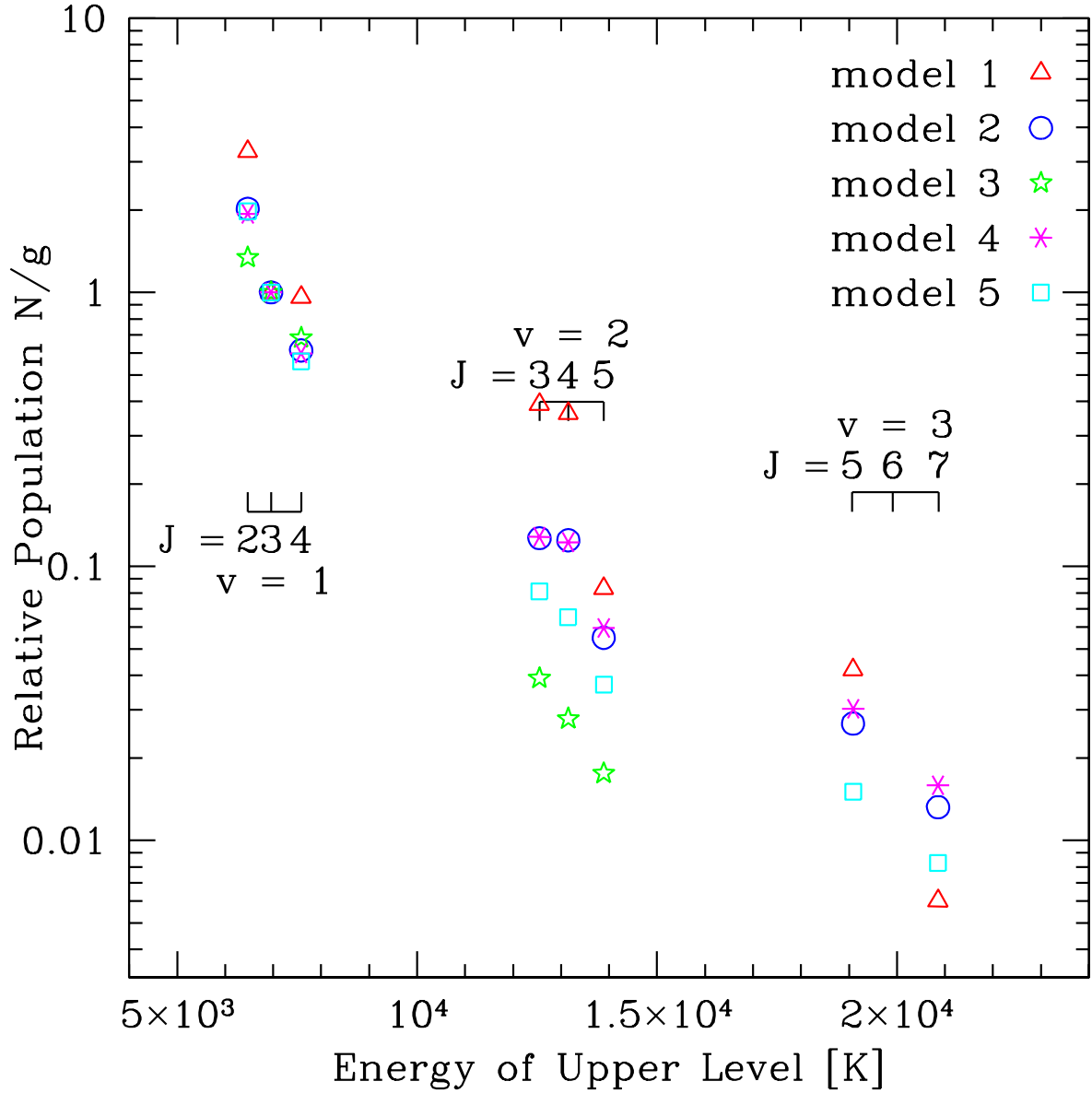}
%\centerline{\psfig{file=f08.eps,width=70mm}}
\caption{Excitation diagrams for hot molecular hydrogen
  ratios in the five models we consider, covering a wide range of
  gas density and FUV intensity parameters.
See text for details of the models.}
\label{fig:poplev-model}
\end{figure}

The 5 models cover a range of illuminating
FUV field intensities $\chi$,
relative to the
FUV field in the local interstellar medium 
\citep[$2.1\times10^{11}$\,photons\,s$^{-1}$\,m$^{-2}$ over
  912--1130\,\AA,][]{dra78}, and total hydrogen particle densities
$n_{\rm H} = n({\rm H}) + 2n({\rm H_2})$ in the clouds.
In all of the models except model 1 (for which we keep the temperature
constant at $T=100$\,K), we assume that the gas temperature varies
with cloud depth and set the temperature at the cloud edge.
The gas temperature is a maximum in the outer atomic zone and declines
significantly to $\sim 25$\,K as the gas becomes molecular, a
behaviour that is consistent with both theoretical expectations for,
and empirical measurements of, dense PDRs.
The details of the models and a description of their salient features is
given by \cite{dav03} and are not repeated here.
Instead, we summarise the parameters 
($\chi$, $n_{\rm H}$, $T_{\rm max}$) in Table~\ref{tab:models}.
The excitation diagrams for the models are plotted in
Fig.~\ref{fig:poplev-model}, which shows the extent of their coverage
of the relative population parameter space.

\begin{figure}
\plotone{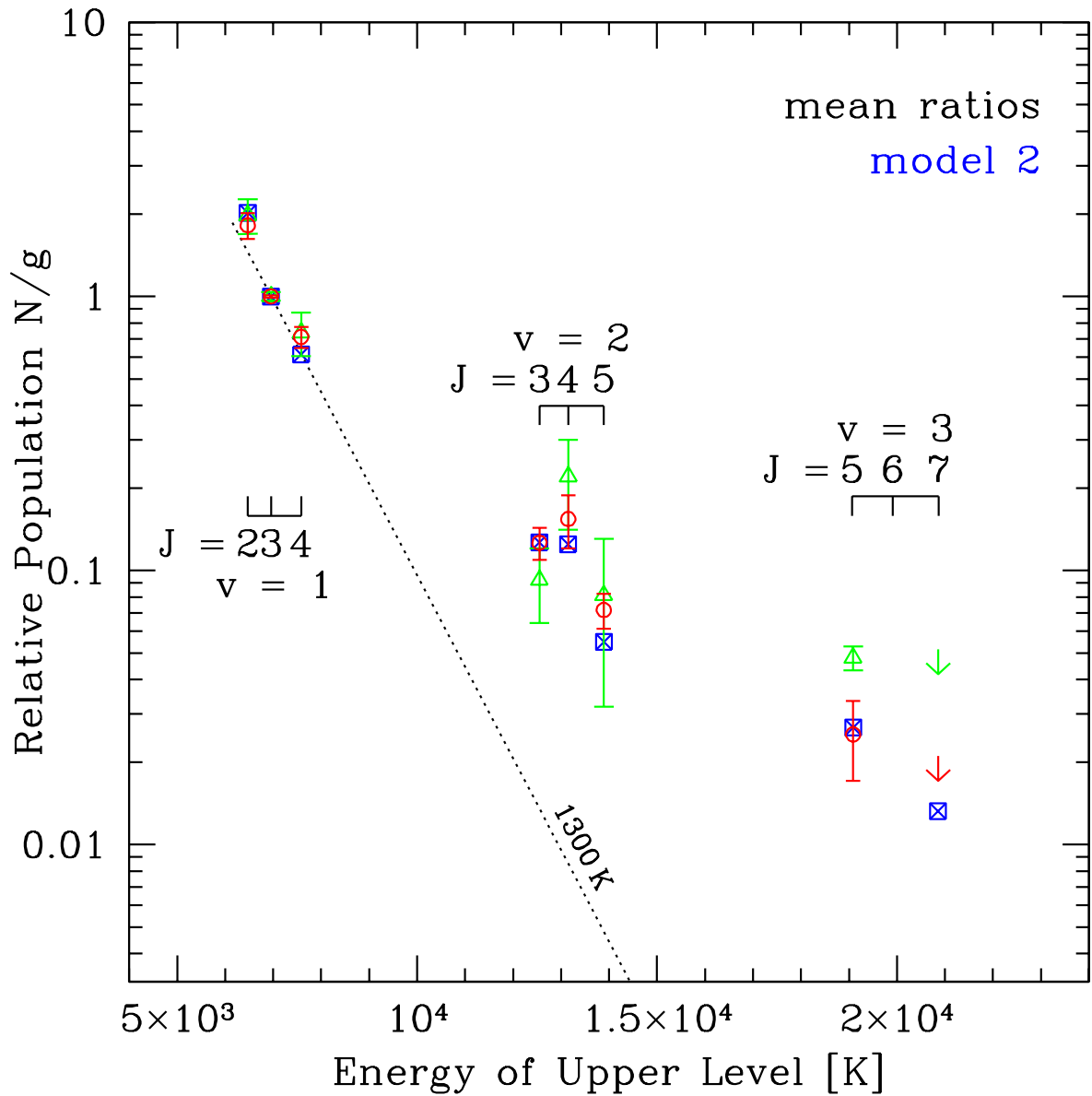}
%\centerline{\psfig{file=f09.eps,width=70mm}}
\caption{Excitation diagrams for the mean hot molecular hydrogen
  ratios.
Points are plotted where the lines were detected in at least 3
  objects; otherwise 3$\sigma$ upper limits are shown.
The error bars indicate the standard deviation of the ratios among the
  objects.
Green triangles correspond to the mean nuclear H$_2$ populations for all
  objects;
red circles to the mean circumnuclear populations in the nearby
  ($\langle D \rangle = 25$\,Mpc) objects.
Blue squares denote the populations for our PDR model 2.
The dotted line indicates where the populations would lie for an
isothermal cloud at 1300\,K, a reasonable match to the $\nu=1-0$ levels.}
\label{fig:poplev-mean}
\end{figure}

What is immediately clear from the mean values for the nuclear and
circumnuclear spectra, shown in Fig.~\ref{fig:poplev-mean} (where the
error bars denote the standard devation of the values among the
objects), as well as from the individual values is that the $\nu=1$
levels are well thermalised.
In every case these levels can be reproduced by an isothermal cloud at
1300\,K.
However it is not possible to say from just the $\nu=1$
levels whether the H$_2$ molecules have been excited by a thermal
process such as a shock front, or whether it is due to high gas
density regardless of the mechanism which provides the energy.
The higher levels suggest the latter since in every case they lie
above the prediction for 1300\,K.
To compensate for this, one would need either multiple thermal
components, or a fluorescent component.
The former is not a satisfactory solution because the temperature
needed -- about 3000\,K between the $\nu=1$ and $\nu=2$ levels, and of
order 5000\,K between the $\nu=2$ and $\nu=3$ levels -- would mean
that the molecules are rapidly dissociated.

The possibility that instead a fluorescent excitation mechanism is
responsible for the $\nu=2$ and $\nu=3$ populations is given by the
blue symbols in the figures, which represent our PDR model~2.
In this model, the $\nu=1$ levels are collisionally thermalised as a
result of the high $10^4$\,cm$^{-3}$ gas density. 
In addition, the model exhibits a shift in the ortho-para ratio for the
$\nu=2$ levels: the $j=4$ para level is above the adjacent $j=3$ and $j=5$
ortho levels.
Suppression of the ortho-para ratio below the value of 3 obtained in local
thermal equilibrium in warm gas is a typical signature of excitation by FUV
pumping. 
It occurs because the ortho UV absorption lines have a higher optical depth,
and hence the vibrationally pumped ortho-H$_2$ is suppressed with
respect to para-H$_2$ \citep{stn99}.
It is exactly this effect that can be seen in the H$_2$ population
diagrams of the AGN for which all 3 relevant lines have been detected,
both in the nuclear and circumnuclear regions.

\begin{figure}
\epsscale{.4}
\plotone{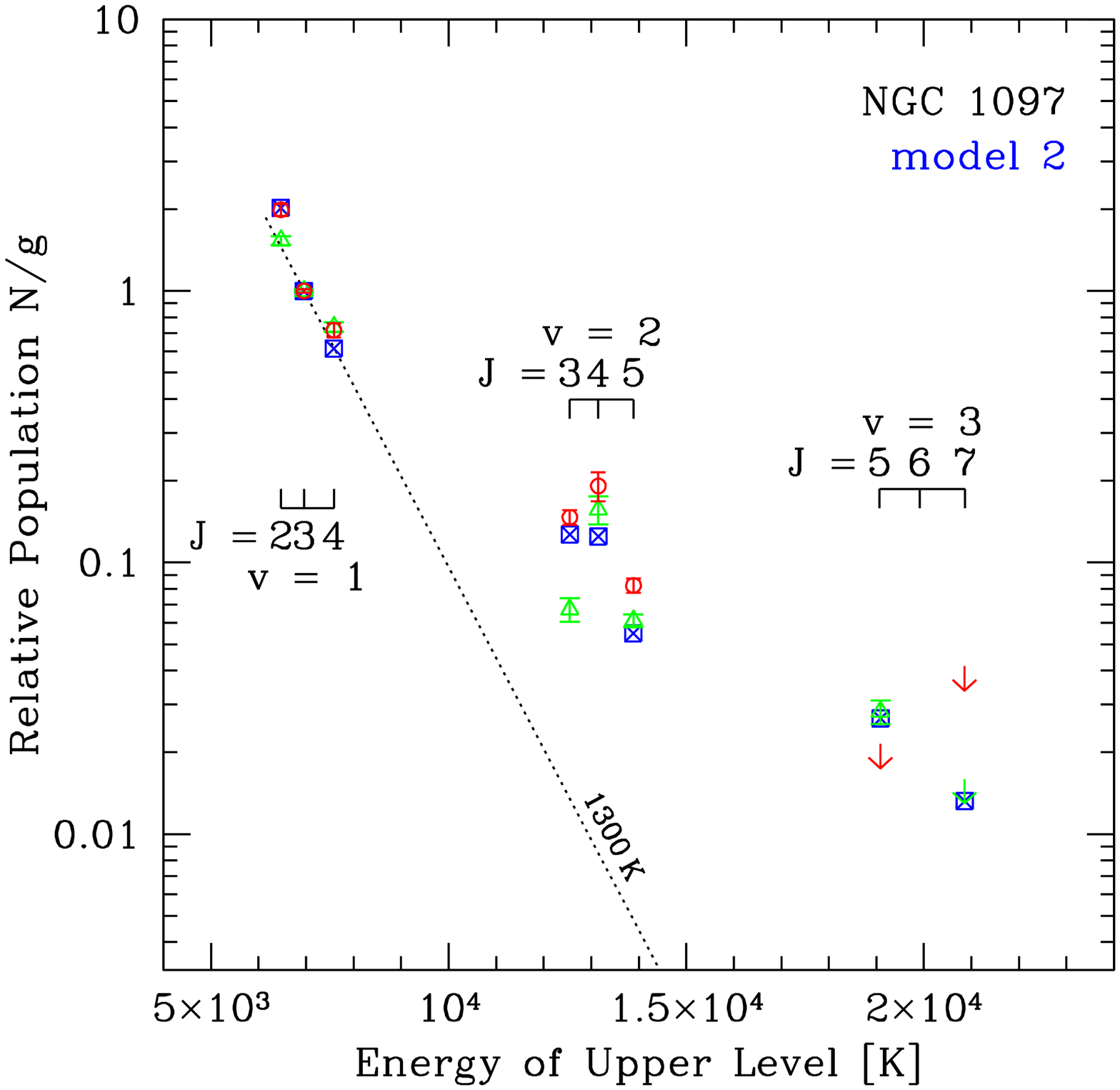}\plotone{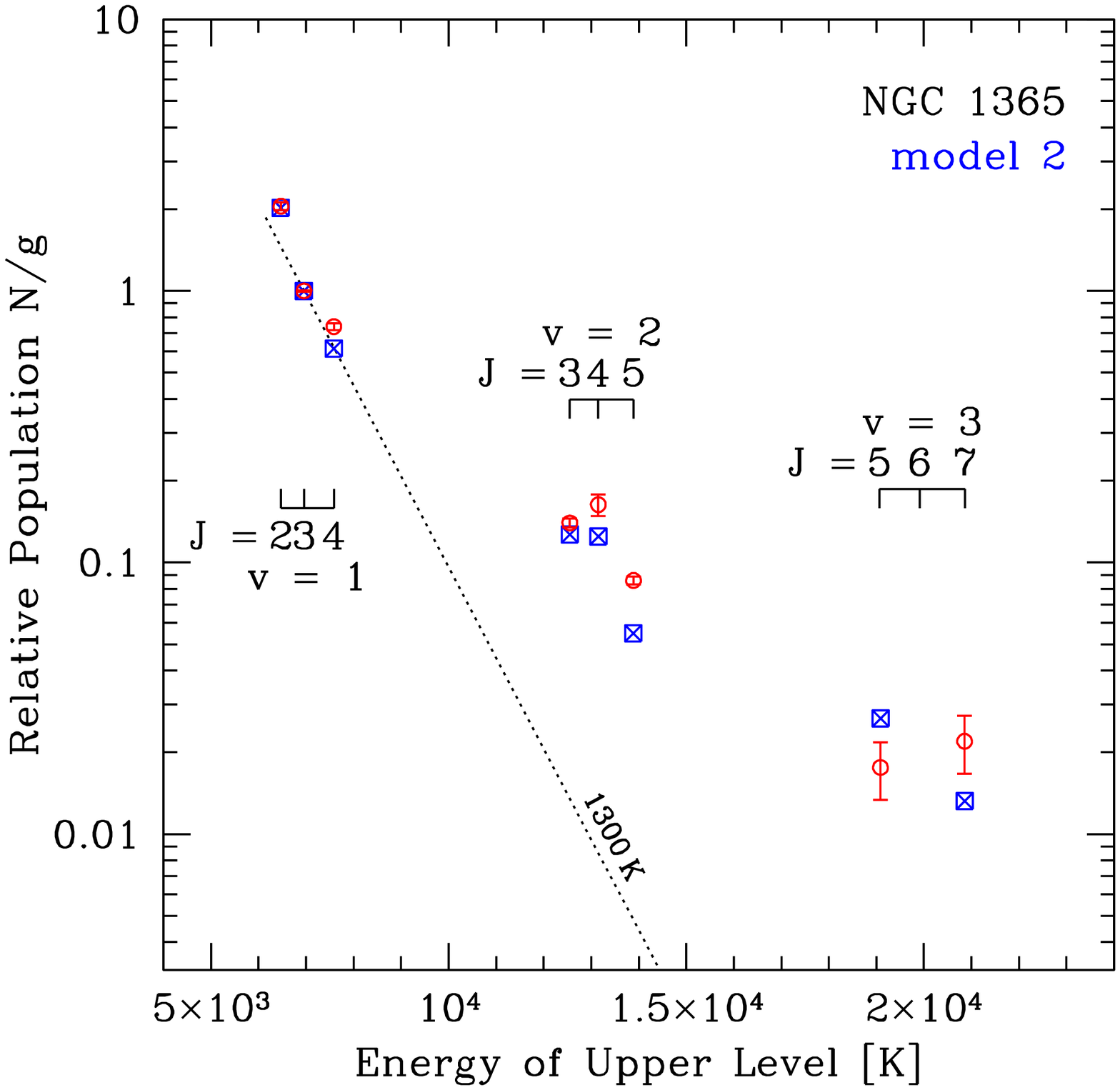}\plotone{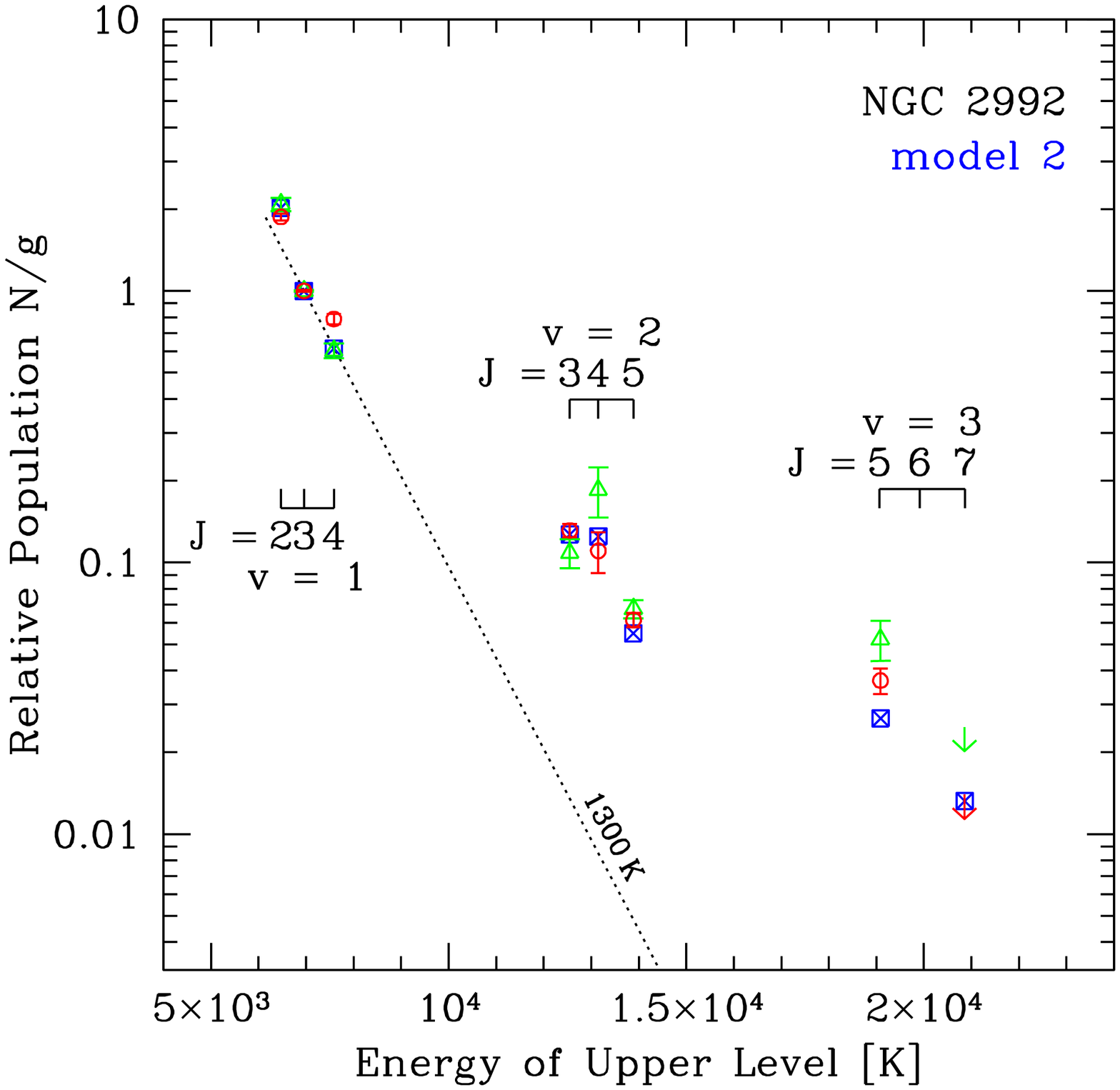}\plotone{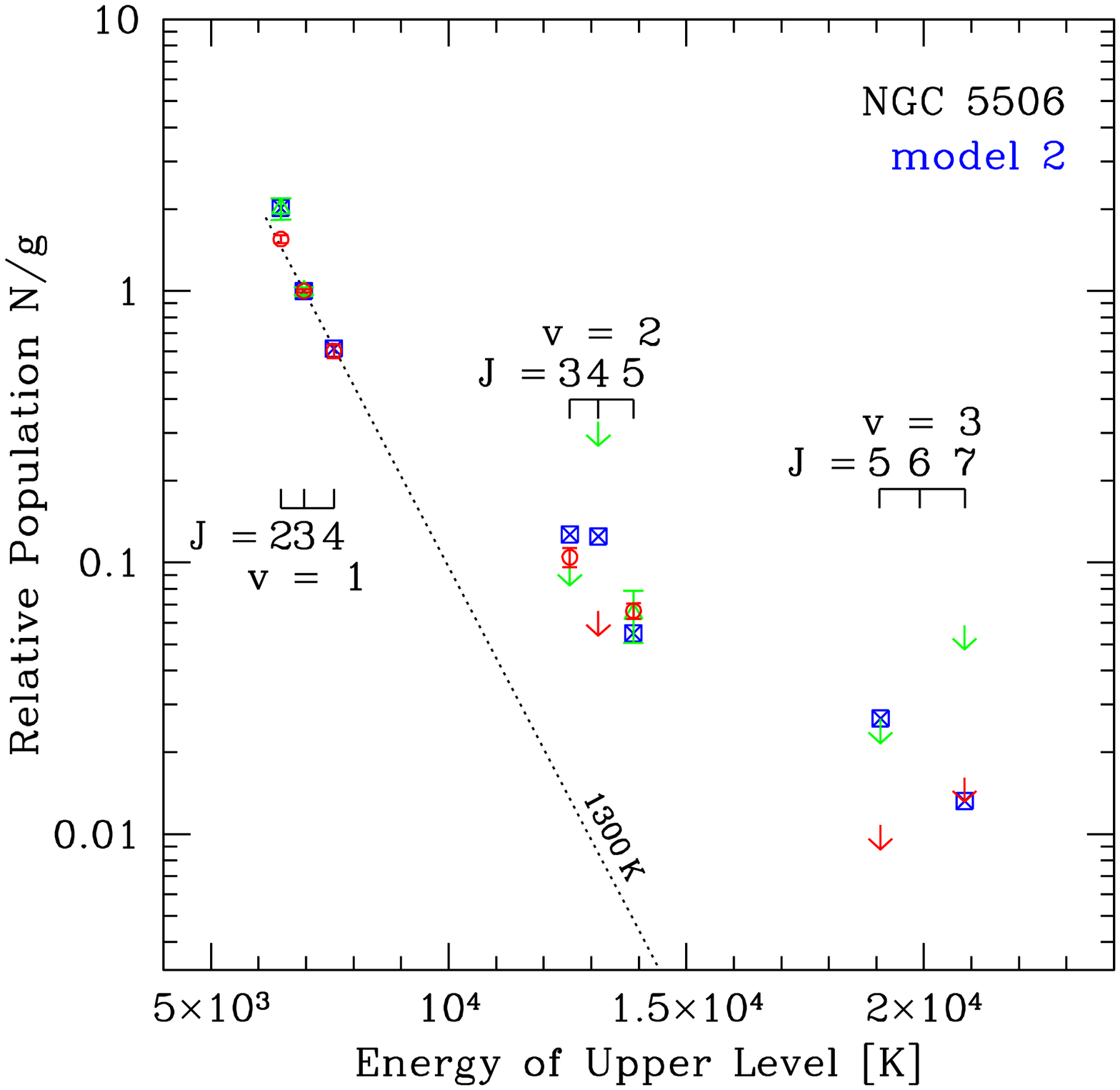}\plotone{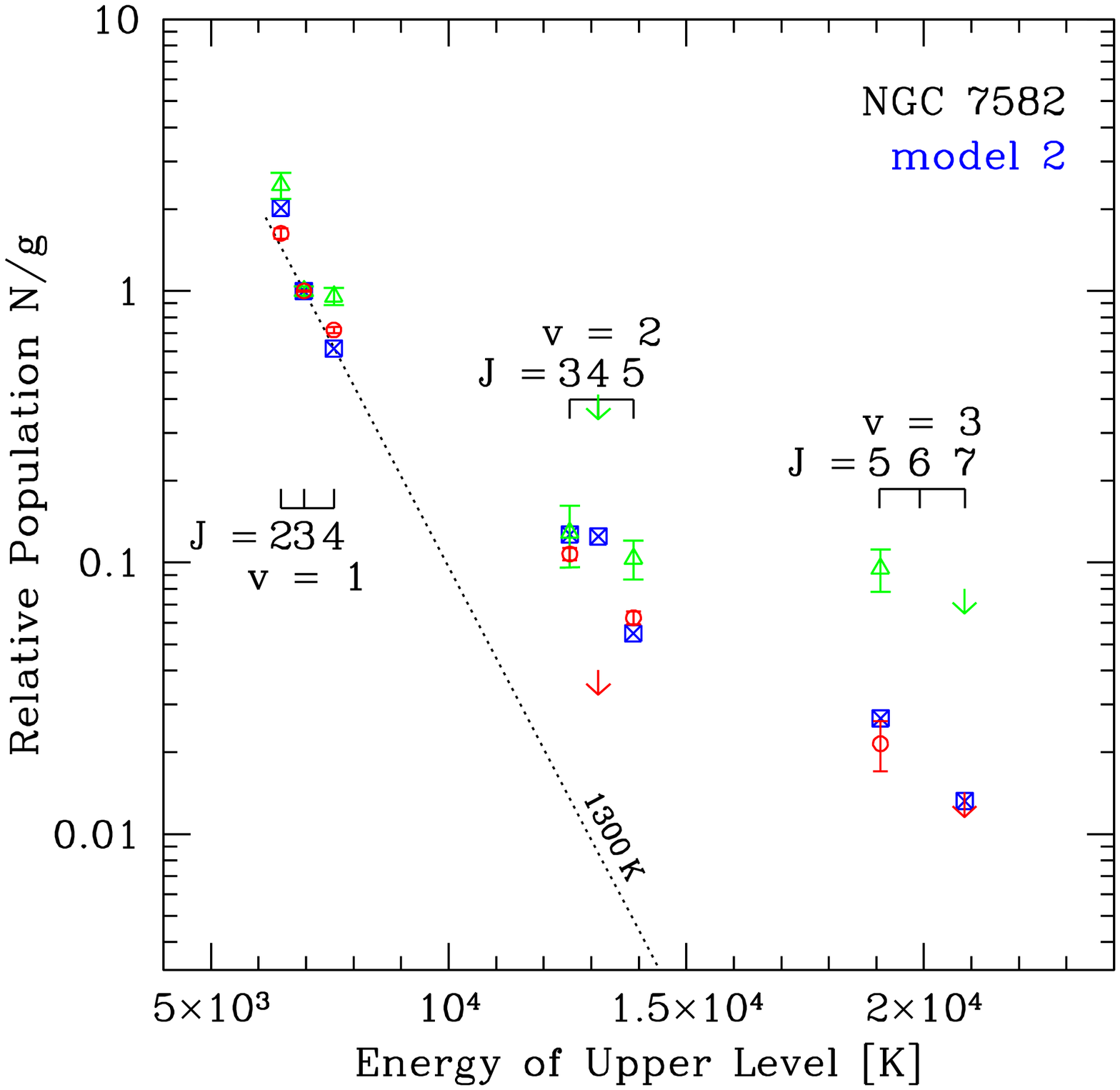}
%\centerline{\psfig{file=f10a.eps,width=70mm}\hspace{5mm}\psfig{file=f10b.eps,width=70mm}}
%\centerline{\psfig{file=f10c.eps,width=70mm}\hspace{5mm}\psfig{file=f10d.eps,width=70mm}}
%\centerline{\psfig{file=f10e.eps,width=70mm}}
\caption{Excitation diagrams for the hot hydrogen molecules in each
of the nearby ($\langle D \rangle = 25$\,Mpc) objects.
Arrows denote $3\sigma$ upper limits, derived from the residual
spectrum after subtraction of the stellar continuum and line emission.
Green triangles correspond to the nuclear H$_2$ populations;
red circles to the circumnuclear region.
Blue squares denote the populations for our PDR model 2.
The dotted line indicates where the populations would lie for an
isothermal cloud at 1300\,K, a reasonable match to the $\nu=1-0$ levels.}
\label{fig:poplev-near}
\epsscale{1.0}
\end{figure}

\begin{figure}
\epsscale{.4}
\plotone{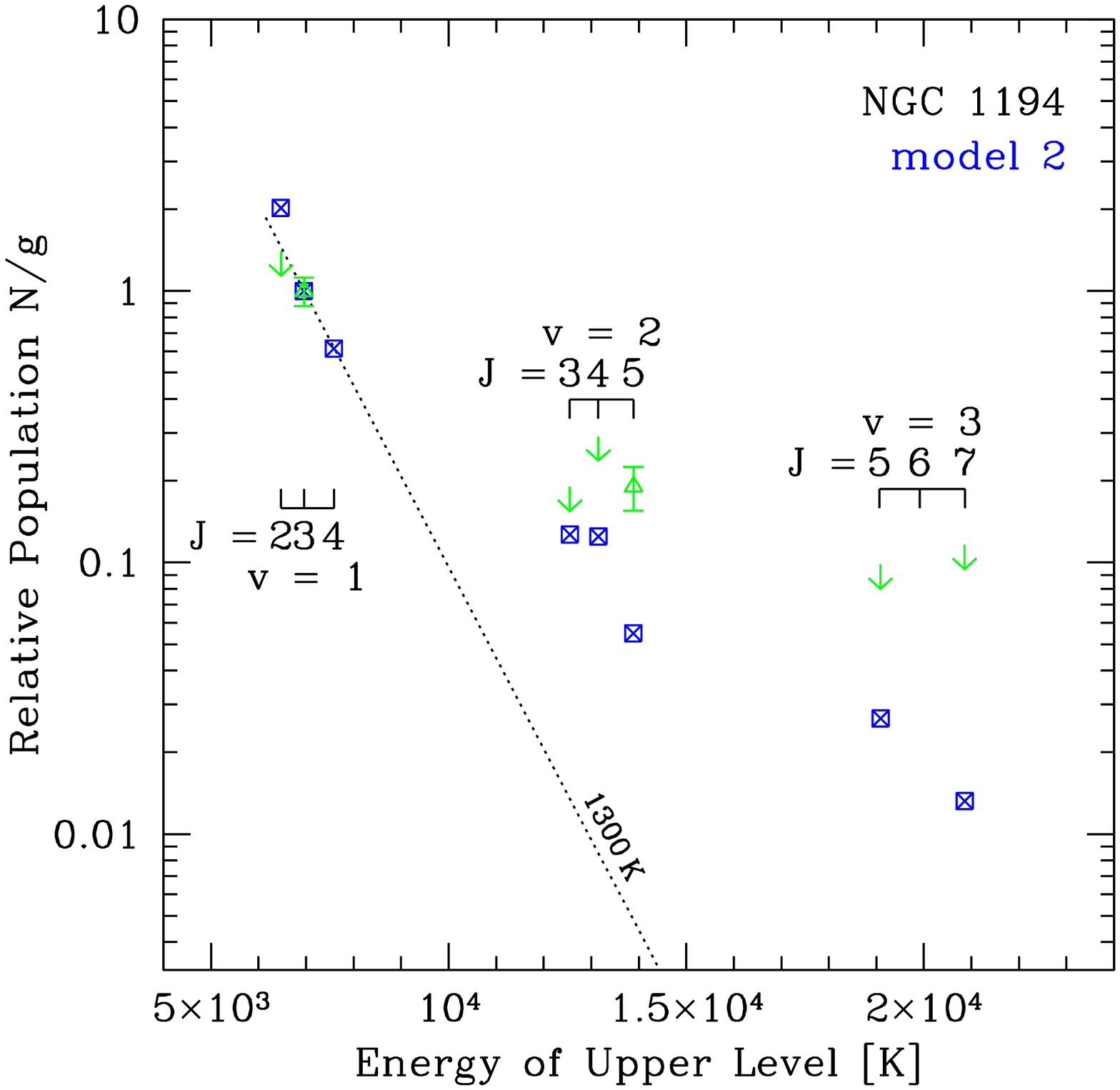}\plotone{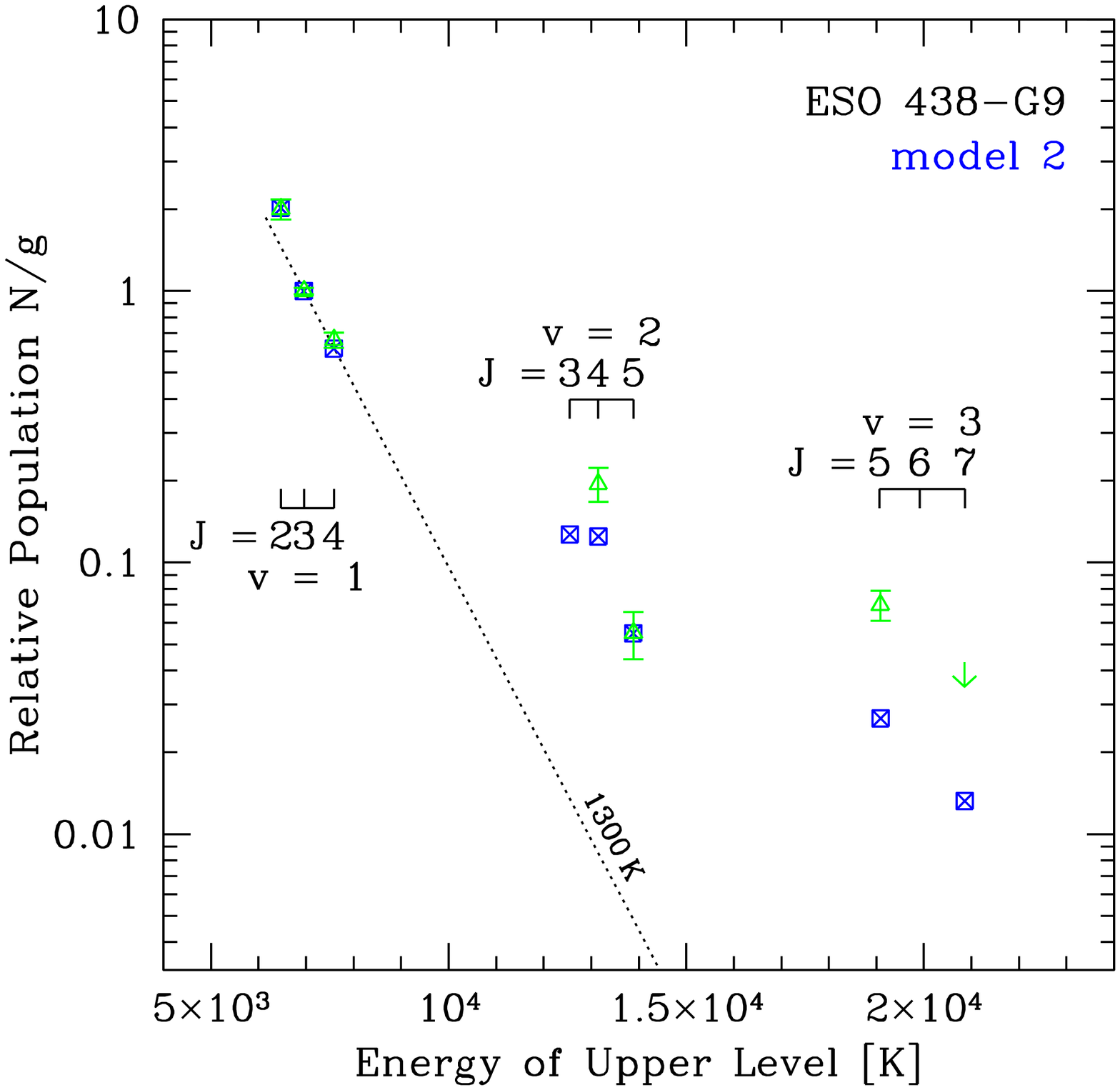}\plotone{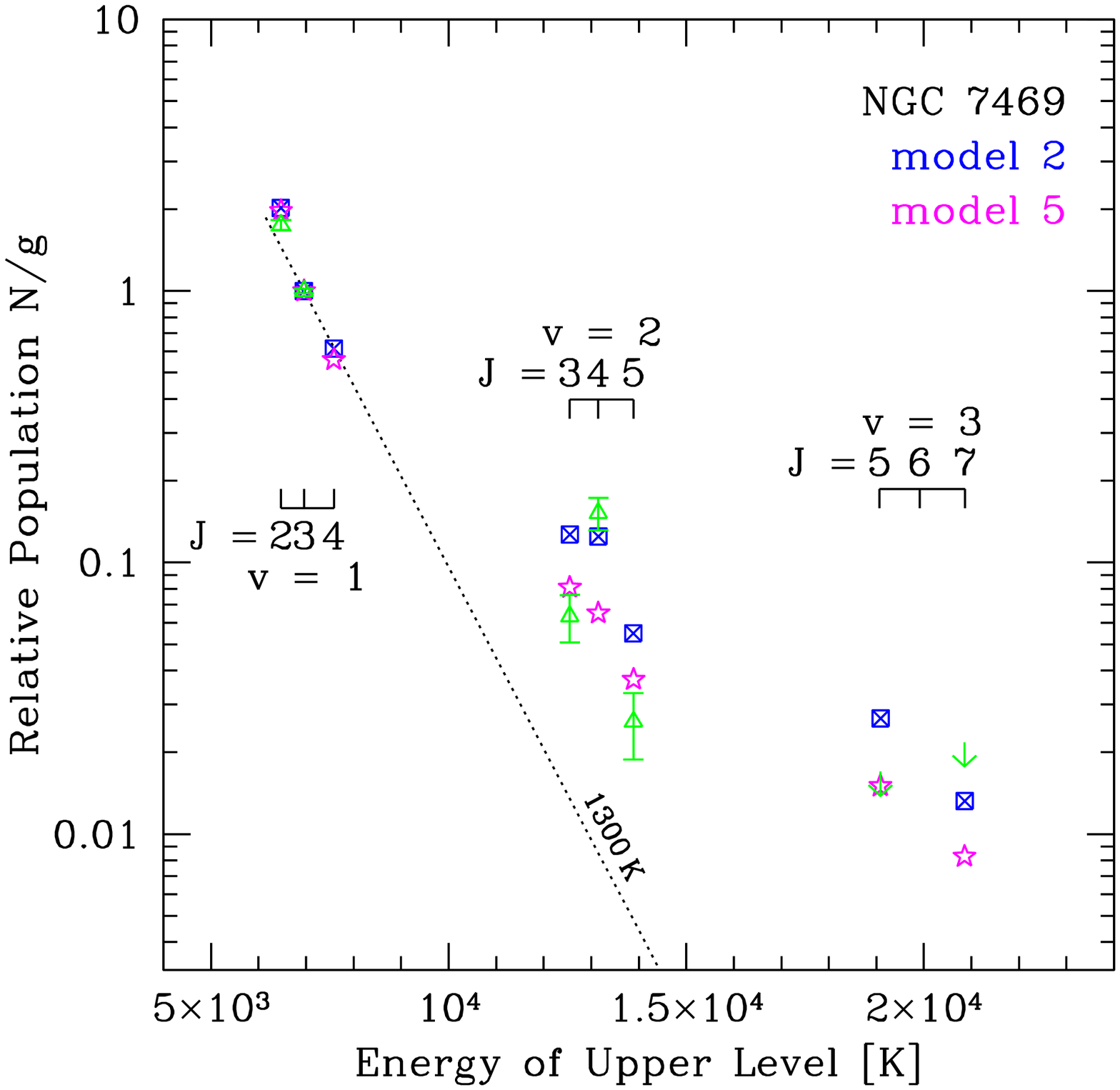}
%\centerline{\psfig{file=f11a.eps,width=70mm}}
%\centerline{\psfig{file=f11b.eps,width=70mm}\hspace{5mm}\psfig{file=f11c.eps,width=70mm}}
\caption{Excitation diagrams for the hot hydrogen molecules in each
of the more distant ($\langle D \rangle = 75$\,Mpc) objects.
Arrows denote $3\sigma$ upper limits, derived from the residual
spectrum after subtraction of the stellar continuum and line emission.
Green triangles correspond to the nuclear H$_2$ populations.
Blue squares denote the populations for our PDR model 2.
For NGC\,7469, 
the magenta stars denote the populations for our PDR model 5.
The dotted line indicates where the populations would lie for an
isothermal cloud at 1300\,K, a reasonable match to the $\nu=1-0$ levels.}
\label{fig:poplev-far}
\epsscale{1.0}
\end{figure}

Of the 5 PDR models we consider -- covering the parameter space
FUV intensity $\chi=10^2$-$10^5$ times that in the local interstellar
medium, 
gas density $n_{\rm H}=10^3$-$10^6$\,cm$^{-3}$, and 
temperature $T_{max}=10^2$-$2\times10^3$\,K -- 
it is model 2 and 4 (differing 
only in FUV intensity, and indistinguishable in terms of their
resulting line ratios) which best match the data.
This indicates that the maximum temperature is $\sim1000$\,K (as
one would expect from the $\nu=1$ levels discussed above), the gas
density is of order $\sim10^4$\,cm$^{-3}$, and the incident FUV
intensity is at least $10^3$ times the local ambient value.
Model 2 provides a reasonable match to all the data in the mean diagram
(Fig.~\ref{fig:poplev-mean}), and is the model we use as a reference
when considering the individual objects, the excitation diagrams for
which are shown in Figs.~\ref{fig:poplev-near} and~\ref{fig:poplev-far}.

\subsection{Circumnuclear line ratios}

In all of the five galaxies belonging to the nearby sample, the
circumnuclear line ratios can be reproduced by model 2 -- and in the
case of NGC\,2992 the match is remarkably good.
There are a few isolated cases where line limits appear to be
inconsistent with the model, but in at least some cases these limits
appear to be unphysical:
for example, in none of our models is the $(\nu,j) = (2,4)$
less than the $(2,3)$ and $(2,5)$ levels.
In this case the 2-1\,(2) line is
significantly weaker than 2-1\,S(1) and 2-1\,S(3) due to its smaller
statistical weight and occurs in the overlap region of the two
spectral segments where small uncertainties in scaling the two
segments can have a disproportionate effect that is not reflected in
the estimation of the line flux uncertainty.
Similarly, the $(3,5)$ level is derived from the 3-2\,S(3) line which
sits close to the edge of the 2.206\,\micron\ Na absorption which can
often be quite deep.
A description of how the uncertainties were estimated is given in
Section~\ref{sec:contfit}, but any such estimate cannot take into
account systematic errors such as incorrect removal of the continuum
features due to mismatch of the stellar templates.
Bearing this in mind, one should be cautious of the  
uncertainties when a weak line or non-detection occurs close to
a region of strong stellar absorption, and not over-interpret them
when assessing the correspondance between model
and data.

The conclusion that in all five cases the H$_2$ lines appear to be 
excited in dense PDRs exposed to high FUV fields is consistent with
what one might expect of regions efficiently forming massive young
stars at a high rate.

\subsection{Nuclear line ratios}

\begin{figure}
\epsscale{.4}
\plotone{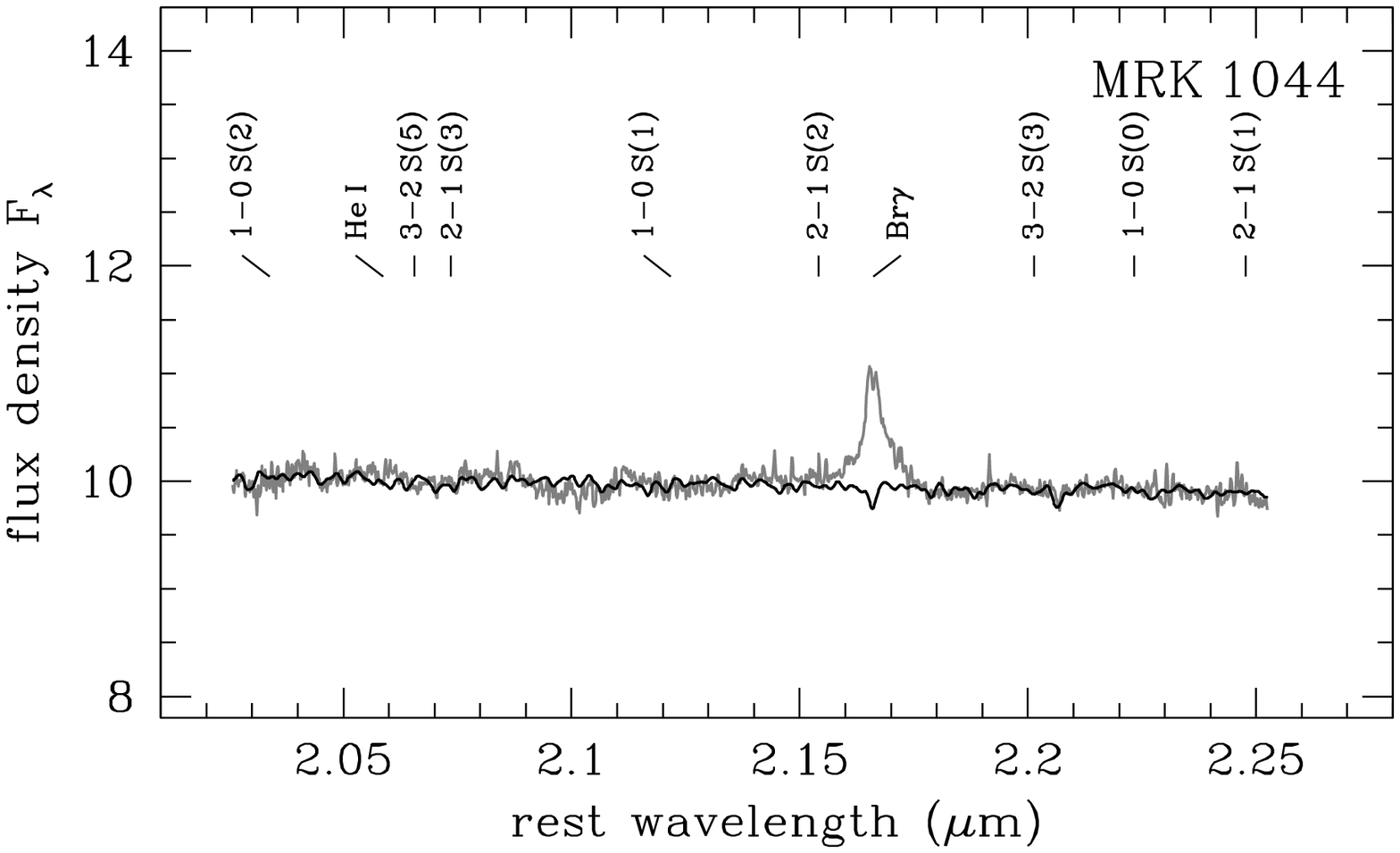}\plotone{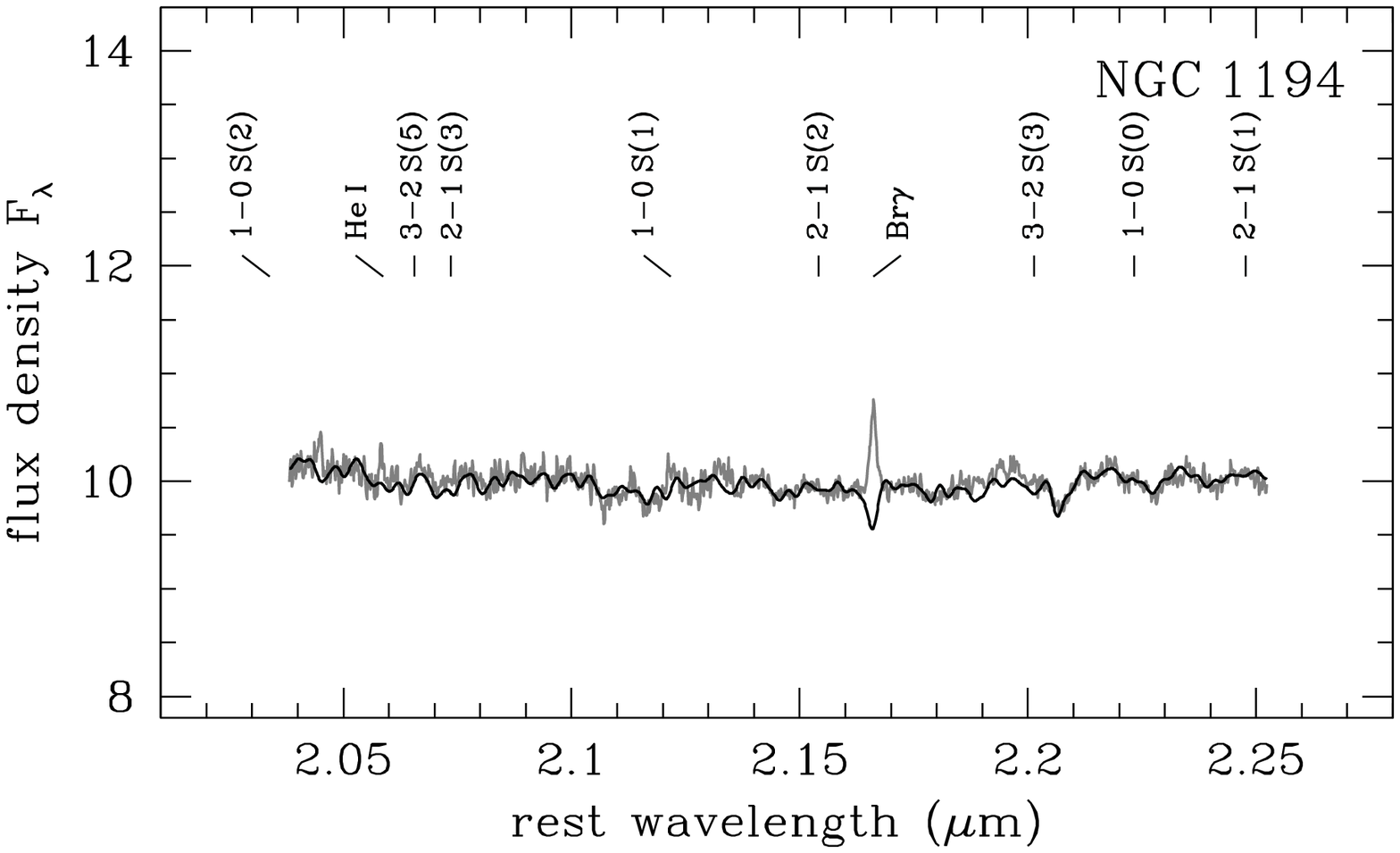}\plotone{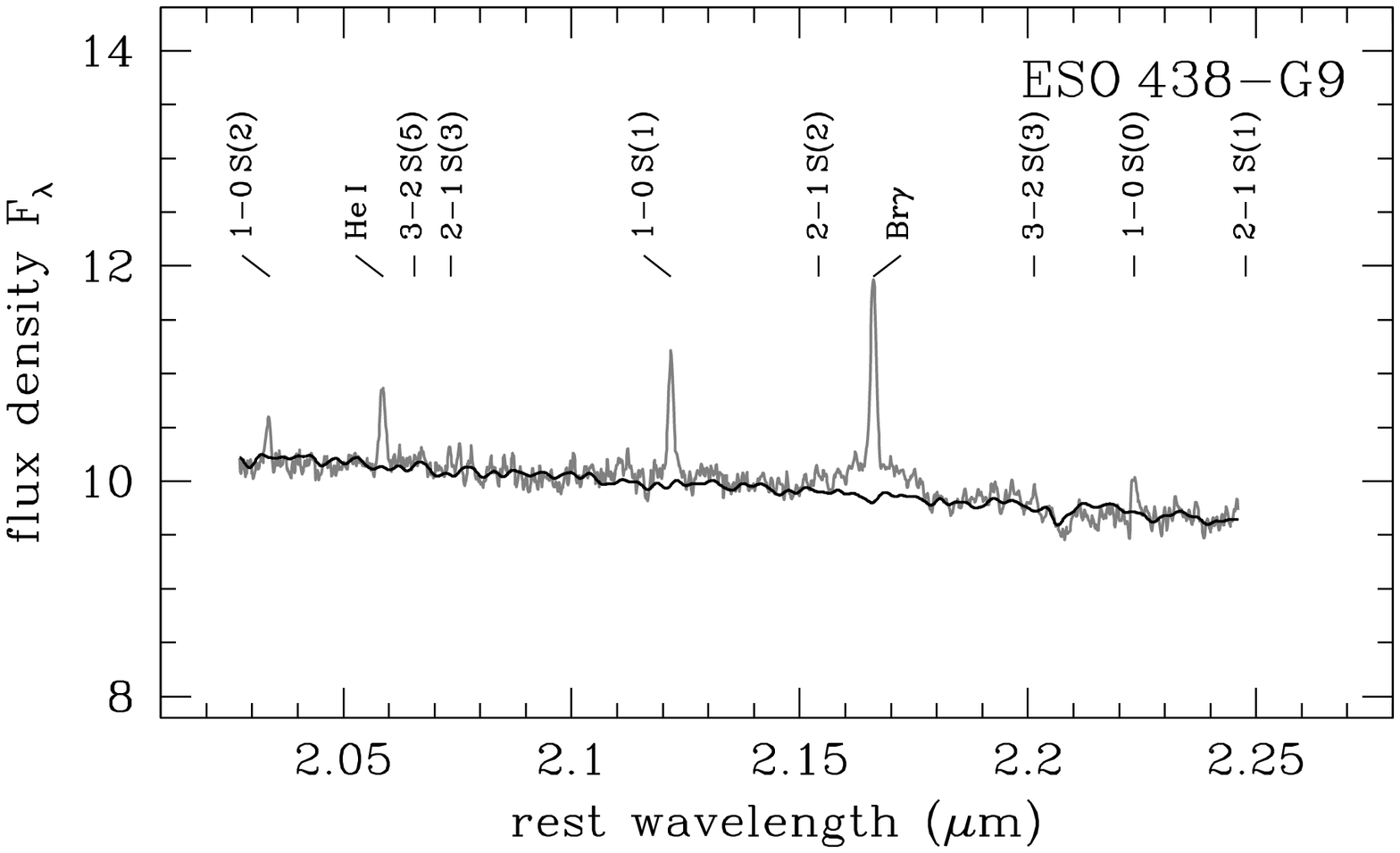}\plotone{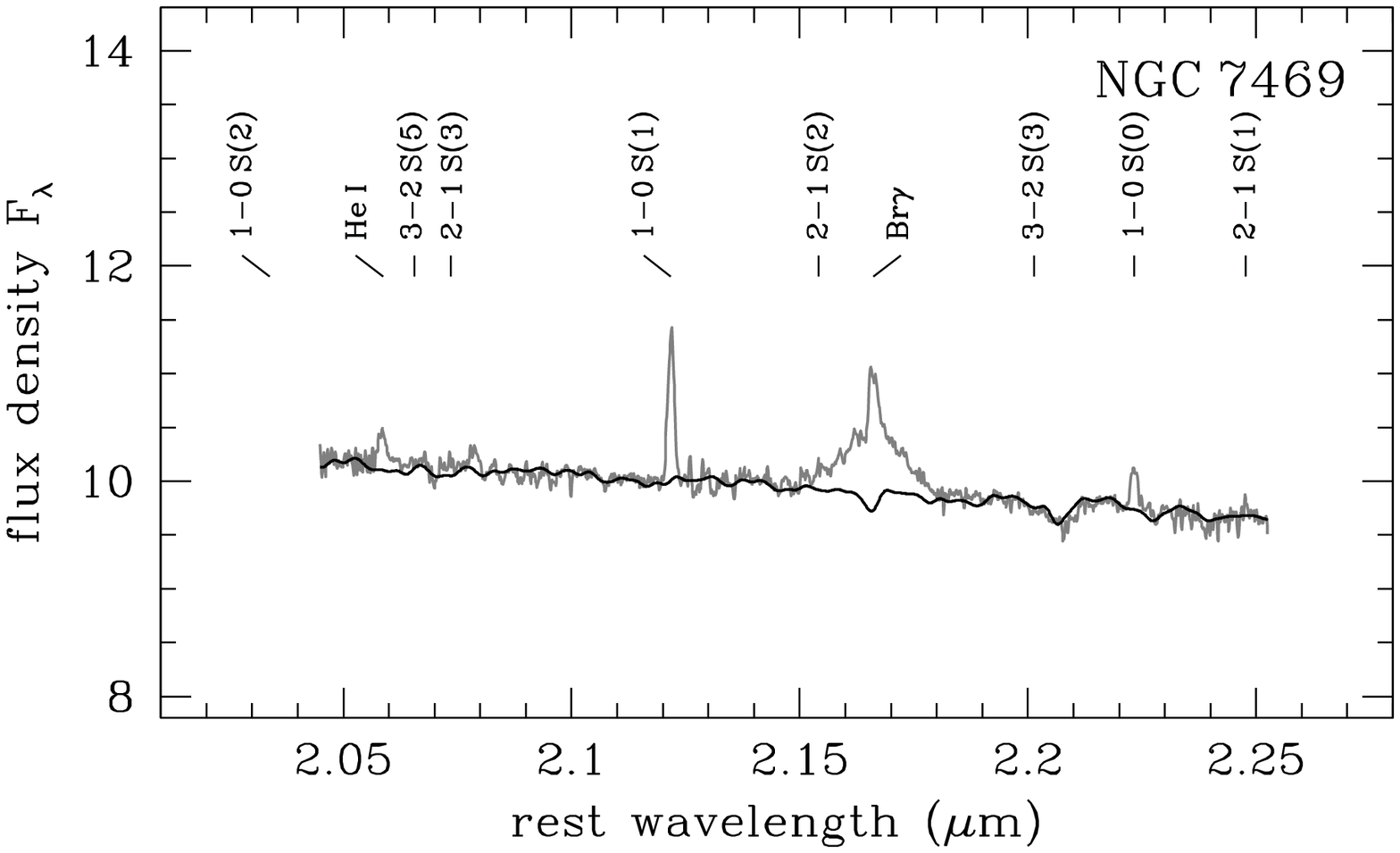}
%\centerline{\psfig{file=f05a.eps,width=90mm}\psfig{file=f05b.eps,width=90mm}}
%\vspace{1mm}
%\centerline{\psfig{file=f05c.eps,width=90mm}\psfig{file=f05d.eps,width=90mm}}
\caption{Normalised spectra of the nuclear region (i.e. central
  1\arcsec) of each of the more distant ($\langle D \rangle =
  75$\,Mpc) objects, plotted at rest wavelength and
  normalised to their mean values.
The dark overlaid line is the best fitting stellar continuum
  constructed from spectra of template stars (see text for details).
The positions (but not necessarily detections) of important emission
  lines are marked.
}
\label{fig:nuclear-spec-far}
\epsscale{1.0}
\end{figure}

The data for the nuclear line fluxes is much less complete, primarily
due to the greatly increased noise in the spectrum from the bright hot
dust continuum associated with the AGN.
Nevertheless, in three of the four nearby galaxies where H$_2$ lines are
detected -- namely NGC\,1097, NGC\,2992, and NGC\,5506 -- their ratios
show a strong similarity to those of the circumnuclear regions, with 
the single exception of the $(\nu,J) = (2,3)$ level in NGC\,1097 which
is a factor of $\sim$2 less than one would expect.
The fourth galaxy, NGC\,7582, is similar for all the levels up to and
including $(2,3)$, but the $(3,5)$ level is
significantly higher than one would expect: none of our PDR models can
reproduce this.
Examination of the region around the 3-2\,S(3) line at 2.2\,\micron\ in
Fig.~\ref{fig:nuclear-spec-far} suggests that in NGC\,7582 it also
seems likely that template mismatch and hence incomplete correction of
the absorption features in the stellar continuum has led to poor
estimation of the weak line flux.

\begin{figure}
\plotone{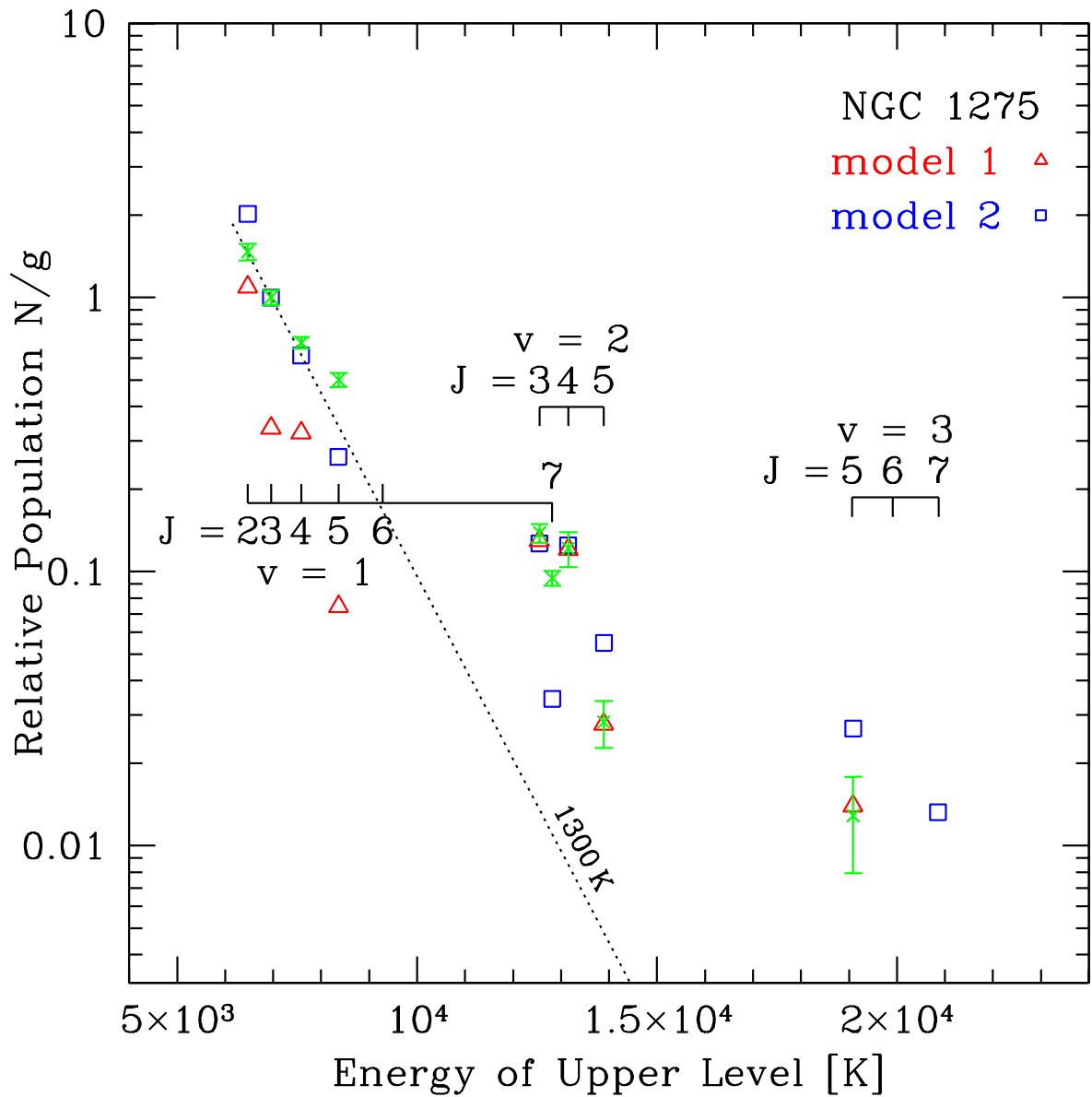}
%\centerline{\psfig{file=f12.eps,width=70mm}}
\caption{Excitation diagram for the hot hydrogen molecules in
  NGC\,1275. The data are taken from Table~2 of \cite{kra00}. 
The dotted line indicates where the populations would lie for an
isothermal cloud at 1300\,K, a reasonable match to the $\nu=1-0$ levels.
Blue squares denote the populations for our PDR model 2.
This model goes some way to accounting for the apparent suppression of
  the 2-1\,S(3) line. 
Fluorescence in lower density gas, such as
  model\,1 (denoted by red triangles) could account for the $\nu=2-1$
  and $\nu=3-2$ levels completely; although
  in such as case thermal excitation -- perhaps non-isothermal -- is
  still needed to account for the $\nu=1-0$ levels.
}
\label{fig:poplev-1275}
\end{figure}

In none of these galaxies do we find evidence for suppression of the
2-1\,S(3) line, which might be expected for AGN if the gas irradiated
by X-rays.
This is a theoretical possibility if there is an excess of
Ly$\alpha$ photons, 
due to an accidental resonance between the 1-2\,P(5)
and 1-2\,R(6) transitions and the Ly$\alpha$ line \citep{bla87}.
An excess of Ly$\alpha$ photons will depopulate the lower 
(i.e. $(\nu,J) = (2,5)$ and $(2,6)$) levels,
resulting in a reduction in the two 2-1\,S(3) and 2-1\,S(4) line strengths.
Such a situation could arise in X-ray irradiated gas, in which the
primary photoelectrons have high enough energies to generate many more
secondary ionisations ($\sim30$ per keV of primary photoelectron
energy; \citealt{mal96}) through collisional excitation.
It has been proposed by \cite{dra90} as the explanation for a previously
reported weakness of that line in NGC\,6240.
However, using higher resolution and signal-to-noise spectra,
\cite{sug97} showed that this line is in fact not suppressed, and
that the $\nu=1$--0 and 2--1 transitions can be well matched by models
of isothermal emission.
A second case is NGC\,1275, which has very luminous
H$_2$ emission ($\log{L_{\rm 1-0S(1)}/L_\odot} \sim 6.8$) mostly
concentrated in a compact ($\sim$0.5\arcsec) nuclear source
\citep{kra00,wil05}. 
Interpreting the line ratios in terms of thermal excitation with two
or more excitation temperatures, \cite{kra00} argued that the flux of
the 2-1\,S(3) line was a factor of at 
least 2.5 less than expected and attributed this to the fact that the
excitation of the gas is dominated by X-ray heating from the AGN.
We plot the data for the central 3\arcsec\ of NGC\,1275 (from their
Table~2) in Fig.~\ref{fig:poplev-1275}.
It is clear that these data tend towards the same trend we have seen
in our Seyfert nuclei: the $\nu=1-0$ transitions appear thermalised
while the higher vibrational transitions indicate an additional
higher temperature thermal or fluorescent (PDR) component.
The levels from our PDR model~2  (overplotted as squares) go some way to
explaining the observed ratios and the apparent weakness of the (2,5)
level.
A PDR model with low density fluorescence (our model~1; overplotted as
triangles), matches the $\nu=2-1$ and $3-2$ levels extremely well if
it is scaled to produce 1/3 of the total 1-0\,S(1) flux.
It does not reproduce the $=1-0$ levels at all;
but it is not unreasonable to suppose that in such a large aperture
(3\arcsec\ is equivalent to 1\,kpc) there may be multiple excitation
processes at work and that the $1-0$ levels could arise from a region
where thermal processes dominate and the higher levels from a region
where non-thermal processes dominate.

In the nuclear specta of our AGN, the line ratios are, as for
the circumnuclear regions, consistent with excitation by intense star
formation in relatively dense PDRs.
This appears to be the case even for NGC\,5506 which has a very
luminous Seyfert nucleus.
On the other hand, determining whether the data are consistent with
X-ray excitation is not straightforward.
From their models of X-ray irradiated gas, both \cite{mal96} and
\cite{tin97} conclude that the 2-1\,S(1) to 1-0\,S(1) ratio should
nearly always be $<0.3$;
most of the diagnostic power of these models appear to lie in
transitions from $\nu \geq 3$ or far infrared lines from other
species.
Instead we consider the 2--10\,keV hard X-ray fluxes, which are listed in
Table~\ref{tab:measdat}.
These indicate that, as also found by \cite{qui99}, there is
absolutely no correlation between X-ray and H$_2$ emission.
This could be either because the two are not causally related,
implying a different excitation mechanism for the H$_2$; or due to the
different timescales for variability.
The hard X-ray luminosities are known to vary on fast 
timescales -- e.g. by a factor of 2 over a period of a few hours
for NGC\,1365 \citep{ris00} and NGC\,7469 \citep{bar86} -- while 
the H$_2$ emission would only vary on
the far longer timescales associated with its physical extent
($\sim100$\,pc implies $\sim$300\,yrs).

For NGC\,1097 nuclear H$_2$ excitation by star formation rather than
an AGN can be understood because the AGN is in general
rather low luminosity (see Section~\ref{sec:notes}).
And in NGC\,1097, NGC\,2992, and NGC\,7582
Fig.~\ref{fig:nuclear-spec-far} indicates that the Na\,I stellar
absorption is clearly visible despite (in the latter 2 cases) strong
dilution by the AGN.
As discussed in Section~\ref{sec:starform}, 
this suggests that in these cases there may be significant active
star formation in the nucleus and as a result any H$_2$ emission from
X-ray excited gas could be hidden by emission from PDRs excited by the
young stars.

Of the four more distant galaxies, H$_2$ emission was detected in three --
namely NGC\,1194, NGC\,7469, and ESO\,438-G9 -- but
the data is sparser still than the nearby ones.
Beyond the statement that the $\nu=1$ levels are thermalised but the
higher levels indicate fluorescent excitation, there is little that we
can conclude.
This in itself suggests that the $\nu=1$ levels are thermalised due to
high gas density rather than purely thermal excitation in shocks,
consistent with the results for the nearby galaxies.
That the excitation could be due to active star formation is evidenced
by the Na\,I absorption which can be seen in these three nuclei
despite the AGN dilution.
However, we can make no statement on whether the gas is also (or
instead) being irradiated by X-rays from the AGN.
In NGC\,7469 the low populations in the $(2,3)$ and $(2,5)$ levels
hint that perhaps even the $\nu=2$ levels are partially thermalised --
i.e. that the gas density may be as high as $n_{\rm H}=10^6$\,cm$^{-3}$
as in model 5, which better reproduces these lower populations.
Given the extremely large cold molecular gas mass in the nuclear region
of this galaxy, inferred from CO\,2-1 observations \citep{dav04a}, very
high gas densities in the clouds themselves are to be expected.

\section{Conclusions}
\label{sec:conc}

We have presented K-band spectra of the central regions of nine AGN,
for five of which we also present circumnuclear spectra.
In these spectra, we
have measured H$_2$ emission line fluxes from the $\nu = 1$, 2, and
3 levels.
We performed two critical steps which enabled us to significantly
increase our sensitivity to the weak lines:
the first is to fit the continuum with stellar templates, allowing us
to remove the absorption features;
the second is to de-rotate the spectra, decorrelating noise from
imperfect telluric correction while simultaneously correlating signal
from the emission lines.
Broad Br$\gamma$ (i.e. FWHM $> 1000$\,\kms) is observed in seven of
the nine nuclear spectra.
We conclude the following:
\begin{enumerate}

\item
Use of the ratios of the $\nu=1$-0 transitions, either between
themselves or with other lines such as 
Br$\gamma$, H$\alpha$, or [O{\sc i}]\,$\lambda6300$\,\AA\ can be misleading
since in starburst and AGN environments, these levels are often
(and perhaps almost always) thermalised
without necessarily implying shock excitation.
Indeed, we have shown strong evidence that fluorescent excitation
plays a major role in the H$_2$ excitation.

\item
The H$_2$ line ratios in the circumnuclear regions are very similar in
all cases. 
While the $\nu=1$ levels are thermalised at $\sim$1000\,K, FUV-pumped
gas is needed to account for the higher levels, consistent with emission
from high density PDRs. 
In our PDR models, the typical parameters that best account for the observed
line fluxes 
are gas density $n_{\rm H}\sim10^4$\,cm$^{-3}$, temperature at the
outer edge of the clouds $T\sim1000$\,K, and illuminating FUV fields
$\chi \gtrsim 10^3$ times more intense than the local interstellar
field.
These parameters are similar to those found for the nuclei of ULIRGs
by \cite{dav03}.

\item
The H$_2$ line ratios in the nuclear regions are also similar in that
the $\nu=1$ levels are thermalised at $\sim$1000\,K but the higher
levels require FUV pumping.
The observed line ratios can be accounted for by H$_2$ excitation
occupying a similar parameter space to that 
found for the PDRs in the circumnuclear regions, 
suggesting that the mechanism dominating the nuclear H$_2$ emission could be
illumination of PDRs by intense young star formation.

\item
We find no evidence for suppression of the 2-1\,S(3) line, which may
occur in X-ray irradiated gas.
We suggest that even in the case of NGC\,1275 \citep{kra00}, the H$_2$
line ratios can be explained without recourse to such a phenomenon.
This does not mean that such suppression does not occur -- simply that
on scales of $\sim100$\,pc, X-ray heating is probably not the dominant
excitation mechanism.

\item
Despite strong dilution by a red continuum, probably associated with hot dust
grains, 
stellar Na\,I absorption at 2.206\,\micron\ is seen in six of the nine 
nuclear spectra, suggesting that even on scales as small as 100\,pc star
formation is an important and active process in AGN.

\end{enumerate}

%----------------------------------------------------------------------

\acknowledgments

The authors are grateful to the staff at the Paranal Observatory for
carrying out in service mode the observations presented in this paper;
and to the referee for a number of comments and suggestions.
AS thanks the Israel Science Foundation (grant 221/03) for support.

%----------------------------------------------------------------------

%----------------------------------------------------------------------

\clearpage

%----------------------------------------------------------------------

\begin{deluxetable}{llrrcrrr}

\tabletypesize{\small}
\tablecaption{Basic Data for the Galaxies Observed\label{tab:basdat}}
\tablewidth{0pt}
\tablehead{

\colhead{object} &
\colhead{AGN\tablenotemark{a}} & 
\colhead{RA\tablenotemark{a}} & 
\colhead{dec\tablenotemark{a}} & 
\colhead{cz\tablenotemark{a}} & 
\colhead{Distance\tablenotemark{b}} & 
\colhead{1\arcsec\tablenotemark{b}} & 
\colhead{seeing\tablenotemark{c}} \\

\colhead{} & 
\colhead{type} & 
\colhead{(J2000)} & 
\colhead{(J2000)} & 
\colhead{(km\,s$^{-1}$)} & 
\colhead{(Mpc)} & 
\colhead{(pc)} & 
\colhead{(\arcsec)} \\

}
\startdata

NGC\,1097   & Sy1   & 02 46 19.0 & $-$30 16 30 & 1275 &   18 &  90 & 0.7, 1.0 \\
NGC\,1365   & Sy1.8 & 03 33 36.4 & $-$36 08 25 & 1640 &   23 & 110 & 0.8, 0.8 \\
NGC\,2992   & Sy2   & 09 45 42.0 & $-$14 19 35 & 2310 &   33 & 160 & 0.7, 0.7 \\
NGC\,5506   & Sy1.9 & 14 13 14.8 & $-$03 12 27 & 1850 &   27 & 130 & 1.4, 0.6 \\
NGC\,7582   & Sy2   & 23 18 23.5 & $-$42 22 14 & 1580 &   23 & 110 & 1.2, 0.8 \\

\noalign{\bigskip}

MRK\,1044   & Sy1   & 02 30 05.4 & $-$08 59 53 & 4930 &   71 & 330 & 0.8, 0.7 \\
NGC\,1194   & Sy1   & 03 03 49.1 & $-$01 06 31 & 4060 &   58 & 270 & 1.3, 0.7 \\
ESO\,438-G9 & Sy1.5 & 11 10 48.0 & $-$28 30 04 & 7010 &  101 & 470 & 0.5, 0.8 \\
NGC\,7469   & Sy1.2 & 23 03 15.6 & $+$08 52 26 & 4890 &   70 & 330 & 0.6, 0.5 \\

\enddata

\tablenotetext{a}{\,AGN types, coordinates, and redshifts are taken
  from the NASA/IPAC Extragalactic Database.}

\tablenotetext{b}{\,Luminosity distance and angular size scale calculated 
using $H_0=70$\,km\,s$^{-1}$\,Mpc$^{-1}$ and $q_0=0.5$.}

\tablenotetext{c}{\,Seeing is measured from the standard star for each
  data set, and indicates that for the short and long wavelength segments respectively.
See the text for a description of how the segments were matched and combined.}

\end{deluxetable}
\clearpage
%----------------------------------------------------------------------

\begin{deluxetable}{lrrrrcclcc}

\tabletypesize{\tiny}
\tablecaption{Measured and Derived Fluxes and Luminosities\label{tab:measdat}}
\tablewidth{0pt}
\tablehead{

\colhead{object} & 
\colhead{S$_{12}$\tablenotemark{a}} & 
\colhead{S$_{25}$\tablenotemark{a}} & 
\colhead{S$_{60}$\tablenotemark{a}} & 
\colhead{S$_{100}$\tablenotemark{a}} &
\colhead{F$_{2-10\,keV}$\tablenotemark{b}} & 
\colhead{K-mag\tablenotemark{c}} & 
\colhead{$F_{S(1)}$\tablenotemark{d}} & 
\colhead{$\log{\frac{L_{S(1)}}{L_\odot}}$} &
\colhead{$\log{\frac{L_{\rm IR}}{L_\odot}}$} \\

\colhead{} & 
\colhead{Jy} & 
\colhead{Jy} & 
\colhead{Jy} & 
\colhead{Jy} &
\colhead{10$^{-15}$\,W\,m$^{-2}$} &
\colhead{} & 
\colhead{10$^{-18}$\,W\,m$^{-2}$} & 
\colhead{} &
\colhead{} \\

}
\startdata

NGC\,1097   &  1.99 &  5.51       & 44.5\phm{0} &   85.3\phm{0}  & \phn2 &       11.8 & \phm{$<$}3.10$\pm$0.04 & \phm{$<$}4.5 & 10.64 \\
NGC\,1365   &  3.37 & 10.8\phm{0} & 76.1\phm{0} &  142.5\phm{0}  & \phn5 & \phm{1}9.8 & $<$0.78                &       $<$4.1 & 11.09 \\
NGC\,2992   &  0.49 &  1.26       &  8.50       &   24.9\phm{0}  & \phn5 &       10.1 & \phm{$<$}8.63$\pm$0.06 & \phm{$<$}5.4 & 10.54 \\
NGC\,5506   &  1.28 &  3.64       &  8.41       &    8.89        &    90 & \phm{1}9.3 & \phm{$<$}6.08$\pm$0.18 & \phm{$<$}5.1 & 10.41 \\
NGC\,7582   &  1.62 &  6.44       & 49.1\phm{0} &   72.9\phm{0}  &    17 &       11.4 & \phm{$<$}1.52$\pm$0.06 & \phm{$<$}4.4 & 10.85 \\

\noalign{\bigskip}

MRK\,1044   &  0.10 &  0.22       &  0.43       &    0.88        &       &       11.3 & $<$0.19                &       $<$4.4 & 10.08 \\
NGC\,1194   &  0.27 &  0.51       &  0.77       &    0.93        &       &       12.5 & \phm{$<$}0.24$\pm$0.03 & \phm{$<$}4.4 & 10.21 \\
ESO\,438-G9 &  0.32 &  0.62       &  3.14       &    4.22        &       &       11.0 & \phm{$<$}1.90$\pm$0.05 & \phm{$<$}5.8 & 11.03 \\
NGC\,7469   &  1.35 &  5.79       & 25.9\phm{0} &   34.9\phm{0}  &    30 &       10.4 & \phm{$<$}6.47$\pm$0.12 & \phm{$<$}6.0 & 11.59 \\

\enddata

\tablenotetext{a}{\,IRAS data is from the Faint Source Catalogue \citep{mos90}.}

\tablenotetext{b}{\,2--10\,keV hard X-ray flux. References: \cite{bas99,tur89}.}

\tablenotetext{c}{K-band magnitude extracted in a 1.0\arcsec\ length of a 
1.0\arcsec\ slit.}

\tablenotetext{d}{\,1-0\,S(1) line flux extracted in a 1.0\arcsec\ length of a
1.0\arcsec\ slit; errors are estimated from the RMS of the residual spectrum 
after subtracing the stellar continuum and emission line fits; they do not 
include calibration uncertainties.}

\end{deluxetable}
\clearpage
%----------------------------------------------------------------------

\begin{deluxetable}{lcccccccc}

\tabletypesize{\tiny}
\tablecaption{Relative H$_2$ Line Fluxes\label{tab:ratios}}
\tablewidth{0pt}
%\rotate
\tablehead{

\colhead{} &
\multicolumn{7}{c}{line and wavelength ($\mu$m)} \\

\colhead{object/model} & 
\colhead{1-0\,S(2)} & 
%\colhead{He\,I} & 
\colhead{3-2\,S(5)} & 
\colhead{2-1\,S(3)} & 
\colhead{1-0\,S(1)} & 
\colhead{2-1\,S(2)} & 
%\colhead{Br$\gamma$} & 
\colhead{3-2\,S(3)} & 
\colhead{1-0\,S(0)} & 
\colhead{2-1\,S(1)} \\

\colhead{} &
\colhead{2.0338} &
%\colhead{2.0587} &
\colhead{2.0656} & 
\colhead{2.0735} & 
\colhead{2.1218} & 
\colhead{2.1542} & 
%\colhead{2.1661} &
\colhead{2.2014} & 
\colhead{2.2233} & 
\colhead{2.2477} \\

}
\startdata

\multicolumn{3}{l}{circumnuclear (objects with $\langle D \rangle = 25$\,Mpc)} \\
NGC\,1097    & 0.367$\pm$0.022 % 2.0338
%             &                 % 2.0587
             & $<$0.119        % 2.0656
             & 0.220$\pm$0.013 % 2.0735
             & 1.000$\pm$0.014 % 2.1218
             & 0.130$\pm$0.016 % 2.1542 
%             &                 % 2.1661
             & $<$0.053        % 2.2014
             & 0.330$\pm$0.018 % 2.2235
             & 0.199$\pm$0.013 % 2.2477
\\
NGC\,1365    & 0.379$\pm$0.011 % 2.0338
%             &                 % 2.0587
             & 0.063$\pm$0.015 % 2.0656
             & 0.230$\pm$0.008 % 2.0735
             & 1.000$\pm$0.007 % 2.1218
             & 0.111$\pm$0.010 % 2.1542 
%             &                 % 2.1661
             & 0.043$\pm$0.010 % 2.2014
             & 0.340$\pm$0.011 % 2.2235
             & 0.189$\pm$0.007 % 2.2477
\\
NGC\,2992    & 0.404$\pm$0.017 % 2.0338
%             &                 % 2.0587
             & $<$0.040        % 2.0656
             & 0.164$\pm$0.010 % 2.0735
             & 1.000$\pm$0.009 % 2.1218
             & 0.075$\pm$0.013 % 2.1542 
%             &                 % 2.1661
             & 0.090$\pm$0.010 % 2.2014
             & 0.310$\pm$0.009 % 2.2235
             & 0.178$\pm$0.010 % 2.2477
\\
NGC\,5506    & 0.308$\pm$0.018 % 2.0338
%             &                 % 2.0587
             & $<$0.046        % 2.0656
             & 0.178$\pm$0.012 % 2.0735
             & 1.000$\pm$0.013 % 2.1218
             & $<$0.045        % 2.1542 
%             &                 % 2.1661
             & $<$0.027        % 2.2014
             & 0.257$\pm$0.008 % 2.2235
             & 0.142$\pm$0.011 % 2.2477
\\
NGC\,7582    & 0.368$\pm$0.010 % 2.0338
%             &                 % 2.0587
             & $<$0.041        % 2.0656
             & 0.168$\pm$0.009 % 2.0735
             & 1.000$\pm$0.006 % 2.1218
             & $<$0.027        % 2.1542 
%             &                 % 2.1661
             & 0.053$\pm$0.011 % 2.2014
             & 0.270$\pm$0.012 % 2.2235
             & 0.146$\pm$0.008 % 2.2477
\\

\noalign{\smallskip}                       %
mean         & 0.365$\pm$0.032 % 2.0338    %
%             &                 % 2.0587   %
             & $<$0.061         % 2.0656           %
             & 0.192$\pm$0.028 % 2.0735    %
             & 1.000$\pm$0.010 % 2.1218    %
             & 0.105$\pm$0.023 % 2.1542    %
%             &                 % 2.1661   %
             & 0.062$\pm$0.020 % 2.2014    %
             & 0.301$\pm$0.033 % 2.2235    %
             & 0.171$\pm$0.023 % 2.2477    %
\\                                         %

\noalign{\bigskip}

\multicolumn{3}{l}{nuclear (objects with $\langle D \rangle = 25$\,Mpc)} \\
NGC\,1097    & 0.378$\pm$0.015 % 2.0338
             & $<$0.046        % 2.0656
             & 0.163$\pm$0.009 % 2.0735
             & 1.000$\pm$0.012 % 2.1218
             & 0.107$\pm$0.013 % 2.1542 
             & 0.069$\pm$0.007 % 2.2014
             & 0.254$\pm$0.009 % 2.2235
             & 0.091$\pm$0.009 % 2.2477
\\
NGC\,1365    & \multicolumn{7}{c}{no H$_2$ lines detected} 
\\
NGC\,2992    & 0.310$\pm$0.019 % 2.0338
             & $<$0.071        % 2.0656
             & 0.180$\pm$0.014 % 2.0735
             & 1.000$\pm$0.007 % 2.1218
             & 0.126$\pm$0.026 % 2.1542 
             & 0.128$\pm$0.022 % 2.2014
             & 0.346$\pm$0.019 % 2.2235
             & 0.147$\pm$0.018 % 2.2477
\\
NGC\,5506    & ---             % 2.0338
             & $<$0.168        % 2.0656
             & 0.173$\pm$0.039 % 2.0735
             & 1.000$\pm$0.030 % 2.1218
             & $<$0.225        % 2.1542 
             & $<$0.065        % 2.2014
             & 0.333$\pm$0.030 % 2.2235
             & $<$0.138        % 2.2477
\\
NGC\,7582    & 0.490$\pm$0.036 % 2.0338
             & $<$0.228        % 2.0656
             & 0.277$\pm$0.045 % 2.0735
             & 1.000$\pm$0.039 % 2.1218
             & $<$0.283        % 2.1542 
             & 0.233$\pm$0.041 % 2.2014
             & 0.406$\pm$0.045 % 2.2235
             & 0.175$\pm$0.045 % 2.2477
\\

\noalign{\smallskip}

%mean         & 0.378$\pm$0.058 % 2.0338   %
%%             &                 % 2.0587  %
%             & $<$0.126        % 2.0656   %
%             & 0.213$\pm$0.084 % 2.0735   %
%             & 1.000$\pm$0.025 % 2.1218   %
%             & $<$0.162        % 2.1542   %
%%             &                 % 2.1661  %
%             & 0.140$\pm$0.076 % 2.2014   %
%             & 0.324$\pm$0.052 % 2.2235   %
%             & 0.133$\pm$0.042 % 2.2477   %
%\\                                        %
%                                          %
%\noalign{\bigskip}                        %

\multicolumn{3}{l}{nuclear (objects with $\langle D \rangle = 75$\,Mpc)} \\
MRK\,1044    & \multicolumn{7}{c}{no H$_2$ lines detected} 
\\
NGC\,1194    & ---             % 2.0338
             & $<$0.333        % 2.0656
             & 0.509$\pm$0.093 % 2.0735
             & 1.000$\pm$0.120 % 2.1218
             & $<$0.199        % 2.1542 
             & $<$0.242        % 2.2014
             & $<$0.232        % 2.2235
             & $<$0.258        % 2.2477
\\
ESO\,438-G9  & 0.339$\pm$0.022 % 2.0338
             & $<$0.122        % 2.0656
             & 0.147$\pm$0.029 % 2.0735
             & 1.000$\pm$0.027 % 2.1218
             & 0.133$\pm$0.019 % 2.1542 
             & 0.172$\pm$0.022 % 2.2014
             & 0.332$\pm$0.029 % 2.2235
             & ---             % 2.2477
\\
NGC\,7469    & ---             % 2.0338
             & $<$0.062        % 2.0656
             & 0.069$\pm$0.019 % 2.0735
             & 1.000$\pm$0.019 % 2.1218
             & 0.104$\pm$0.014 % 2.1542 
             & $<$0.042        % 2.2014
             & 0.290$\pm$0.013 % 2.2235
             & 0.086$\pm$0.017 % 2.2477
\\

\noalign{\smallskip}                      %
mean         & 0.379$\pm$0.068 % 2.0338   %
%             &                 % 2.0587  %
             & $<$0.147        % 2.0656   %
             & 0.217$\pm$0.132 % 2.0735   %
             & 1.000$\pm$0.036 % 2.1218   %
             & 0.118$\pm$0.012 % 2.1542   %
%             &                 % 2.1661  %
             & 0.150$\pm$0.054 % 2.2014   %
             & 0.327$\pm$0.047 % 2.2235   %
             & 0.125$\pm$0.038 % 2.2477   %
\\                                        %

\noalign{\bigskip}

     thermal:\  1000\,K & 0.27 & 0.00 & 0.00 & 1.00 & 0.00 & 0.00 & 0.27 & 0.01 \\
\phm{thermal:\ }2000\,K & 0.37 & 0.00 & 0.08 & 1.00 & 0.03 & 0.01 & 0.21 & 0.08 \\

     PDR:\  model 1 & 0.49 & 0.02 & 0.22 & 1.00 & 0.25 & 0.10 & 0.54 & 0.53 \\
\phm{PDR:\ }model 2 & 0.32 & 0.04 & 0.15 & 1.00 & 0.08 & 0.07 & 0.33 & 0.17 \\
\phm{PDR:\ }model 3 & 0.35 & 0.00 & 0.05 & 1.00 & 0.02 & 0.00 & 0.22 & 0.05 \\
\phm{PDR:\ }model 4 & 0.31 & 0.05 & 0.16 & 1.00 & 0.08 & 0.07 & 0.32 & 0.17 \\
\phm{PDR:\ }model 5 & 0.29 & 0.02 & 0.10 & 1.00 & 0.04 & 0.04 & 0.33 & 0.11 \\

\enddata

\tablecomments{
The 1$\sigma$ error is measured directly from the local spectrum
around each emission lines as the {\small rms} of the residual after
subtraction of both the stellar continuum and emission lines.
Lines with no ratio given were not within the
wavelength covered; 3$\sigma$ upper limits are given for lines within
this range that were not detected.
For details of the PDR models, see \cite{dav03}.
}

\end{deluxetable}
\clearpage
%----------------------------------------------------------------------

\begin{deluxetable}{lrrrrr}

\tabletypesize{\small}
\tablecaption{Parameters for the H$_2$ PDR Models\label{tab:models}}
\tablewidth{0pt}
\tablehead{

\colhead{model} & 
\colhead{$\chi$\tablenotemark{a}} & 
\colhead{$n_{\rm H}$\tablenotemark{b}} & 
\colhead{$T_{\rm max}$\tablenotemark{c}} & 
\colhead{$I_{\rm 1-0S(1)}$\tablenotemark{d}} &
\colhead{$I_{\rm H_2}$\tablenotemark{e}} \\

\colhead{} & 
\colhead{} & 
\colhead{(cm$^{-3}$)} & 
\colhead{(K)} & 
\colhead{(W\,m$^{-2}$\,sr$^{-1}$)} &
\colhead{(W\,m$^{-2}$\,sr$^{-1}$)} \\

}
\startdata

1 & $10^2$ & $10^3$ & $10^2$ & $1.29\times10^{-9}$ & $7.06\times10^{-8}$ \\
2 & $10^3$ & $10^4$ & $10^3$ & $5.17\times10^{-8}$ & $2.88\times10^{-6}$ \\
3 & $10^3$ & $10^4$ & $2\times10^3$ & $2.00\times10^{-6}$ & $3.22\times10^{-5}$ \\
4 & $10^5$ & $10^4$ & $10^3$ & $7.41\times10^{-8}$ & $4.04\times10^{-6}$ \\
5 & $10^5$ & $10^6$ & $10^3$ & $8.82\times10^{-7}$ & $1.80\times10^{-5}$ \\

\enddata

\tablenotetext{a}{FUV intensity relative to that in the local
  interstellar medium ($2.1\times10^{11}$\,photons\,s$^{-1}$\,m$^{-2}$;
  \cite{dra78})}.

\tablenotetext{b}{gas density $n_{\rm H} = n({\rm H}) + 2n({\rm H_2})$.}

\tablenotetext{c}{For model 1 $T=100$\,K at all cloud depths. 
For the other models, the temperature varies as given in
Eq.~(1) of \cite{dav03}.}

\tablenotetext{d}{1-0\,S(1) intensity predicted by the
model.
Intensities for other lines can be found by using the model line
ratios given in Table~\ref{tab:ratios}.}

\tablenotetext{e}{Total intensity summed over all H$_2$ lines
  predicted by the model.}

\end{deluxetable}

%----------------------------------------------------------------------

\clearpage

%----------------------------------------------------------------------

%----------------------------------------------------------------------

%----------------------------------------------------------------------

%----------------------------------------------------------------------

%----------------------------------------------------------------------

%----------------------------------------------------------------------

%----------------------------------------------------------------------

%----------------------------------------------------------------------

%----------------------------------------------------------------------

%----------------------------------------------------------------------

%----------------------------------------------------------------------

%----------------------------------------------------------------------

\end{document}